\documentclass[usenatbib]{mn2e} 
\usepackage{longtable} 
\usepackage{graphicx}

\def\ubvri{$BVRI$\,}
\def\CLN{ NGC 1767, NGC 1994, NGC 2002, NGC 2003, NGC 2006, SL 538, NGC 2011,
   NGC 2098 and NGC 2136}

\newcommand{\eg}{{\textrm e.g.}}

\def\aj{AJ}
\def\apj{ApJ}
\def\apjs{ApJS}
\def\aap{A\&A}
\def\aaps{A\&AS}
\def\jrasc{JRASC}
\def\mnras{MNRAS}
\def\nar{New A Rev.}
\def\pasp{PASP}

\title[Mass function of LMC star clusters]
      {CCD photometric and mass function study of
        9 young Large Magellanic Cloud star clusters
        \thanks{Based on observations collected at the European Southern 
                Observatory, Chile.}
      }

\author[Kumar et al.]
   { B. Kumar$^{1,2}$, 
     R. Sagar$^{2}$ and 
     J. Melnick$^{3}$ \\
     $^{1}$ Departamento de F\'isica, Universidad de 
      Concepci\'on, Casilla 160-C, Concepci\'on, Chile (bkumar@astro-udec.cl)\\
     $^{2}$ Aryabhatta Research Institute of Observational Sciences, 
            Manora Peak, Nainital 263 129, India (sagar@aries.ernet.in)\\ 
     $^{3}$ European Southern Observatory, Alonso de C\'{o}rdova 3107,
           Casilla 19001, Vitacura, Santiago, Chile (jmelnick@eso.org)
   } 

\begin{document}

\date{\today}

\pagerange{\pageref{firstpage}--\pageref{lastpage}} \pubyear{2007}

\maketitle

\label{firstpage}

\begin{abstract}
   We present CCD photometric and mass function study of 9 young Large 
   Magellanic Cloud star clusters namely \CLN. The \ubvri data reaching 
   down to $V \sim$ 21 mag, are collected from 3.5-meter NTT/EFOSC2 in 
   sub-arcsec seeing conditions. For NGC 1767, NGC 1994, NGC 2002, NGC 2003, 
   NGC 2011 and NGC 2136, broad band photometric CCD data are presented for 
   the first time. Seven of the 9 clusters have ages  between 16 to 25 Myr 
   while remaining two clusters have ages $32\pm4$ Myr (NGC 2098) and 
   $90\pm10$ Myr (NGC 2136). For 7 younger clusters, the age estimates 
   based on a recent model and the integrated spectra are found to be 
   systematically lower ($\sim$ 10 Myr) from the present estimate. In the 
   mass range of $\sim 2 - 12$ $M_{\odot}$, the MF slopes for 8 out of nine 
   clusters were found to be similar with the value of $\gamma$ ranging 
   from $-1.90\pm0.16$ to $-2.28\pm0.21$. For NGC 1767 it is flatter 
   with  $\gamma = -1.23\pm0.27$. Mass segregation effects are observed 
   for NGC 2002, NGC 2006, NGC 2136 and NGC 2098. This is consistent with 
   the findings of \citet{kontizas98} for NGC 2098. Presence of mass 
   segregation in these clusters could be an imprint of star formation 
   process as their ages are significantly smaller than their dynamical 
   evolution time. Mean MF slope of $\gamma = -2.22\pm0.16$ derived for a 
   sample of 25 young ($\le 100$ Myr) dynamically unevolved LMC stellar 
   systems provide support for the universality of IMF in the intermediate 
   mass range $\sim 2-12\ M_{\odot}$.
\end{abstract}



\section{Introduction} \label{sec:intro}

  The distribution of stellar masses that form in one star-formation event in 
  a given volume of space is called initial mass function (IMF). Some 
  theoretical studies consider that the IMF should vary with the pressure and 
  temperature of the star-forming cloud in such a way that higher-temperature 
  regions ought to produce higher average stellar masses while others have 
  exactly opposite views (see \citealt{larson98, elmegreen00} and references 
  therein). Detailed knowledge of the IMF shape in different star forming
  environments is therefore essential. One would like to know whether it is 
  universal in time and space or not. For this, in a galaxy, its young (age 
  $\le$ 100 Myr) star clusters of different ages, abundance etc. need to be 
  observed, as they contain dynamically unevolved, (almost) coeval sets 
  of stars at the same distance with the same metallicity. For a number of 
  such reasons, populous young star clusters of the Large Magellanic Clouds 
  (LMC) are the most suitable objects for investigating the IMF. They provide 
  physical conditions not present in our Galaxy e.g. stellar richness, 
  metallicity and mass ranges (see \citealt{sagar93, sagar95} and references 
  therein). Unlike the galactic counterparts, where corrections for 
  interstellar absorption are not always trivial since it could be large 
  as well as variable \citep{sagar87, yadav01, kumar04}, for LMC star 
  clusters it is relatively small. Its treatment is therefore not a problem. 
  Furthermore, choosing young (age $\le 100$ Myr) clusters reduces the effects 
  of dynamical evolution on their MF. The present day mass function of these 
  stellar systems can therefore be considered as the IMF. The study of young 
  LMC star clusters is thus important for providing the answer to the 
  question of universality of the IMF. Both ground and Hubble Space Telescope 
  (HST) observations have therefore been obtained (see \citealt{sagar91a, 
  brocato01, matteucci02} and references therein) for a few of the large 
  number of young LMC star clusters \citep{bica99}. The potential offered by 
  them  has not been fully utilised as still a large fraction of them are 
  unobserved. 

  \begin{figure}
  \centering
  \includegraphics[width=9cm]{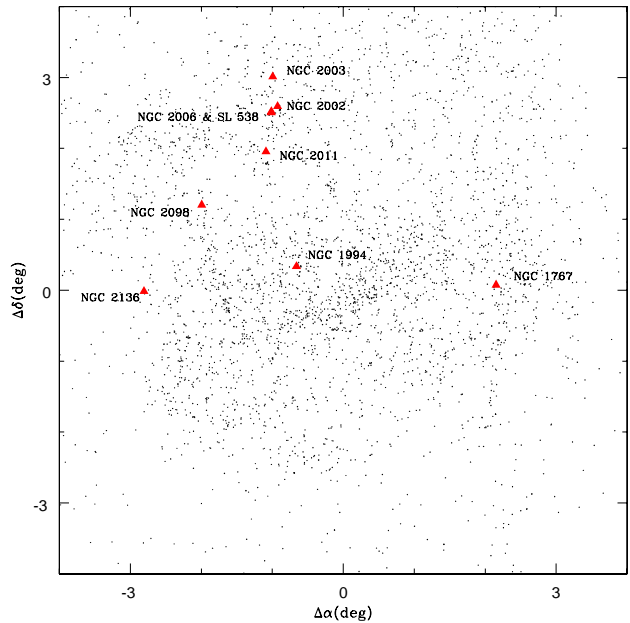}
  \caption{Small dots show the location of identified LMC star clusters 
     from the catalog of \citet{bica99}. A sky area of about 
     8\degr $\times$ 8\degr is shown centered around the optical center 
     ($\alpha_{\rm J2000} =5^h 20^{m} 56^{s}$, 
     $\delta_{\rm J2000} = -69^{\degr} 28^{\arcmin} 41^{\arcsec}$) of the LMC. 
     The bar region is clearly seen. The target clusters are shown with 
     filled triangles.}
  \label{fig:radec}
  \end{figure}

  In this paper we derive mass function (MF) slopes using new broad band 
  $BVRI$ CCD photometric observations of the stars in 9 young LMC star clusters
  namely \CLN. Their integrated photometric colours indicate that all of them 
  belong to SWB \citep{searle80} class of 0 or I and hence are very young with 
  ages $\le 30$ Myr \citep{elson95, bica96} except NGC 2136 which belongs to 
  SWB class of III indicating and age between 70\,-\,200 Myr. 
  Table \ref{tab:sample} lists the relevant information available prior to 
  this study. All the clusters are rich indicating higher stellar 
  density \citep{bica95} and thus most suitable for the MF study. Except 
  NGC 2011, all are elliptical in size with major axes diameters ranging 
  from 1\farcm3 to 2\farcm8. Except for NGC 2002 and NGC 2098, all others 
  are candidate members of either a pair or a multiple
  system (see \citealt{dieball02}). Locations of the target clusters are
  shown in Fig. \ref{fig:radec}. Most of them lie towards north-east side 
  of the LMC bar harboring young star forming regions in contrast to the 
  intermediate age (1 - 3 Gyr) cluster field of the bar. NGC 1767 lies 
  south-west region of the LMC bar. Being spread over a wide region 
  ($\sim$ 5\degr $\times$ 10\degr), the sample may reflect 
  different star forming environments. It is therefore suitable for testing
  the universality of the IMF. When CCD observations were carried 
  out in 1990, no detailed photometric observations and MF studies had been 
  published. However, in the mean time some CCD photometric observations have 
  been published for a few of the clusters under study. A brief description of 
  the previous work on the clusters under study is given below.

   \begin{table*}
   \centering
   \caption{Preliminary information about the clusters under study. Cluster 
     identifications are from \citet{sulentic73} with acronym NGC and 
     \citet{shapley63} with acronym SL. The coordinates, major and 
     minor diameters and position angle (P.A.) are taken from \citet{bica99} 
     while age and membership of pair (mP) or multiple (mM) system along with 
     group number given inside the bracket in the last column are taken 
     from \citet{dieball02}. All the clusters are of type C indicating 
     higher stellar density \citep{bica95}.}
   \label{tab:sample}

   \begin{tabular}{lccc crc r}      \hline
      Clusters & $\alpha_{\rm J2000}$&$\delta_{\rm J2000}$& D$_{\rm maj}$  
      & D$_{\rm min}$ & P.A. & age(Myr) & Remarks \\  \hline
      NGC1767&4$^h 56^{m} 27^{s}$& $-69\degr 24\arcmin 
      12\arcsec$ &1\farcm30 & 1\farcm20& 30\degr &0-30&mM(39)\\
      NGC1994 & 5~ 28~ 21
      & $-69~ 08~ 30$ &  1.60 & 1.50& 170~ & 10-30 & mM(275) \\
      NGC2002 & 5~ 30~ 21
      & $-66~ 53~ 02$ &  1.90 & 1.70& 20~ &10-30 & \\
      NGC2003 & 5~ 30~ 54
      & $-66~ 27~ 59$ &  1.70 & 1.40& 70~ & 10-30& mP(297) \\
      SL538 & 5~ 31~ 18
      & $-66~ 57~ 28$ &  1.40 & 1.40&  & 18$\pm$2&mP(303)  \\
      NGC2006& 5~ 31~ 19
      & $-66~ 58~ 22$ &  1.60 & 1.40& 140~ & 22.5$\pm$2.5&mP(303)  \\
      NGC2011& 5~ 32~ 19
      & $-67~ 31~ 16$ &  1.00 & 1.00&  & 10 - 30&mP(320)  \\
      NGC2098 & 5~ 42~ 30
      & $-68~ 16~ 29$ &  2.20 & 2.00&140~  & 10 - 30 & \\
      NGC2136 & 5~ 52~ 59
      & $-69~ 29~ 33$ &  2.80 & 2.50&140~  & 70-200&mP(456)  \\ \hline
   \end{tabular}
   \end{table*}

\subsection{Previous work}

  \begin{itemize}

  \item  {\bf NGC 1767.} This, a member of triple star cluster system, is  
  located in the OB association LH 8. Integrated ($U-B$) and ($B-V$) colors
  indicate that the cluster is young with an age of $\sim$ 10 Myr.
 
  \item {\bf NGC 1994.} This, located in LMC DEM 210 region, is a member of a 
  5-cluster system. Its irregular size is largest amongst them. An age of about 
  5 - 30 Myr has been derived for the cluster from its integrated photometric 
  colour observations.
 
  \item {\bf NGC 2002.} This single cluster is  located in the OB association 
  LH 77 in the supergiant shell LMC 4 region. The cluster center is 
  condensed,  but the outer part is resolved. Integrated light observations 
  indicate an age of $\sim$ 10 - 30 Myr along with the presence of a few red 
  supergiants \citep{bica96}.
 
  \item {\bf NGC 2003.} Integrated photometric observations indicate an age of
  10 - 30 Myr for this cluster located in the Shapley III region of the LMC. 
  Its shape on the photographic image is elongated with resolved outer parts.
 
  \item {\bf NGC 2006 and SL 538.} This binary star cluster is located in the 
  northwestern  part of the OB association LH 77 in supergiant shell LMC 4. 
  The clusters are separated by $\sim 55''$ on the sky corresponding to a 
  linear separation of 13.3 pc at the distance of LMC. Integrated photometric
  observations obtained by \citet{bhatia92} and \citet{bica96} indicate a 
  similar age for both the clusters. Using low-resolution objective prism 
  spectra and integrated IUE spectra, \citet{kontizas98} suggested that this 
  binary cluster may merge in $\sim$ 10 Myr. Broad band and H$_{\alpha}$ CCD 
  photometric observations were obtained  by \citet{dieball98}. Based on the 
  colour-magnitude diagrams (CMDs) of the clusters they derived an age of 
  $18\pm2$ Myr for SL 538 and of $22.5\pm2.5$ Myr for NGC 2006. The MF slopes 
  obtained for both the clusters were consistent with that 
  of \citet{salpeter55} and indicated similar total masses. These studies 
  thus indicates near-simultaneous formation of the cluster pair in the 
  same giant molecular cloud.
  
  \item  {\bf NGC 2011.} This is located in the OB association region LH 75.  
  Its age estimated from the integrated photometric observations is between
  10 to few tens of Myr. Its photographic image indicates that it is elongated, 
  fairly condensed and partly resolved cluster. A recent analysis of its
  stellar content using HST observations reveal that it has two parallel main 
  sequence branches, and may be a binary system \citep{gouliermis06}. However,
  the analysis also indicate that both populations might have formed in a 
  single star forming event as the redder stars are situated in the central 
  half arcmin region and are thought to be embedded in the dust and gas, while 
  the blue stars are spread in the outer region up to 1 arcmin.  
 
  \item {\bf NGC 2098.} This is another single cluster amongst the objects 
  under study. The first $BR$ broad band CCD photometric observations have been 
  presented by \citet{kontizas98}. They derived an age of 63 - 79 Myr and
  found strong evidence for mass segregation in agreement with their earlier
  studies based on the photographic observations. However, poor quality of 
  their CCD data was indicated by the authors.
 
  \item  {\bf NGC 2136.} This is the brighter component of the young binary 
  globular cluster NGC 2136/ NGC 2137 in the LMC. The angular separation 
  between the components is about 1\farcm3. \citet{hilker95} using Stromgren 
  CCD photometry of the clusters indicates their common origin. They indicate 
  an age of 80 Myr and metallicity [Fe/H] = $-0.55\pm0.06$ dex for the cluster 
  while \citet{dirsch00} derive an age of $100\pm20$ Myr but the same 
  metallicity. The cluster contains a number of Cepheids as well as red giants.

  \end{itemize}
 
  The present CCD observations, in combination with earlier observations, have
  been used to estimate and/or interpret the interstellar reddening to the 
  cluster regions, ages and mass functions of the clusters. Section 2 deals 
  with the observational data, reduction procedures and comparisons with the 
  published photometric data. In section 3, we analyse the stellar surface 
  density profiles, CMDs and MFs of the sample clusters. Last section is 
  presented with the results and discussions. 

  \begin{table}
  \centering
  \caption{Observing Log of the CCD data taken for nine young LMC clusters 
  during 1990. The suffix 'F' in the "Object" column 
  refers to the field region. The last column provides the number of stars (N) 
     measured in different passbands. The seeing refers to the mean Gaussian 
     FWHM of the stars.}
  \label{tab:obslog}

  \begin{tabular}{lllrrr}  \hline
     Object   & Date            &  & Exp  &  Seeing  & N \\  
              & (Year 1990)     &        & (s)& (\arcsec) & \\ \hline
     NGC 1767 & Jan 10/11 & B & 20             &1.05& 632  \\
              &           & V & 10             &0.77&  \\
              &           & R & 10             &0.81&  \\ 
     NGC 1994 & Jan 10/11 & B & 20             &1.02&1156 \\
              &           & V & 10,20          &0.84& \\
              &           & R & 10,20$\times$3 &0.81& \\
              &           & I & 5              &0.81& \\ 
     NGC 2002 & Jan 10/11 & B & 40             &0.98& 799\\
              &           & V & 20             &0.72& \\
              &           & R & 15             &0.88& \\ 
              &           & I & 10,20          &0.86& \\
     NGC 2002F& Jan 10/11 & B & 20             &1.00& 337\\
              &           & V & 10             &0.84& \\
              &           & R & 10             &0.72& \\ 
     NGC 2003 & Jan 10/11 & B & 60             &0.81& 729\\
              &           & V & 20             &0.74& \\
              &           & R & 20             &0.72& \\ 
              &           & I & 10             &0.72& \\ 
     NGC 2006 & Jan 10/11 & B & 20             &0.93& 747\\
      and     &           & V & 10             &0.86& \\
     SL 538   &           & R & 10             &1.02& \\ 
              & Jan 12/11 & B & 45             &0.91& \\
              &           & V & 20, 30$\times$2&1.00& \\
              &           & R & 8$\times$2     &0.88& \\ 
     NGC 2011 & Jan 10/11 & B & 60             &0.91& 610\\
              &           & V & 20             &0.84& \\
              &           & R & 15             &0.81& \\ 
     NGC 2098 & Jan 10/11 & B & 180            &1.02& 686\\
              &           & V & 60             &0.86& \\
              &           & R & 15             &0.81& \\
     NGC 2136 & Jan 12/13 & B & 60             &1.00& 1266\\
              &           & V & 30             &0.93& \\
              &           & R & 15             &0.86& \\ 
     NGC 2136F& Jan 13/14 & B & 60             &1.09& 270\\
              &                 & V & 30             &0.88& \\
              &                 & R & 15$\times$2    &0.86& \\ \hline 
  \end{tabular}
  \end{table}


\section{Observations and Data reductions} \label{sec:obs}

   The observations, procedures for data reductions are described in this 
   section along with the photometric accuracy and comparison with published 
   photometry.
 
\subsection{Photometric data}

  The broad band $BVRI$ CCD photometric observations were carried out at
  the European Southern Observatory (ESO), La Silla, Chile, in 1990 between 
  January 10 and 13 using the ESO Faint Object Spectrograph and 
  Camera-2 (EFOSC-2) mounted at the Nasmyth focus of the 3.5-m New Technology 
  Telescope (NTT). The filters used in these observations were standard Bessel 
  BVR (ESO\#583, 584, 585) and Gunn {\it i} (ESO\#618). At the focus 
  of the telescope, a 27$\mu$m square pixel of the 512$\times$512 size 
  Tektronix CCD (\#16) chip corresponds to $\sim$ 0\farcs23 and entire chip 
  covers a square area of side $\sim$ 2\farcm0 on the sky. The read out noise 
  for the system was 14 e with a gain factor of 5.5 e/ADU. During our 
  observations the seeing varied from about 0\farcs7 to 1\farcs2 (see 
  Table \ref{tab:obslog}) with a mean value of 1\farcs0 
  for $B$ band and about 0\farcs8 for $V, R$ and $I$ bands. We obtained only 
  one image for all clusters, as the CCD size was large enough to cover the 
  entire region of the compact clusters. In the case of NGC 2002 and the binary 
  cluster NGC 2136, we have also imaged a field region located about $3\arcmin$ 
  away from the center of clusters. Table \ref{tab:obslog} 
  lists the log of observations. Most of the observations were taken during 
  commissioning phase of the EFOSC2 when the instrument rotator was still not 
  available. As the image on NTT rotates during exposures at rates 
  which depends upon the position on the sky, only exposures of up to at 
  most few minutes were possible. Bias frames were taken intermittently. 
  Flat-field exposures were made of the twilight sky. Dark current frames 
  were also secured.
 
  Nine \citet{landolt92} standards covering a range in brightness 
  ($11.4 < V < 13.1$) as well as in colour ($-0.13 <(V-R)< 0.67$) were 
  observed for calibration purposes. The excellent photometric quality of 
  the sky during the observations have ensured the accuracy of the data 
  presented here.

\subsection {Reductions}

  \begin{table}
  \caption{Internal photometric errors in magnitudes as a function of 
      brightness in NGC 2002 cluster region. $\sigma$ is the standard 
      deviation per observations in magnitude.}
  \label{tab:photerr}
  \centering

  \begin{tabular}{cc ccc} \hline 
     Magnitude range& $\sigma_B$& $\sigma_V$& $\sigma_R$& $\sigma_I$ \\ \hline
     $\le 14.0$ &  0.001 & 0.003 & 0.012 & 0.02  \\
     14.0$-$15.0&  0.005 & 0.02 & 0.03 & 0.04 \\
     15.0$-$16.0&  0.01 & 0.02 & 0.03 & 0.04 \\
     16.0$-$17.0&  0.02 & 0.03 & 0.04 & 0.04 \\
     17.0$-$18.0&  0.04 & 0.04 & 0.05 & 0.05 \\
     18.0$-$19.0&  0.05 & 0.05 & 0.07 & 0.07 \\
     19.0$-$20.0&  0.07 & 0.08 & 0.08 & 0.08 \\ \hline
  \end{tabular}
  \end{table}

  \begin{table*}
  \centering
  \caption[width=\textwidth]{Relative positions (X, Y), CCD magnitude ($V$) and colors 
  ($B-V$, $V-R$ and  $V-I$) of all the
  stars measured in cluster and nearby field regions are presented sequentially for the
  clusters NGC 1767, NGC 1994, NGC 2002, NGC 2002F, NGC 2003, NGC 2006 (including SL 538), NGC 2011,
  NGC 2098, NGC 2136 and NGC 2136F. Along with the star ID, the cluster or field region
  identification is also provided in the first column. The full version of the table 
  is available as online material.}
  \label{tab:photmag}

  \begin{tabular}{rrrrrrr} \hline 
    ID&      X&      Y& $V$&  $B-V$& $V-R$&           $V-I$\\ 
  \noalign{\smallskip} \hline
n1767\_1    &  241.02& 163.95& $12.98\pm 0.01$& $ 0.22\pm 0.01$& $ 0.14\pm 0.01$&-\\
n1767\_2    &  251.32&  91.73& $13.04\pm 0.01$& $ 2.02\pm 0.01$& $ 1.02\pm 0.01$&-\\
n1767\_3    &  256.84& 147.23& $13.30\pm 0.01$& $ 1.91\pm 0.02$& $ 0.95\pm 0.01$&-\\
n1767\_4    &   60.10&  34.90& $13.46\pm 0.01$& $ 0.08\pm 0.01$& $ 0.12\pm 0.01$&-\\
    -&      -&      -& -&  -& -&           -\\ 
    -&      -&      -& -&  -& -&           -\\ 
  \noalign{\smallskip} 

  \hline
  \end{tabular}
  \end{table*}

  \begin{figure}
  \centering
  \includegraphics[width=8cm]{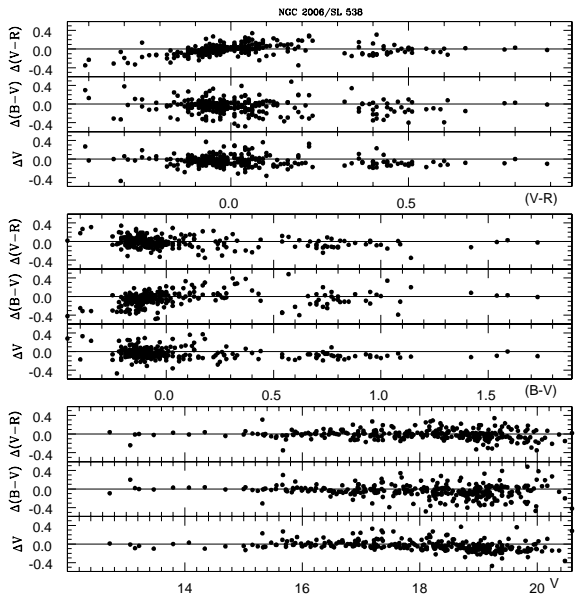}
  \caption{Comparison of the present photometry in NGC 2006/SL 538 with 
     those of \citet{dieball98}. The differences denote present minus 
     literature data.}
  \label{fig:n2006dif}
  \end{figure}

  The data were reduced using computing facilities available at the 
  ARIES Observatory, Nainital. Initial processing of the data frames was done 
  in the usual manner using the IRAF/MIDAS data reduction package. The 
  flat-field frames were summed for each colour band. The evenness of 
  flat fields is better than a few percent in all the filters.
 
  The magnitude estimate of a star on the data frames has been done
  using DAOPHOT software \citep{stetson87, stetson92}. Further processing and
  conversion of these raw instrumental magnitudes into the standard
  photometric system have been done using the procedure outlined by
  \citet{stetson92}. The image parameters and errors provided by DAOPHOT
  were used to reject poor measurements. About 10\% of the stars were
  rejected in this process. The DAOMASTER program was used for cross
  identifying the stars measured on different frames of a cluster
  region. In those cases where brighter stars are saturated on deep
  exposure frames, their magnitudes have been taken only from the short
  exposure frames. Most of the stars brighter than V $\sim$ 10.5 mag
  could not be measured because they are saturated even on the shortest
  exposure frames. 

  In deriving the colour equations for the CCD system and evaluating the
  zero-points for the data frames, we have used mean values of atmospheric 
  extinction coefficients of the site viz 0.3, 0.2, 0.15 and 0.1 mag
  for $B, V, R$ and $I$ band respectively. The colour equations 
  for the CCD system are determined by performing aperture photometry on the 
  photometric standards. By fitting least square linear regressions in the 
  observed aperture magnitudes as a function of the standard photometric 
  indices, following colour equations are derived for the system:

  \[ B-V  = 1.219\pm0.024 (b-v)  - 1.113\pm0.028 \]
  \[ V-R  = 1.065\pm0.019 (v-r)  - 0.116\pm0.016 \]
  \[ V-I  = 1.062\pm0.010 (v-i)  + 1.101\pm0.014 \]
  \[ V-v  = 0.032\pm0.017 (V-R)  - 1.145\pm0.013 \]
 
  \noindent
  where $B, V, R$ and $I$ are the standard magnitudes provided 
  by \citet{landolt92}. The $b, v, r$ and $i$ are the CCD aperture 
  magnitudes. The RMS deviations of the Landolt standards around the 
  fitted magnitudes are found to be 0.033, 0.035, 0.027 and 0.026 mag 
  respectively for $B, V, R$ and $I$. For establishing the local standards, 
  we selected about 30 isolated stars in each field and used the DAOGROW 
  program for construction of the aperture growth curve required for 
  determining the difference between aperture and profile-fitting magnitudes. 
  These differences, together with the differences in exposure times and 
  atmospheric extinction, were used in evaluating zero-points for local 
  standards in the data frames. The zero-points are uncertain 
  by $\sim$ 0.013 mag in $B, V, R$ and $I$. 

  The internal errors estimated from the scatter in the individual measures 
  of different exposures in NGC 2002 cluster region are listed in 
  Table \ref{tab:photerr} as a function of magnitude 
  for all filters. The errors become large ($\ga 0.10$ mag) for stars fainter 
  than 20 mag. They can be considered as representative of the accuracy of 
  our photometry in all the cluster and field regions under study. The number 
  of stars measured in different photometric passbands in an imaged region 
  are given in Table \ref{tab:obslog}. The X and Y pixel coordinates as well 
  as $V, (B-V), (V-R)$, and $(V-I)$ CCD magnitudes of the stars observed 
  in the regions of \CLN\  are listed in Table \ref{tab:photmag}. Stars 
  observed by others have been identified in the last column of the table. 
  Only sample table is presented here; the entire table is available in 
  electronic version of the article and also from the authors.

  \begin{figure}
  \centering
  \includegraphics[width=8cm]{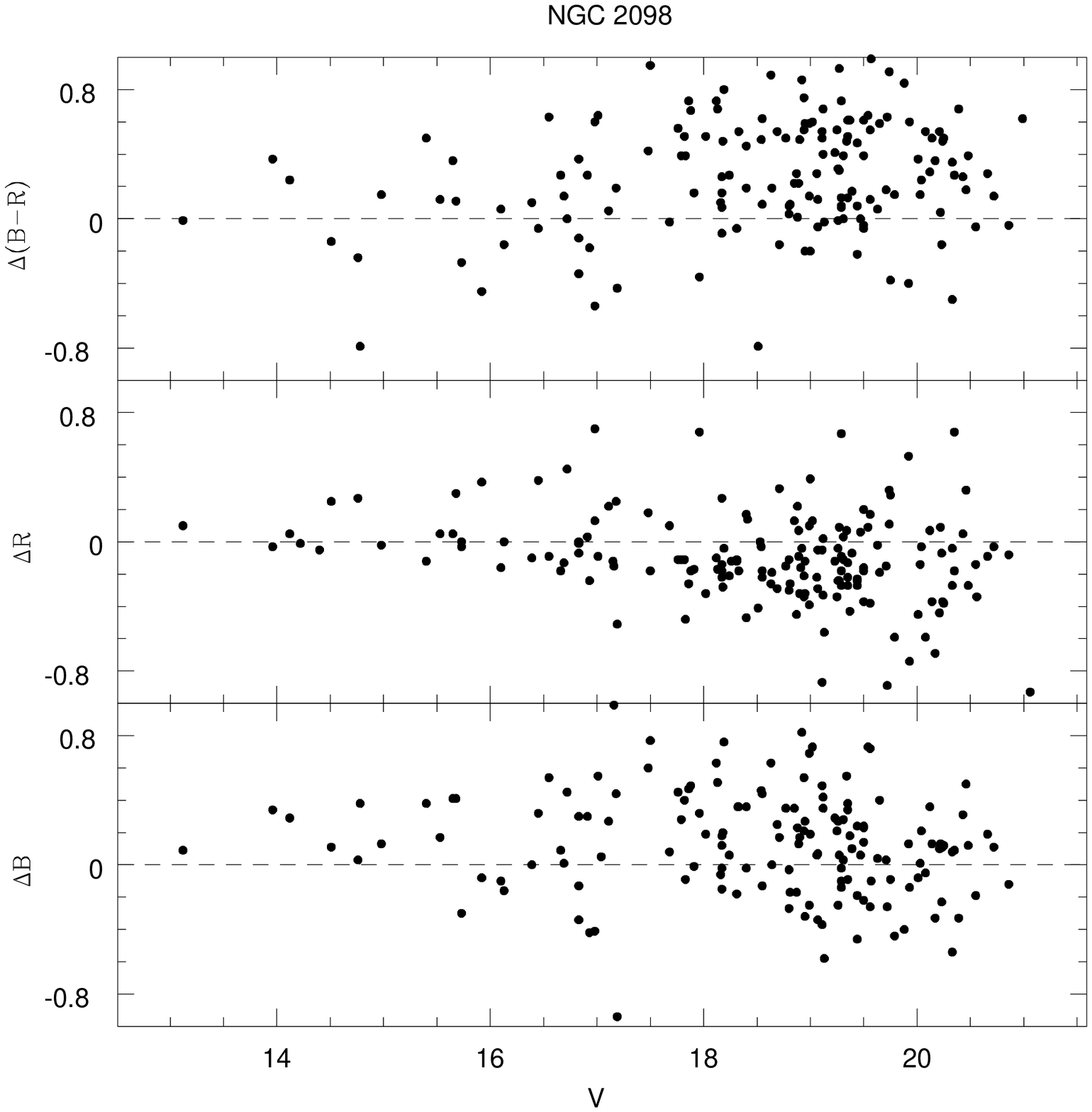}
  \caption{Comparison of the present photometry in NGC 2098 with those of
     \citet{kontizas98}. The differences denote present minus literature data.}
  \label{fig:n2098dif}
  \end{figure}

  \begin{table*}
  \centering
  \caption{Estimated coordinates of the cluster center ($X_{\rm c}, Y_{\rm c}$),
    cluster radii ($r_{\rm c}$) and the radii limits of the regions are tabulated.
    All the units are in pixels and one pixel corresponds to $\sim$ 0.23 arcsec on 
    the sky.} 
   \label{tab:radden}

  \begin{tabular}{ccc ccc c } \hline 
    Cluster &$X_{\rm c}$& $Y_{\rm c}$& $r_{\rm c}$ & Core & Ring 1 & Ring 2  \\ \hline
    NGC 1767 & 255 & 355 & 150 & $r < 30$& $30 \le r < 80$&  $80 \le r < 150$  \\
    NGC 1994 & 240 & 275 & 240 & $r < 30$& $30 \le r < 80$&  $80 \le r < 240$  \\
    NGC 2002 & 250 & 280 & 240 & $r < 30$& $30 \le r < 80$&  $80 \le r < 240$  \\
    NGC 2003 & 265 & 240 & 220 & $r < 15$& $15 \le r < 65$&  $65 \le r < 220$  \\
    NGC 2006 & 285 & 90  & 120 & $r < 10$& $10 \le r < 60$&  $60 \le r < 120$  \\
    SL 538   & 230 & 310 & 110 & $r < 20$& $20 \le r < 60$&  $60 \le r < 110$  \\
    NGC 2011 & 240 & 245 & 220 & $r < 30$& $30 \le r < 75$&  $75 \le r < 220$  \\
    NGC 2098 & 260 & 250 & 160 & $r < 25$& $25 \le r < 80$&  $80 \le r < 160$  \\
    NGC 2136 & 280 & 275 & 260 & $r < 35$& $35 \le r < 90$&  $90 \le r < 260$  \\ \hline
  \end{tabular}
  \end{table*}

\subsection {Photometric Comparisons}

  We compare the present CCD photometry with the published ones for the 
  clusters NGC 2006, SL538 and NGC 2098 in the following subsections. 
 
  {\bf NGC 2006 and SL 538.} The present photometry has 297 stars in common 
  with the CCD photometric data given by \citet{dieball98}. The plot 
  of the differences between the two data sets (see Fig. \ref{fig:n2006dif}) 
  indicate agreement between the V magnitudes of the two data sets and the 
  present photometry is $\sim$ 0.04 mag brighter in $V$ while the $B-V$ and 
  $V-R$ colour agrees fairly well. The RMS scatter in $\Delta V$, 
  $\Delta (B-V)$ and $\Delta (V-R)$ was found to be 0.08 mag, 0.12 mag and 
  0.09 mag respectively and it can be understood in terms of the error 
  present in both the photometries. There are a few outliers which appear 
  to be mostly stars located in the nucleus region of the cluster and were 
  treated as blended multiple stars in one of the photometry. We also note 
  that the plot of differences with the color show a small systematic trend 
  in $\Delta (V-R)$ and it is apparent also to some extent in $\Delta (B-V)$. 
  This may arise due to a second order color term from the B filter or due 
  to a minor calibration uncertainties present in one of the photometry. 

  {\bf NGC 2098.} There are 174 stars common between present photometry and 
  the $BR$ data given by \citet{kontizas98}. The differences between these 
  data are plotted in Fig. \ref{fig:n2098dif}. They indicate that there is 
  a constant difference between $B$ and $R$ magnitudes of the two data sets. 
  We suspect that poor observing conditions during the \citet{kontizas98} 
  observations may be responsible for the observed differences.

 
\section {Data Analysis} \label{sec:analysis}

  The photometric data of the clusters under study have been used to study 
  extent of the clusters along with their colour-magnitude diagrams (CMDs) 
  and mass function in the following sub sections.

  \subsection {Radial density profiles} \label{sec:radden}

  The spatial surface density profile of stars can be used to determine 
  cluster radius, $r_{\rm c}$, which is taken as the distance from the 
  cluster center where the average cluster 
  contribution becomes negligible with respect to the background stellar field. 
  It can also be used to estimate extent of field star contamination in the 
  cluster region. For this, first we derive the cluster center iteratively 
  by calculating average X and Y positions of stars within 150 pixels from 
  an eye estimated centre, until it converged to a constant value. An error 
  of about 10 to 20 pixels is expected in locating the cluster centre. The 
  (X,Y) pixel coordinates of the cluster centres are given in 
  Table \ref{tab:radden}. For determining the radial surface density of stars 
  in a cluster, the imaged area has been divided into a number of concentric 
  circles with respect to the above estimated cluster centre, in such a way 
  that each zone contains statistically significant numbers of stars. The 
  number density of stars, $\rho_i$, in $i^{th}$ zone has been evaluated as 
 
   $\rho_i = \frac{N_i}{A_i}$ 
 
  \noindent 
  where $N_i$ is the number of stars up to $V \sim$ 20 mag and $A_i$ is the 
  area of the $i^{th}$ zone. Wherever the zones cover only part of the imaged 
  cluster area, it has been accounted far in the determination of $A_i$.
  The assumed concentric circles and the stellar surface densities derived 
  in this way for the clusters under discussion are shown in Figs 4 to 6 and 8
  to 12. Presence of clear radius-density variation confirms the relatively 
  small diameters (compactness) of the star clusters under study.

  \begin{figure*}
  \centering
  \includegraphics[width=7cm]{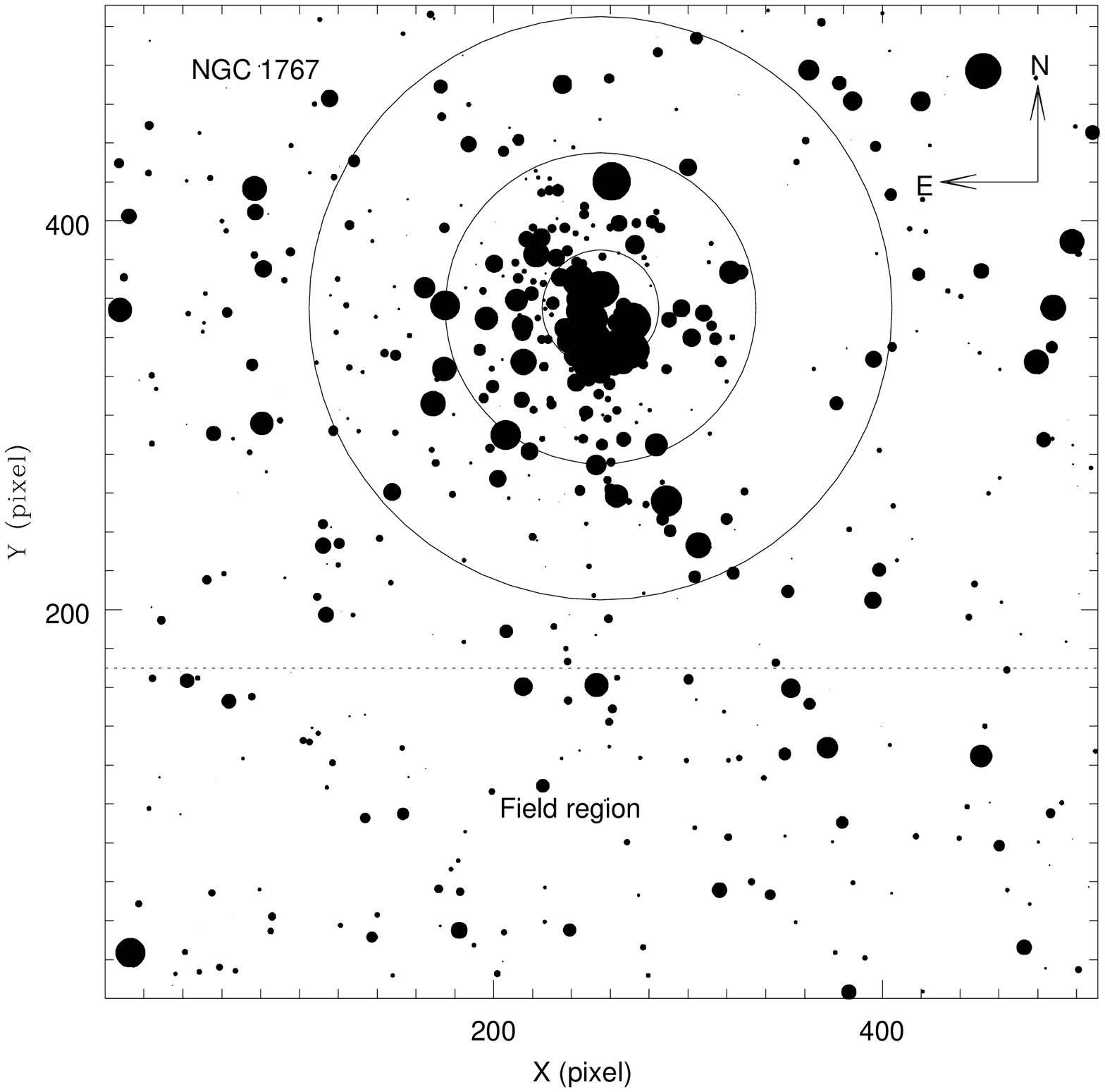}
  \includegraphics[width=7cm]{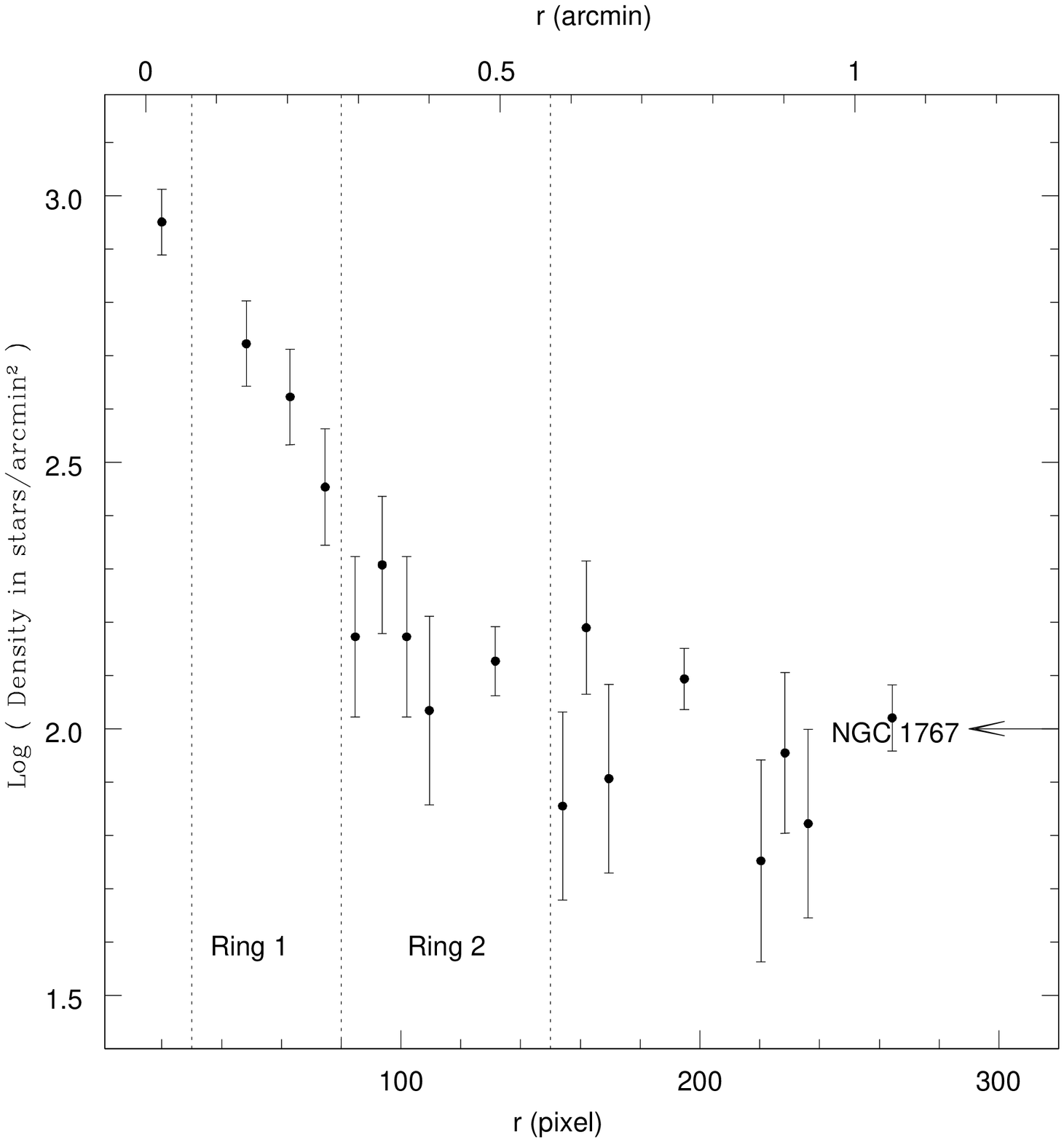}
  \includegraphics[width=7cm]{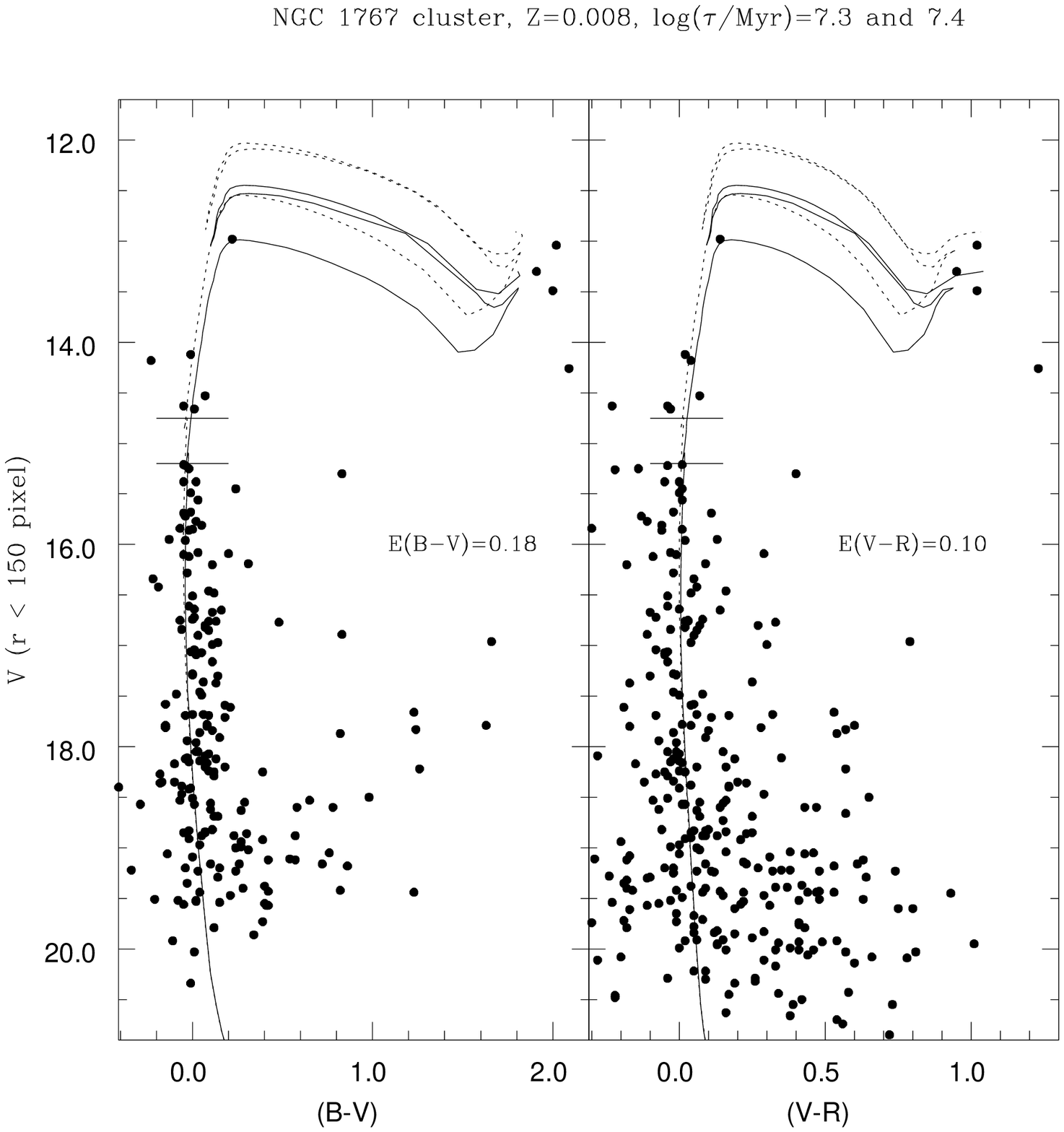}
  \includegraphics[width=7cm]{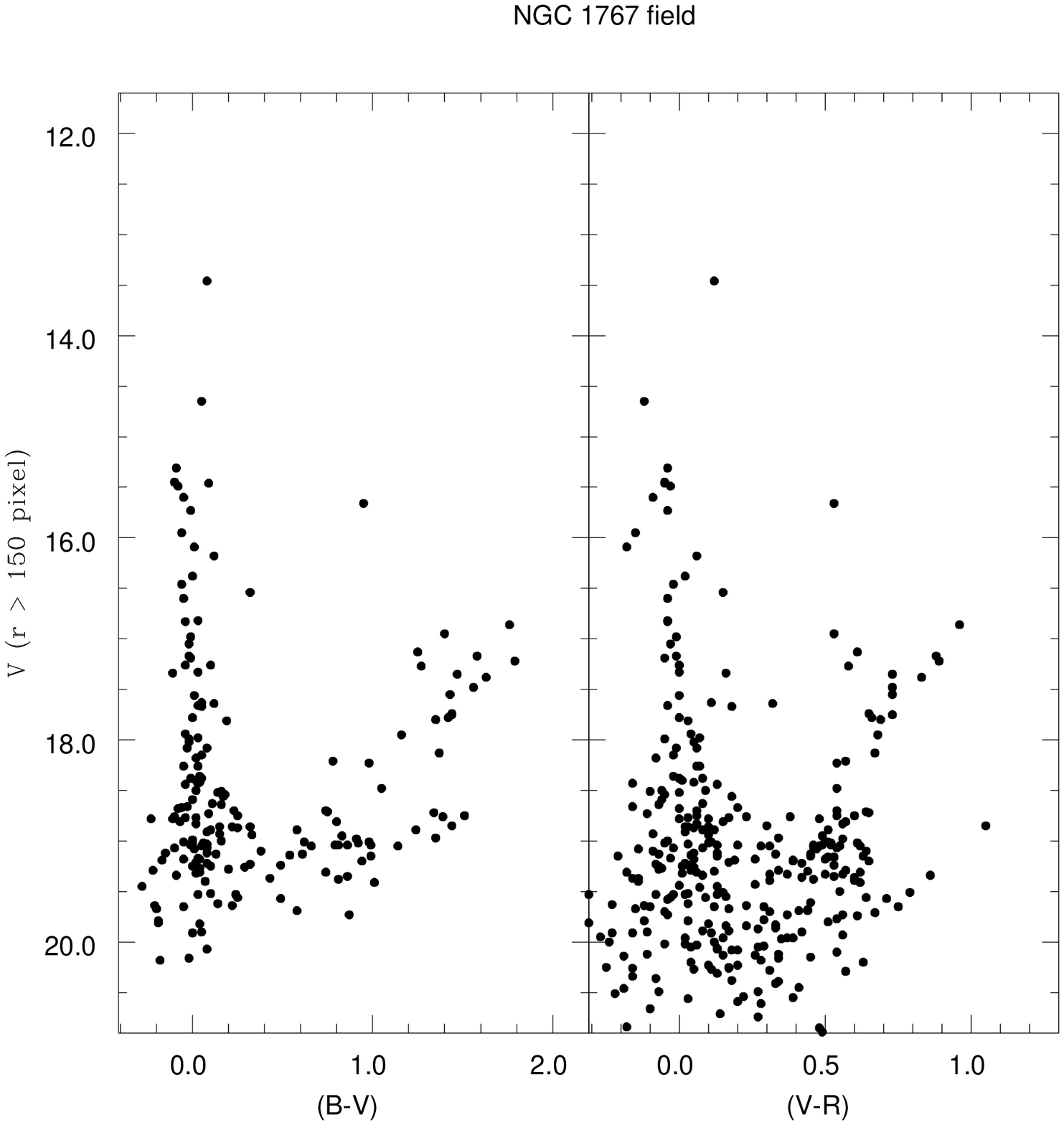}
  \caption{NGC 1767. {\it Top left} : Identification chart for the 
     observed $\sim 2\arcmin \times 2\arcmin$ region of NGC 1767 by EFOSC2/NTT. 
     The size of the filled circle is proportional to apparent magnitude in 
     such a way that brighter stars have larger sizes. North is 
     up and east is to left. Innermost ring defines core region while the 
     outermost ring represents cluster boundary (see Table \ref{tab:radden}). 
     Except NGC 2002 and NGC 2136, the field region is considered beyond the 
     outermost ring. {\it Top right} : Stellar surface density of stars 
     around cluster center is shown. Poisson errors are shown with vertical 
     bars. First and third dotted vertical lines represent core and cluster 
     radius respectively. Ring 1 and Ring 2 region are used for 
     MF determination. The horizontal arrow at the rightmost 
     corner of the plot shows field star density. {\it Bottom left} : The 
     CMDs for the cluster region ($r < r_{\rm c}$) and the two isochrones from 
     \citet{girardi02} confining the best age estimates are shown by dotted 
     (younger ishochrone) and solid (older isochrone) continuous curves. 
     LMC distance modulus of 18.5 and normal reddening law has been assumed. 
     The resulting color excess parameters are posted in the respective CMD 
     panel. The MS gaps (see Table \ref{tab:age}) are
     marked with horizontal bars. {\it Bottom right} : The CMDs for field
     region alone ($r > r_{\rm c}$) are shown. Reduced stellar density
     for MS as well as the red clump around $(V-R) \sim 0.5$ mag, 
     $(B-V) \sim 0.95$ mag and $V \sim 19.5$ mag are clearly seen.}  
     \label{fig:n1767}
  \end{figure*}

  \begin{figure*}
  \centering
  \includegraphics[width=7cm]{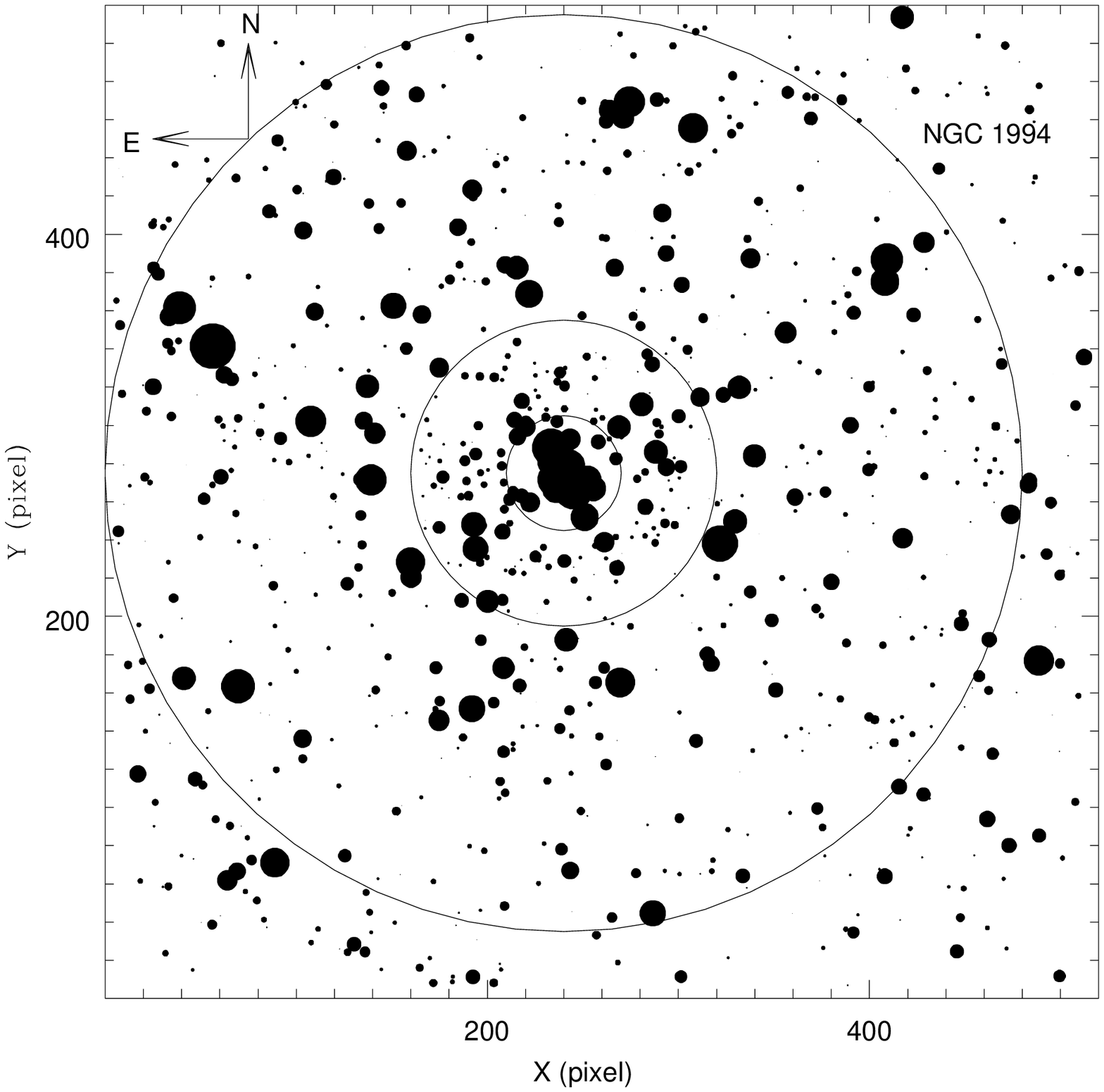}
  \includegraphics[width=7cm]{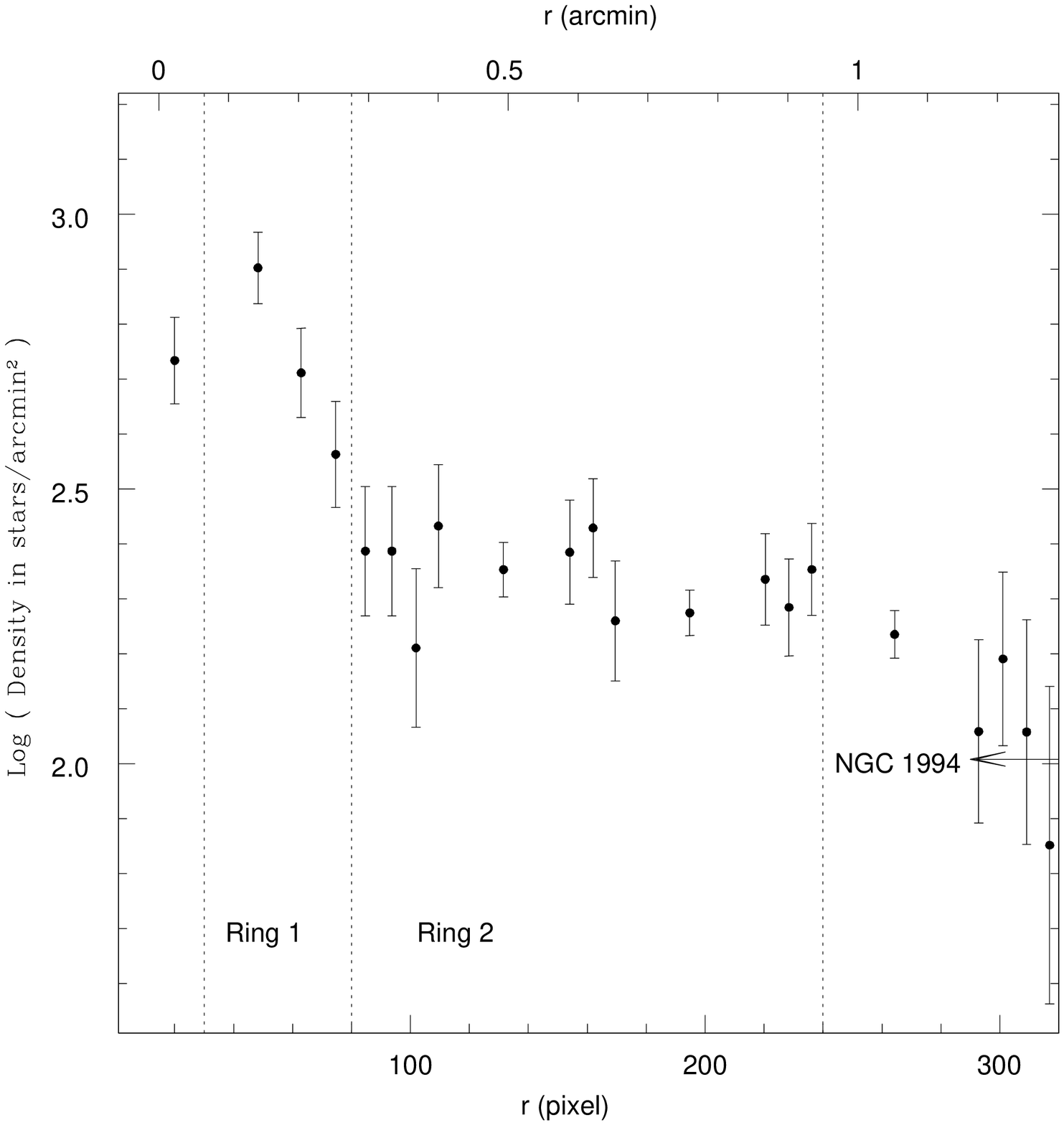}
  \includegraphics[width=7cm]{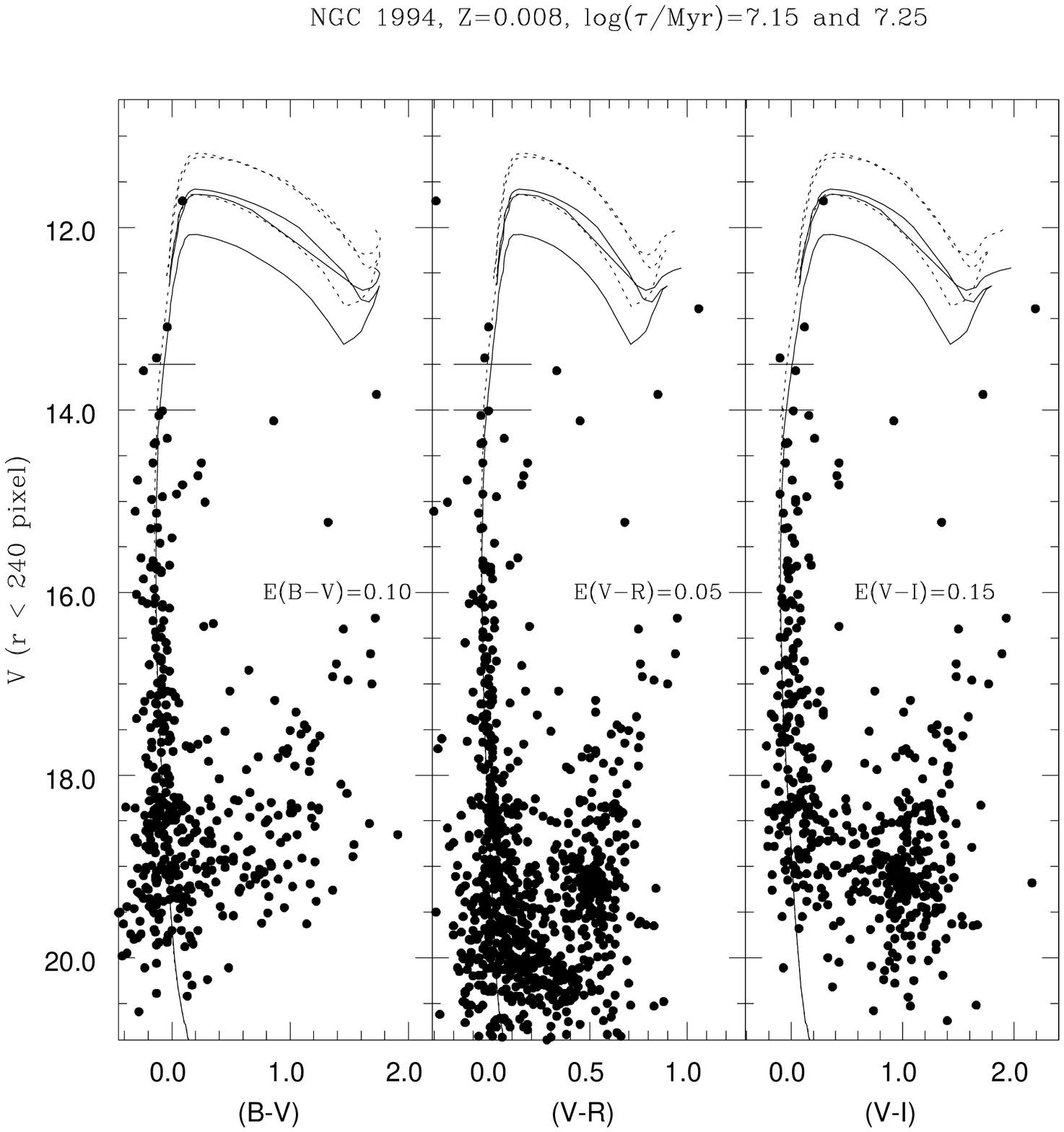}
  \includegraphics[width=7cm]{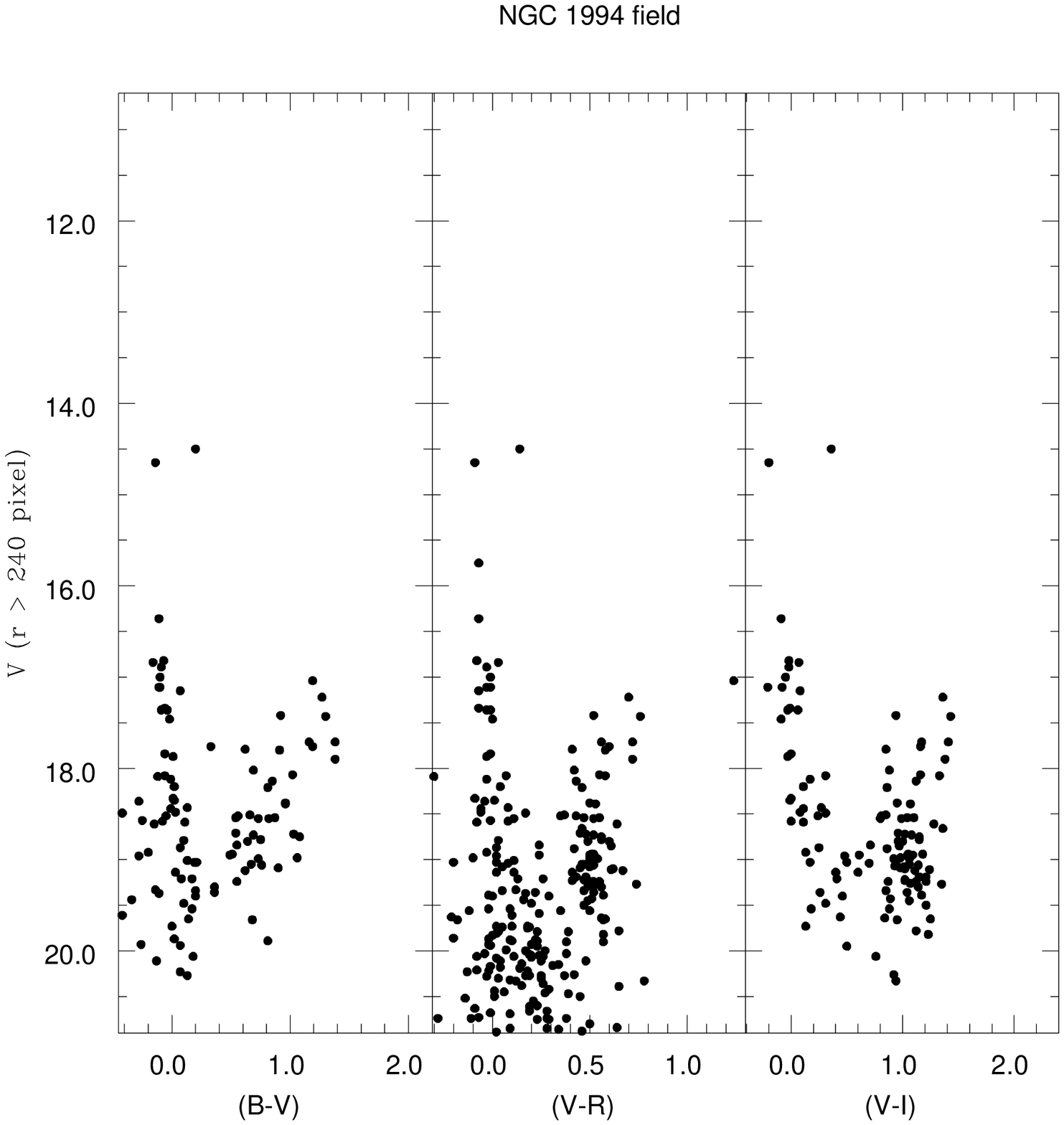}
  \caption{NGC 1994. Other descriptions are the same as for NGC 1767 
           in Fig. \ref{fig:n1767}.}
  \label{fig:n1994}
  \end{figure*}

  \begin{figure*}
  \centering
  \includegraphics[width=7cm]{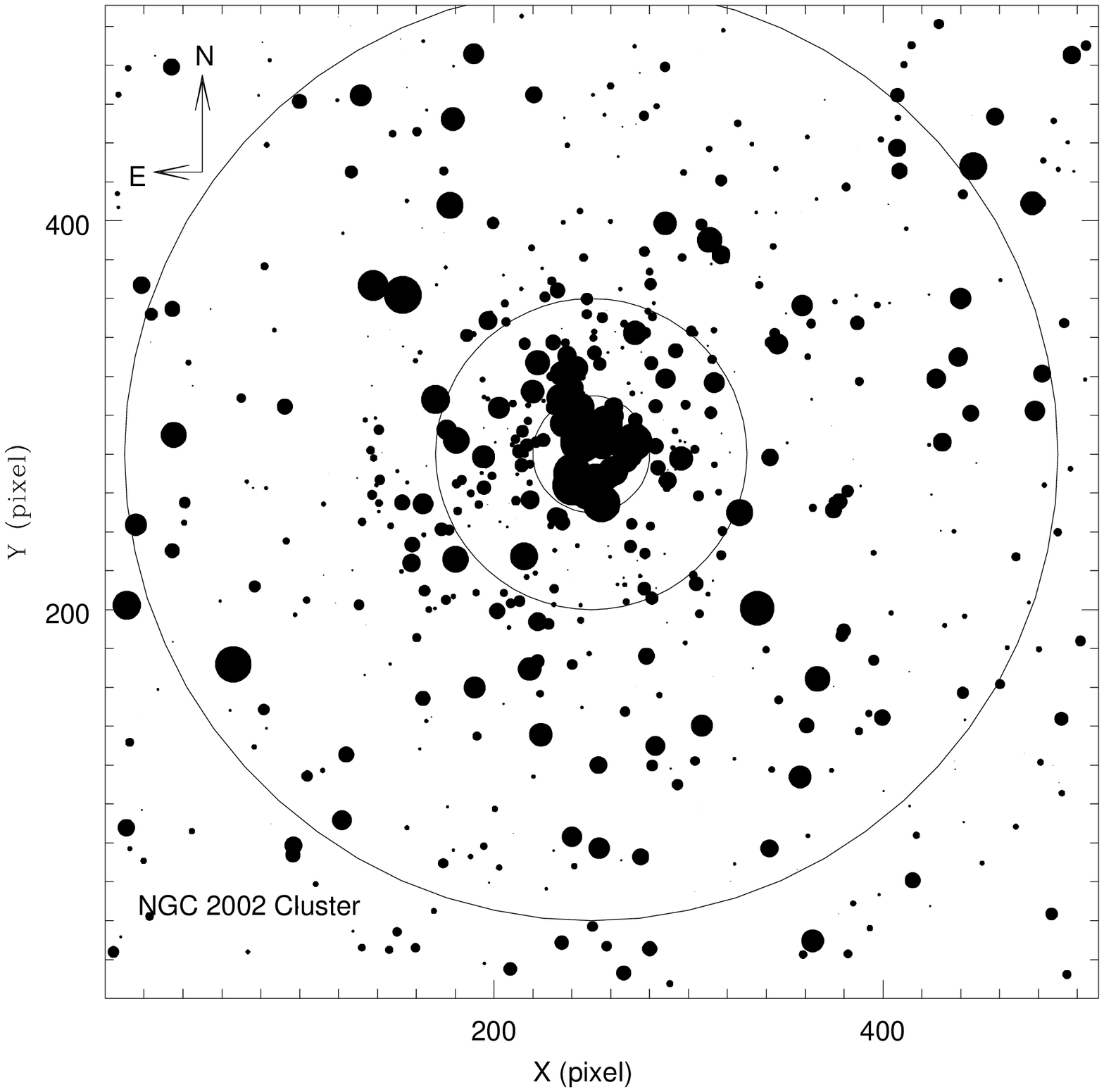}
  \includegraphics[width=7cm]{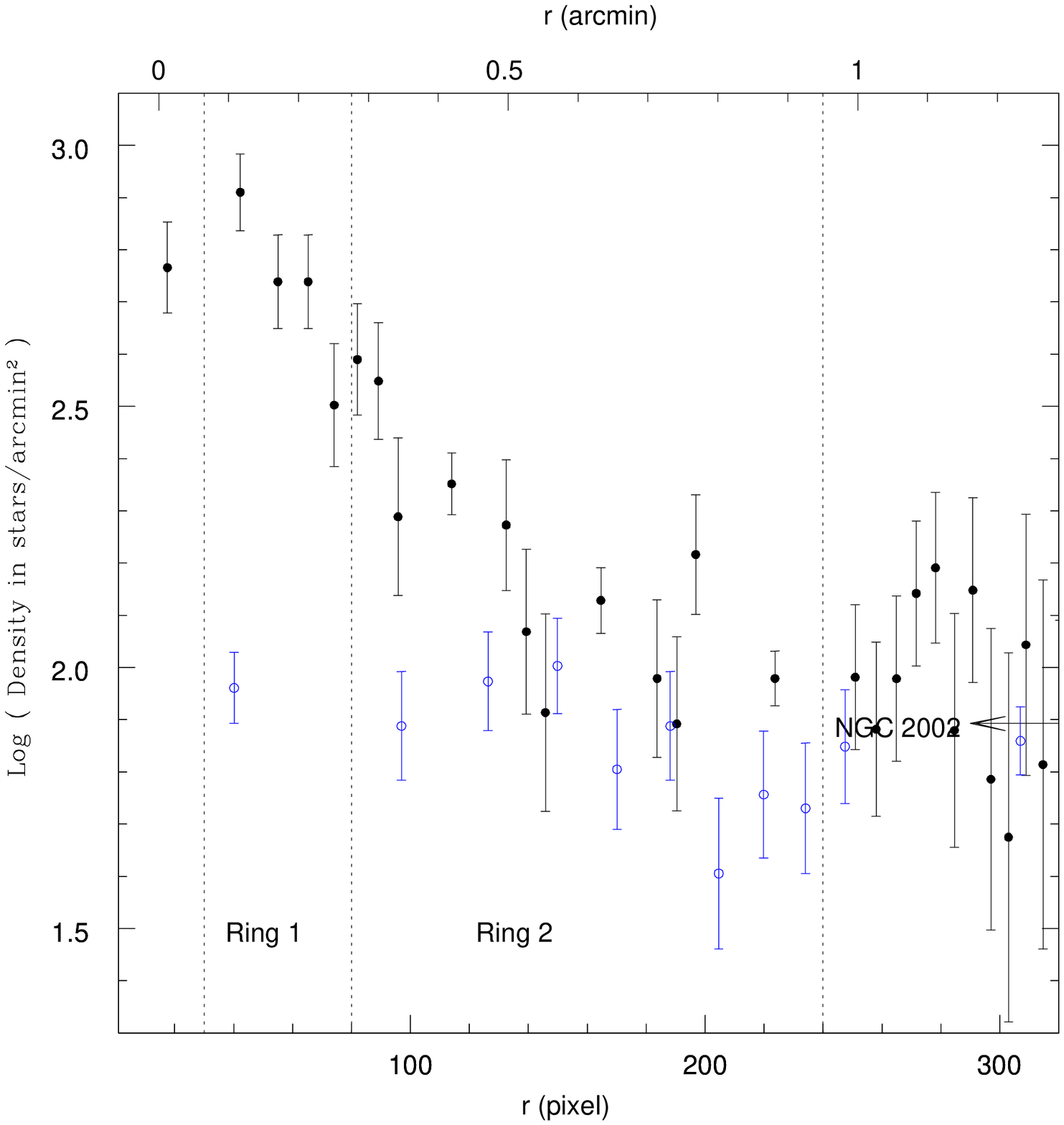}
  \includegraphics[width=7cm]{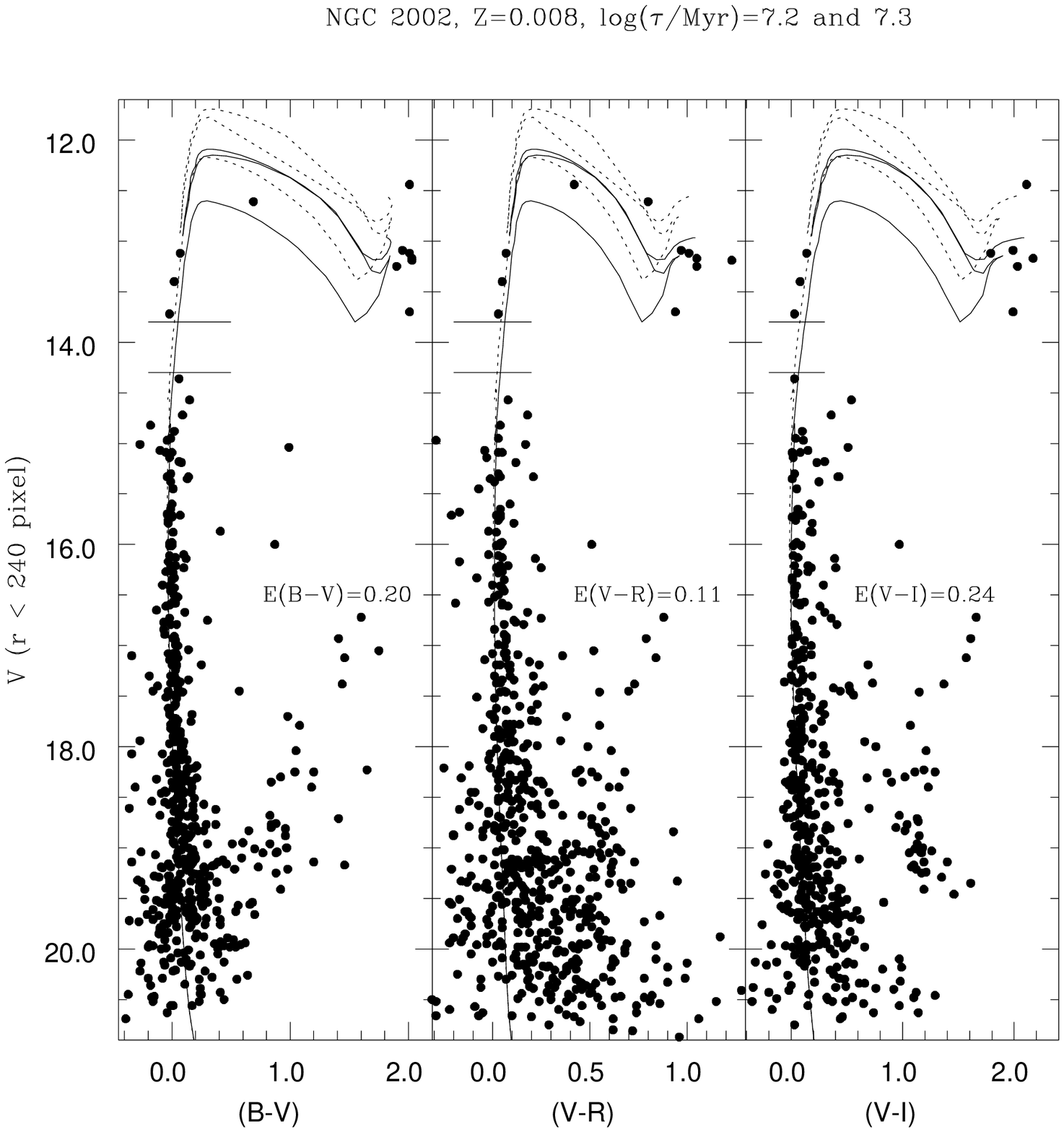}
  \includegraphics[width=7cm]{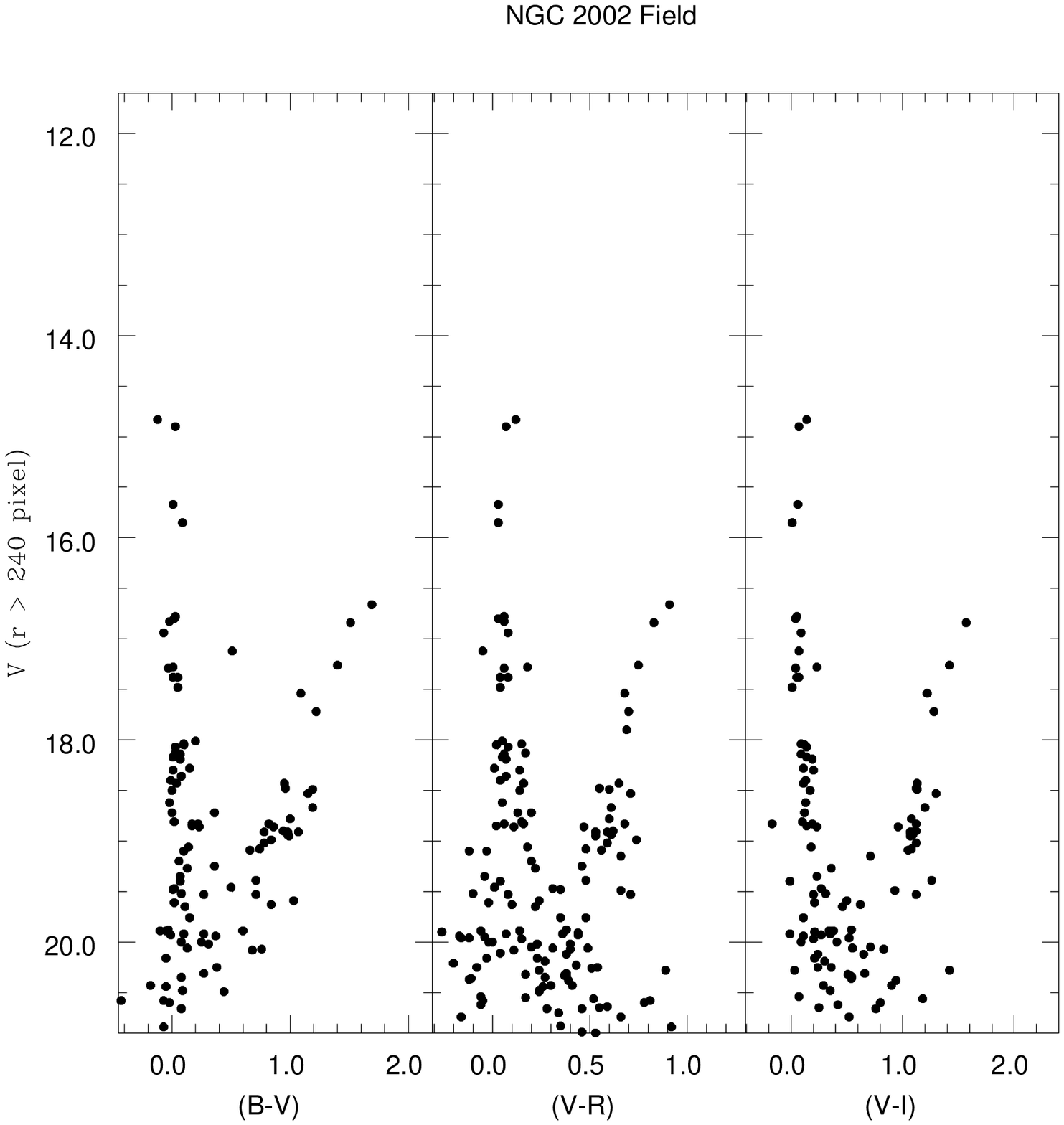}
  \caption{NGC 2002. Other descriptions are the same as for NGC 1767 
       in Fig. \ref{fig:n1767}. The top right figure also shows the
       radial star density (open circles) of the field region imaged
       about 3\arcmin away from the cluster center and shown in 
       Fig. \ref{fig:f2002}, the center is selected arbitrarily at X=256
       and Y=256 pixel.} 
  \label{fig:n2002}
  \end{figure*}

  \begin{figure*}
  \centering
  \includegraphics[width=7cm]{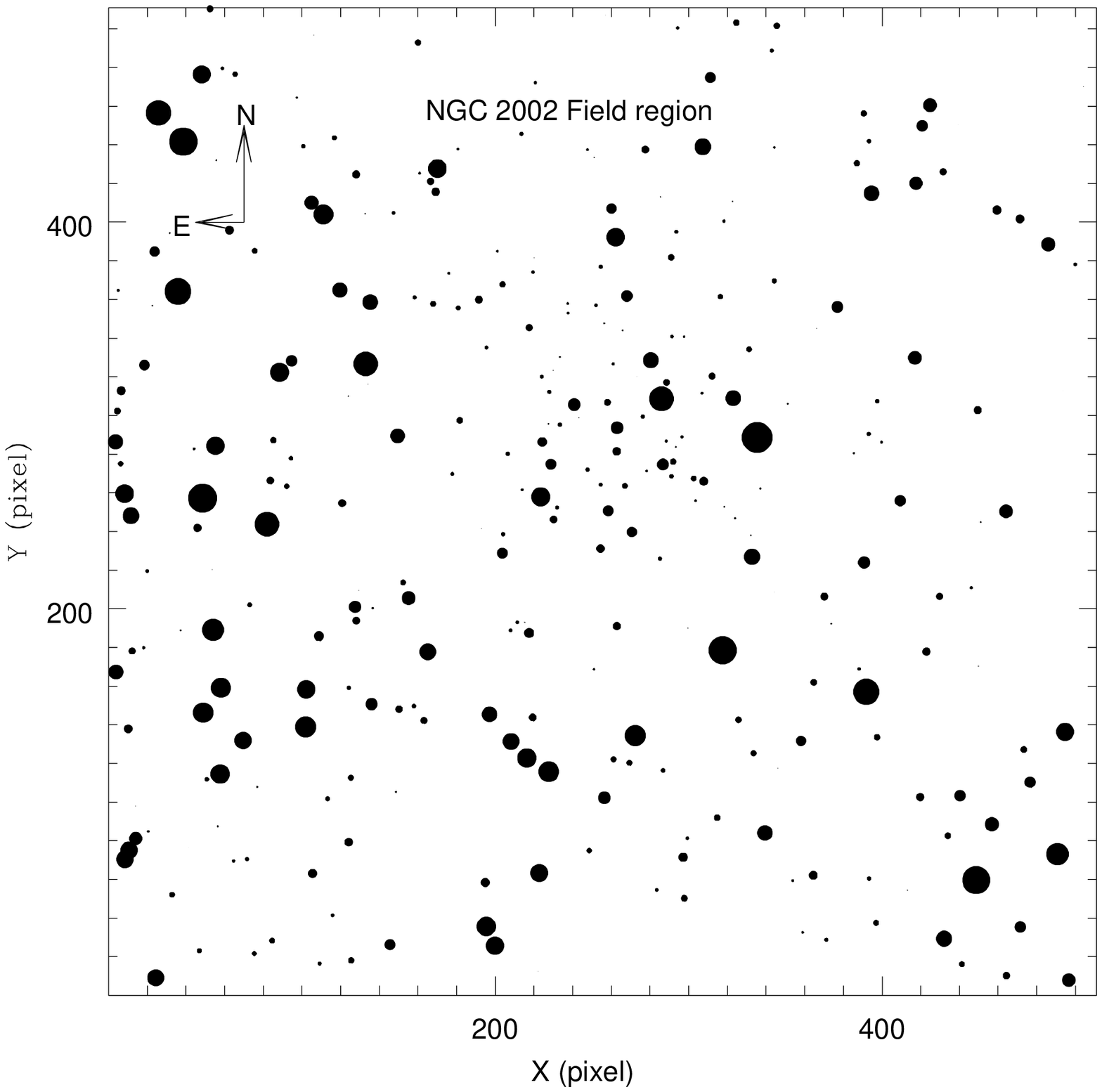}
  \includegraphics[width=7cm]{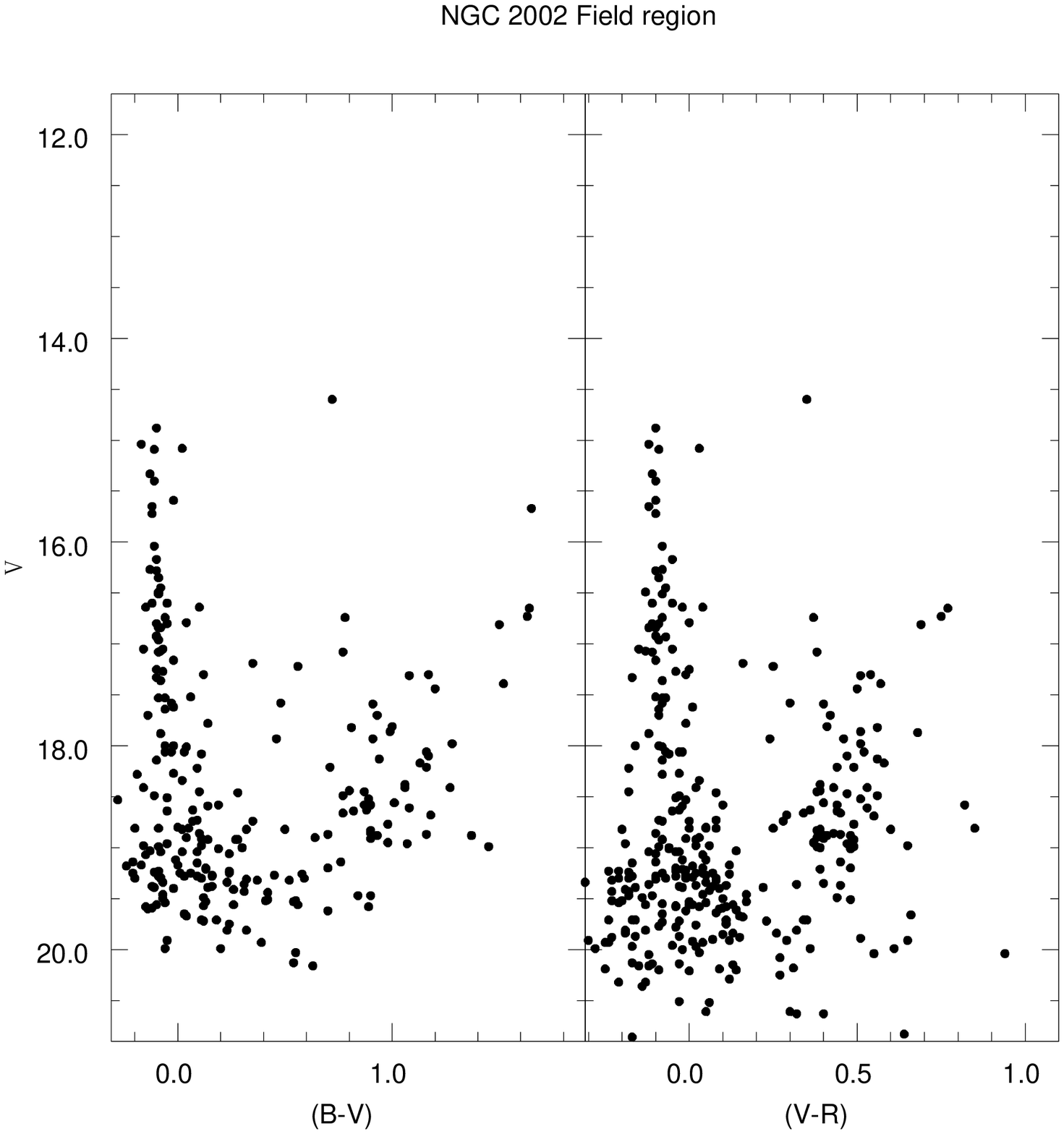}
  \caption{Identification chart and CMDs for NGC 2002 field region.}
  \label{fig:f2002}
  \end{figure*}

  \begin{figure*}
  \centering
  \includegraphics[width=7cm]{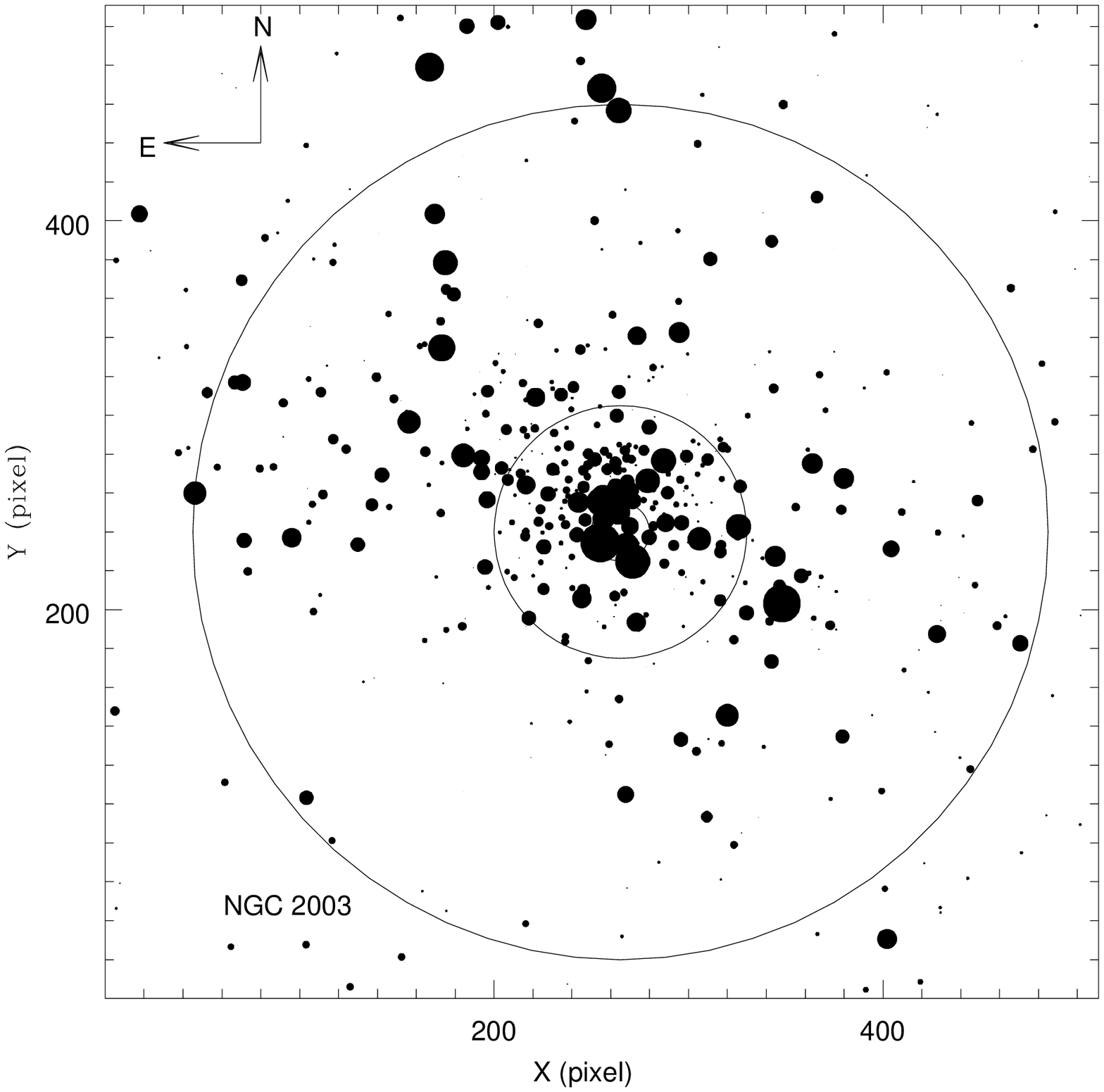}
  \includegraphics[width=7cm]{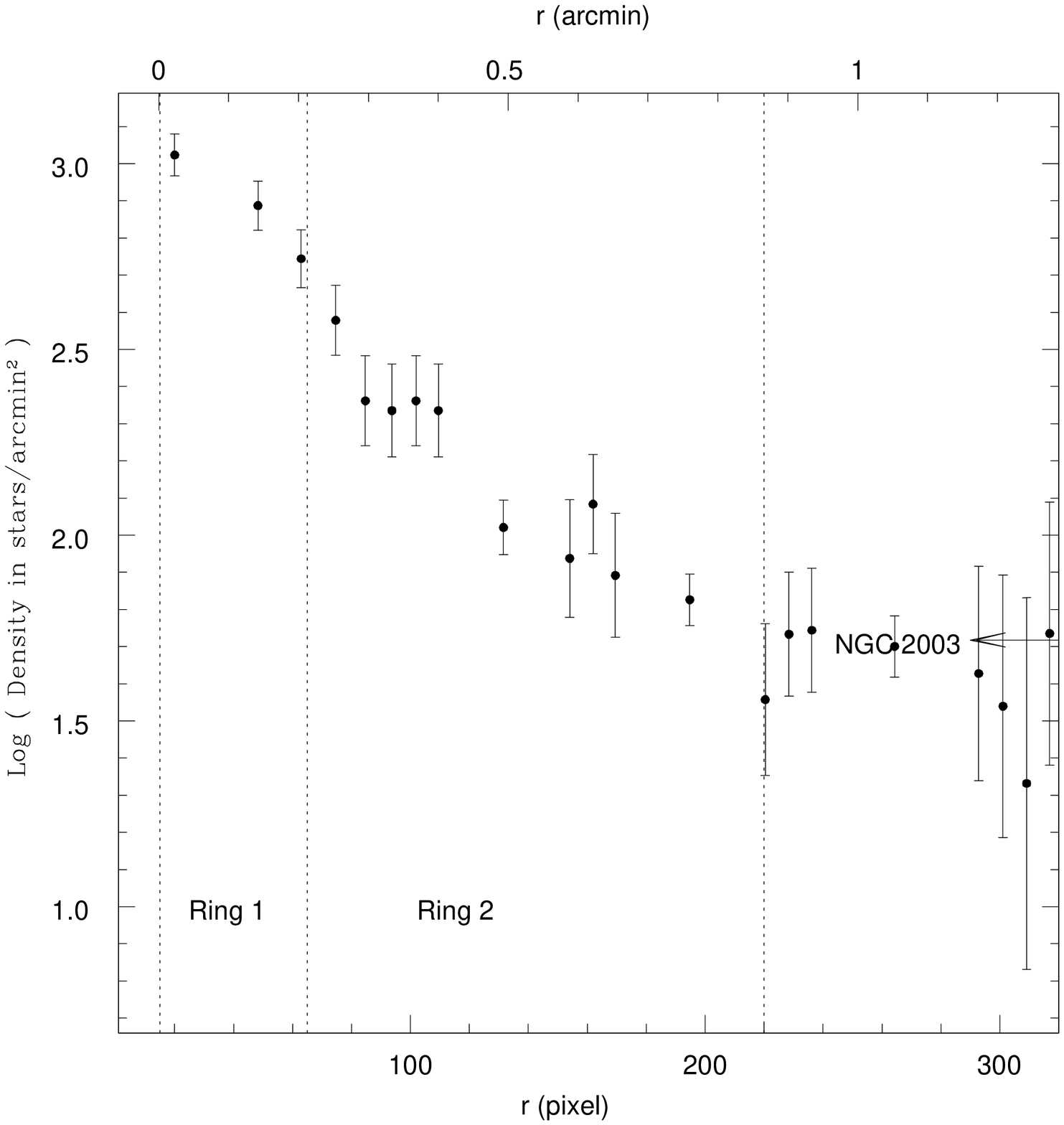}
  \includegraphics[width=7cm]{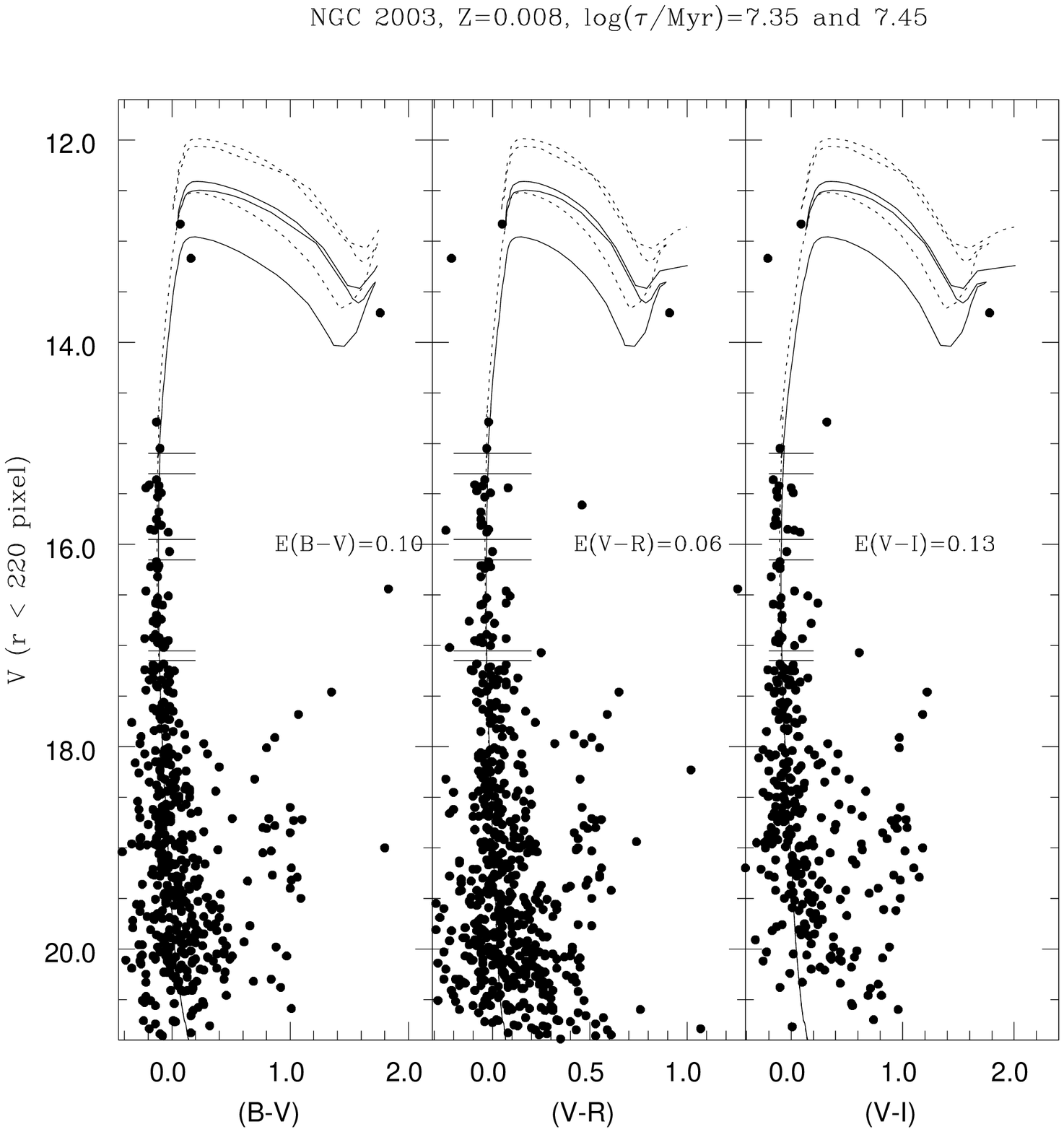}
  \includegraphics[width=7cm]{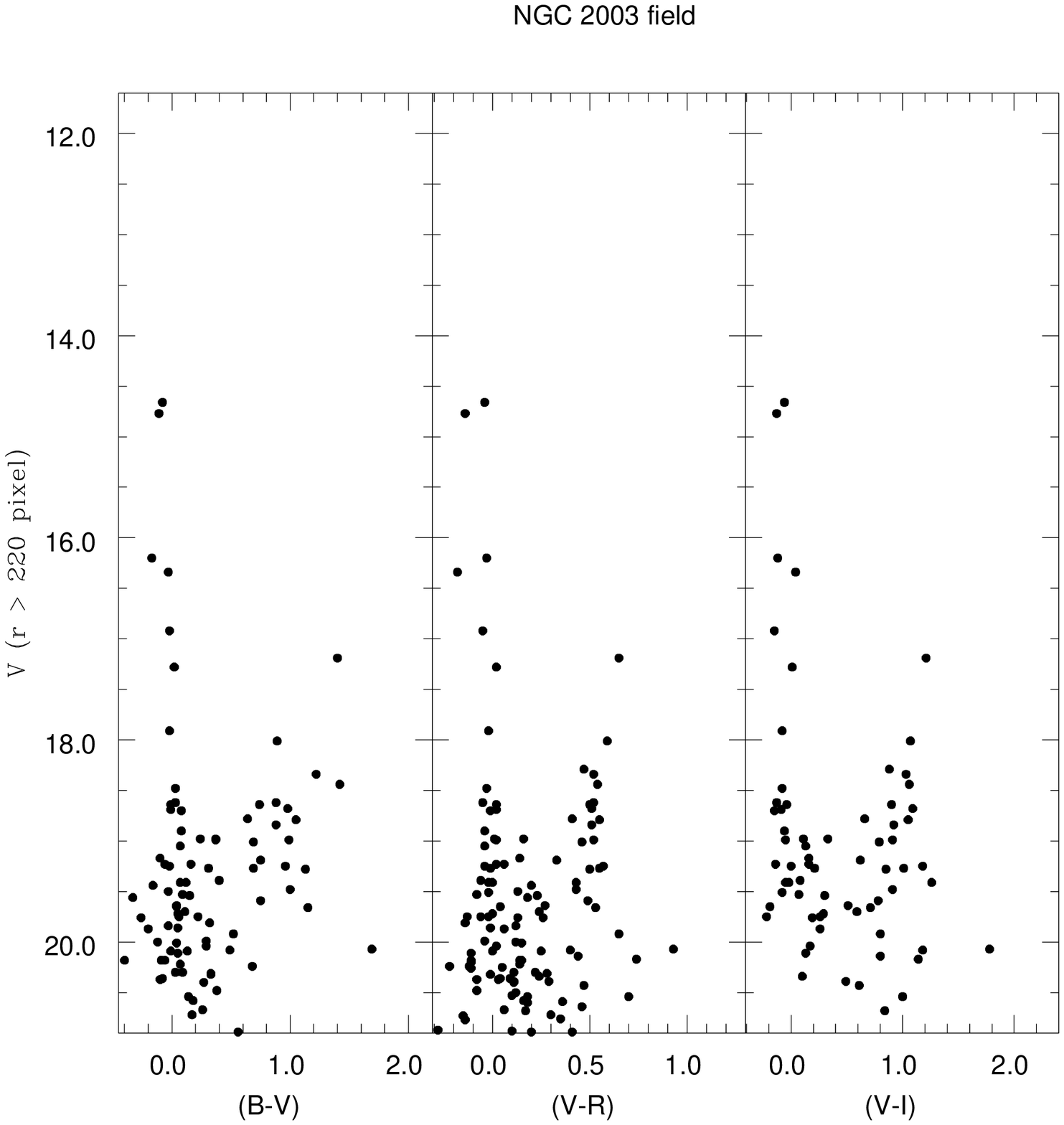}
  \caption{NGC 2003. Other descriptions are the same as for NGC 1767 
           in Fig. \ref{fig:n1767}.}
  \label{fig:n2003}
  \end{figure*}

  \begin{figure*}
  \centering
  \includegraphics[width=7cm]{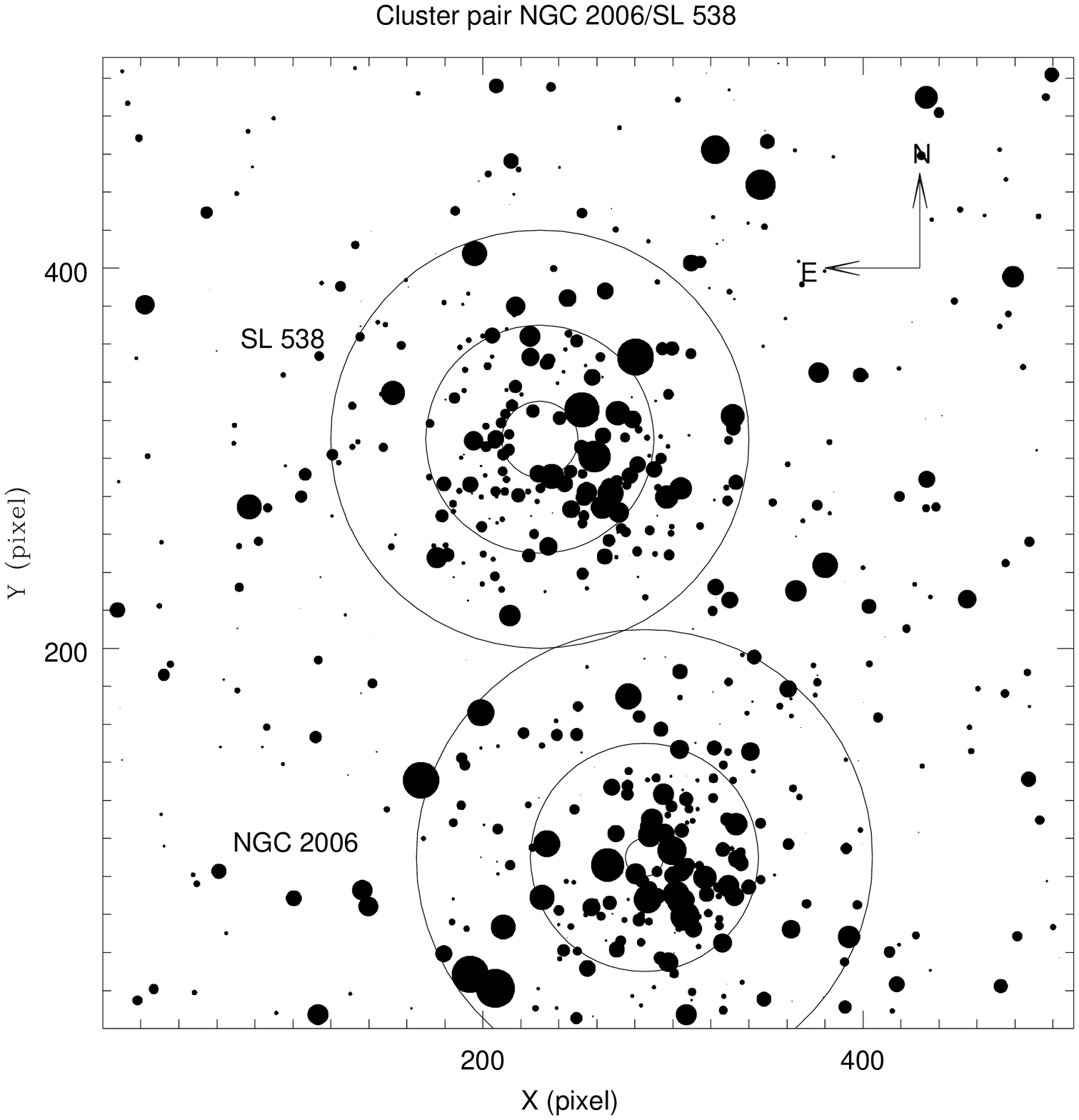}
  \includegraphics[width=7cm]{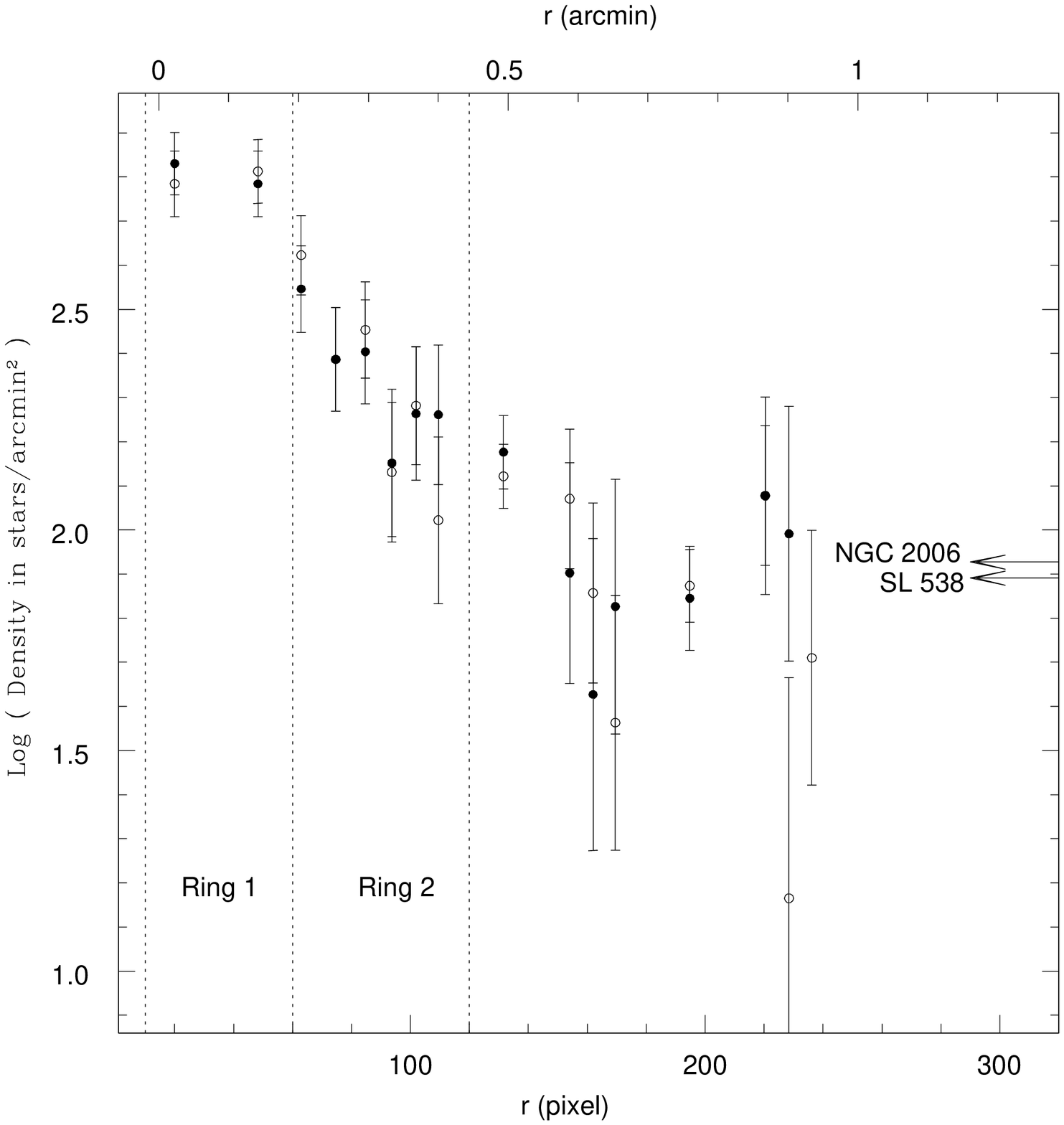}
  \includegraphics[width=7cm]{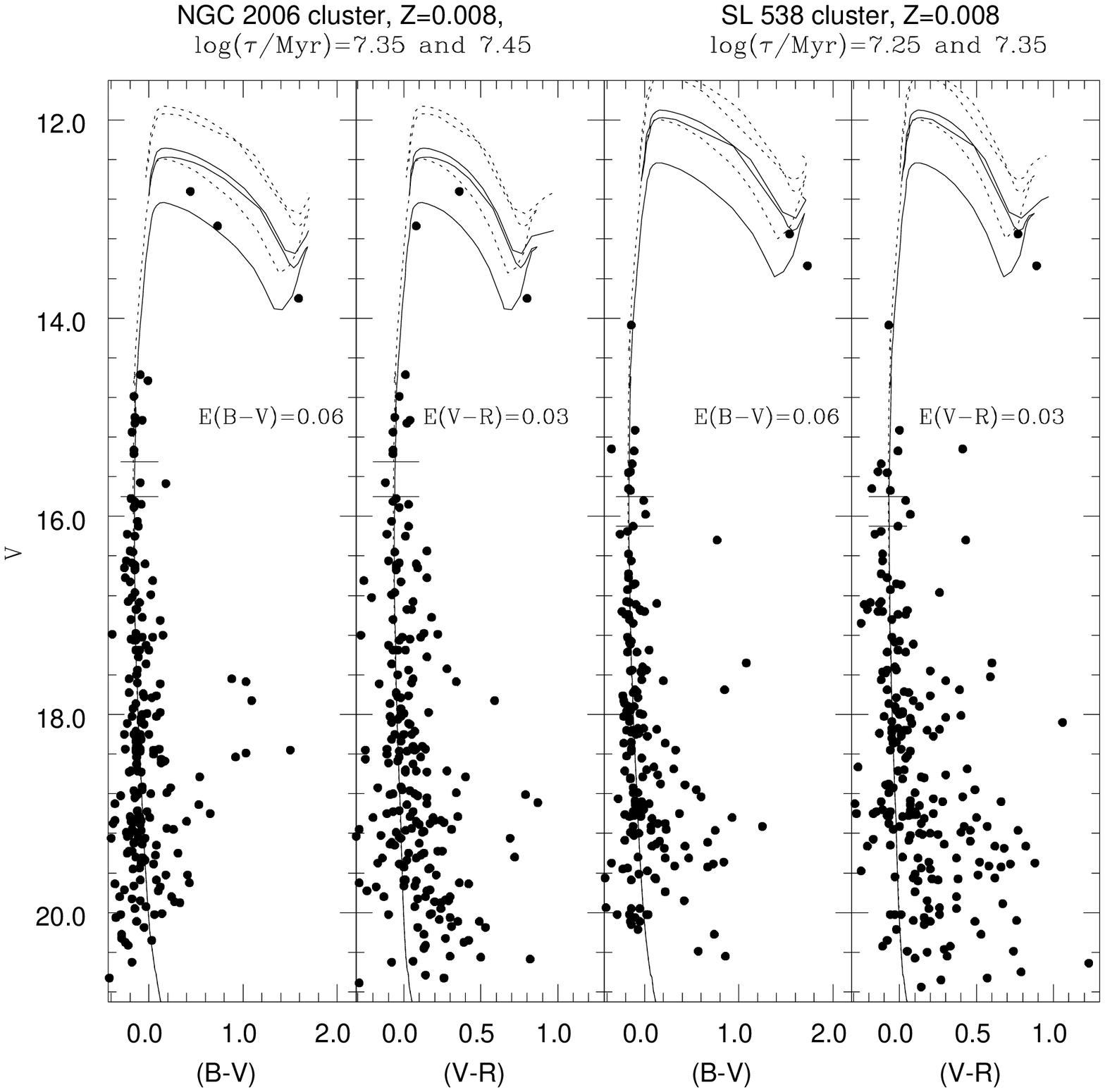}
  \includegraphics[width=7cm]{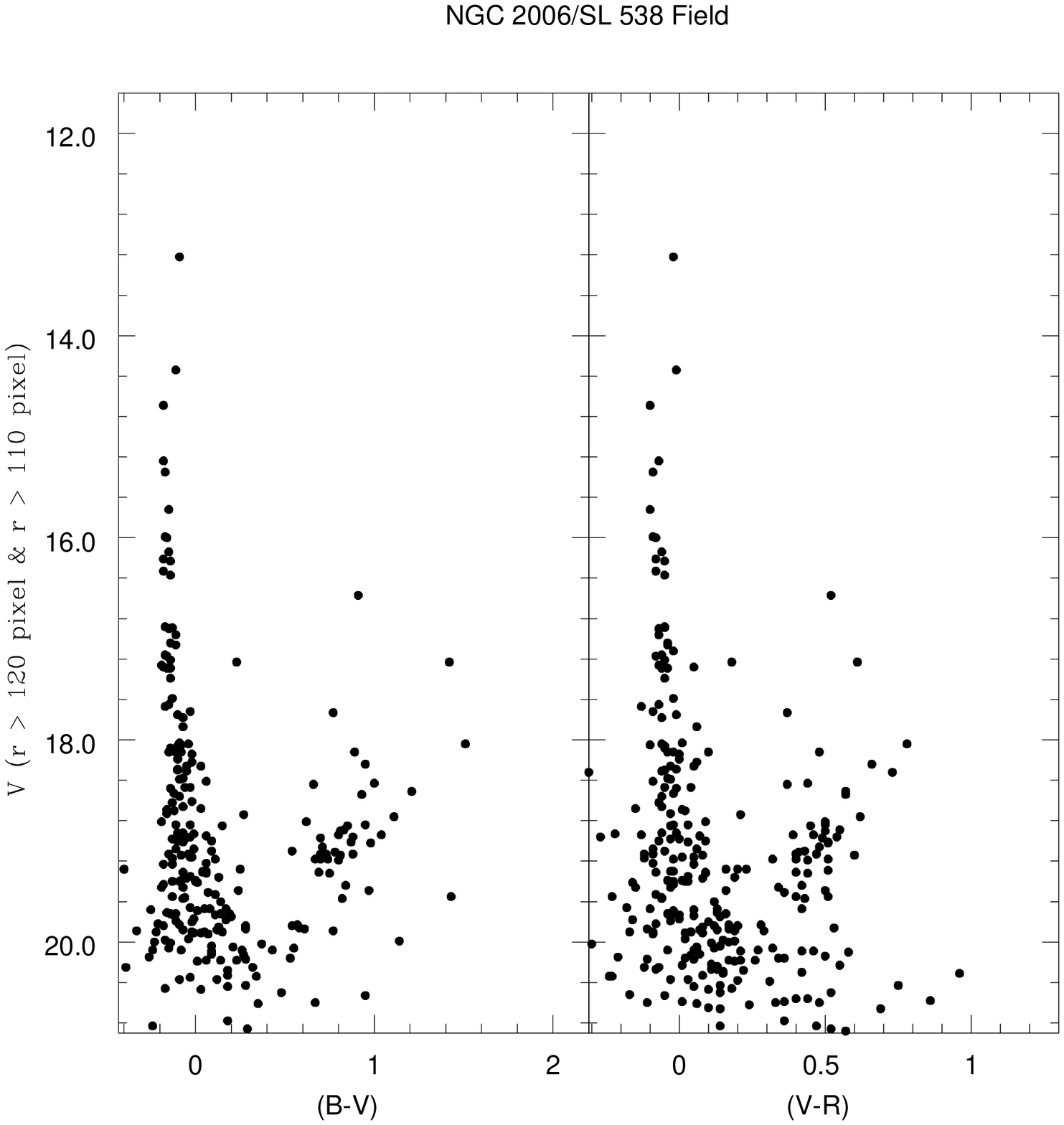}
  \caption{NGC 2006 and SL 538. The two left panels in the bottom left figure
           correspond to NGC 2006 and the stars are shown with $r < 120$ pixel
           while the two right panels belong to SL 538 with $r < 110$ pixel.
           Other descriptions are the same as for NGC 1767 in Fig. 
           \ref{fig:n1767}.}
  \label{fig:n2006}
  \end{figure*}

  \begin{figure*}
  \centering
  \includegraphics[width=7cm]{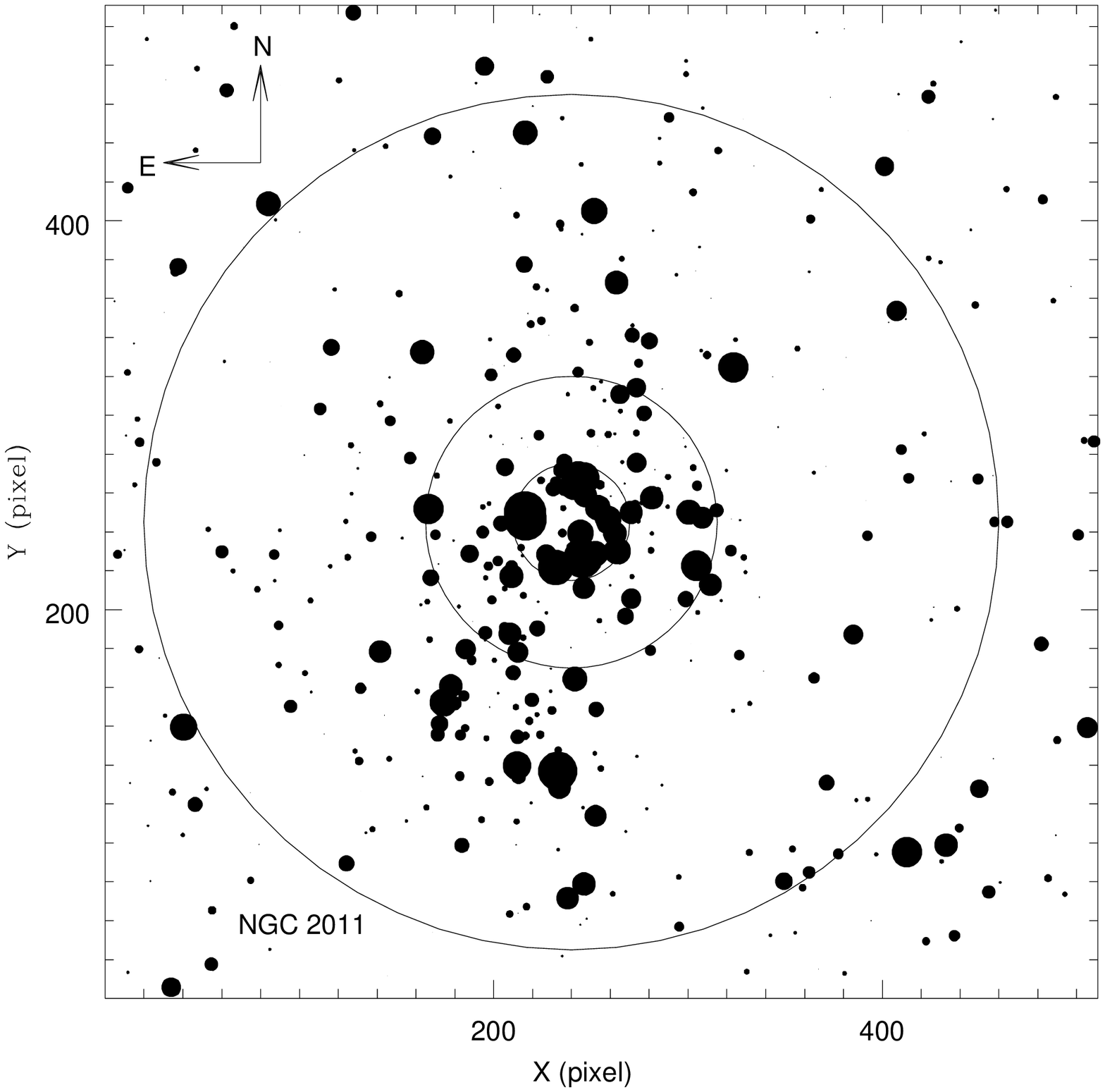}
  \includegraphics[width=7cm]{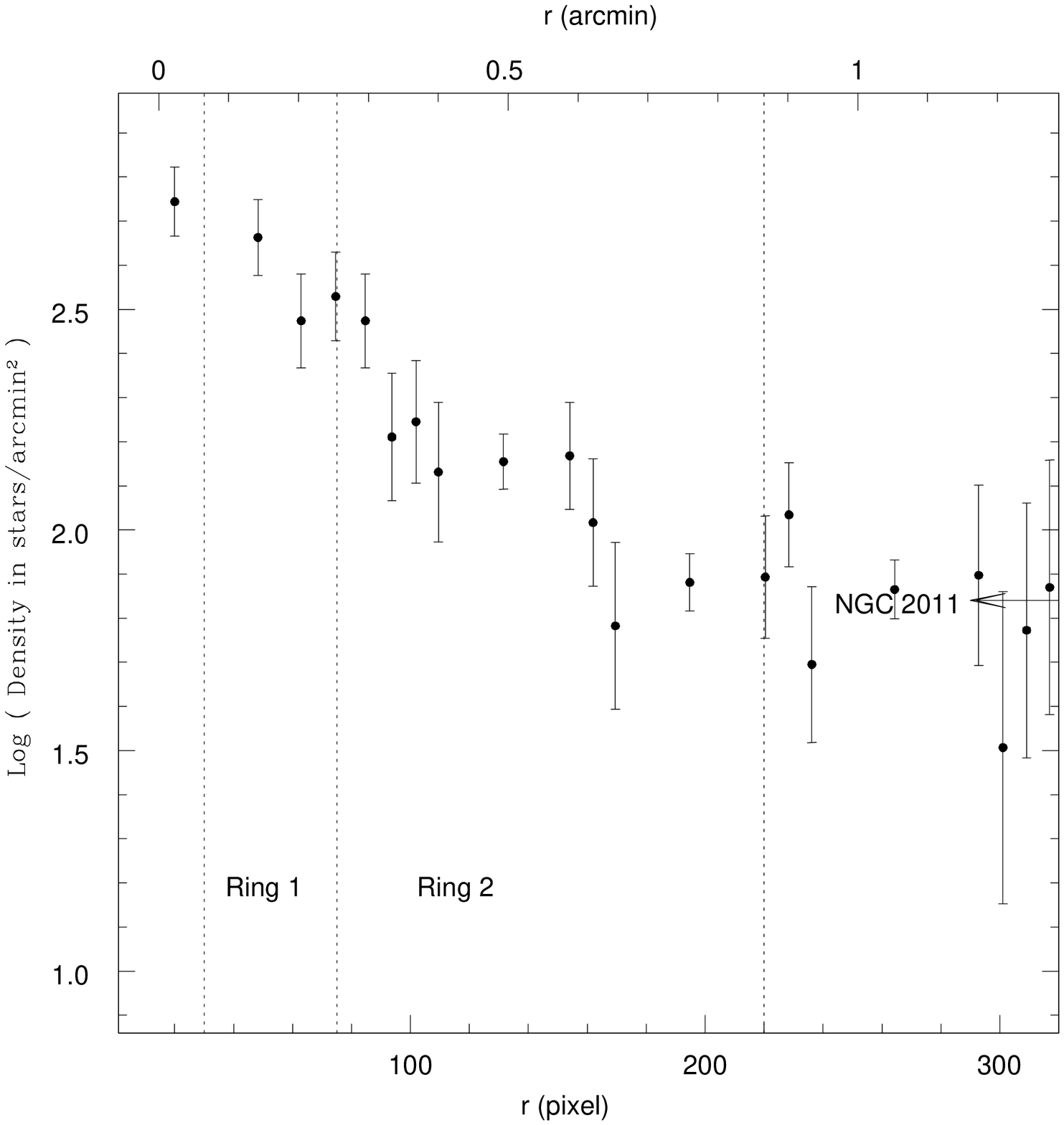}
  \includegraphics[width=7cm]{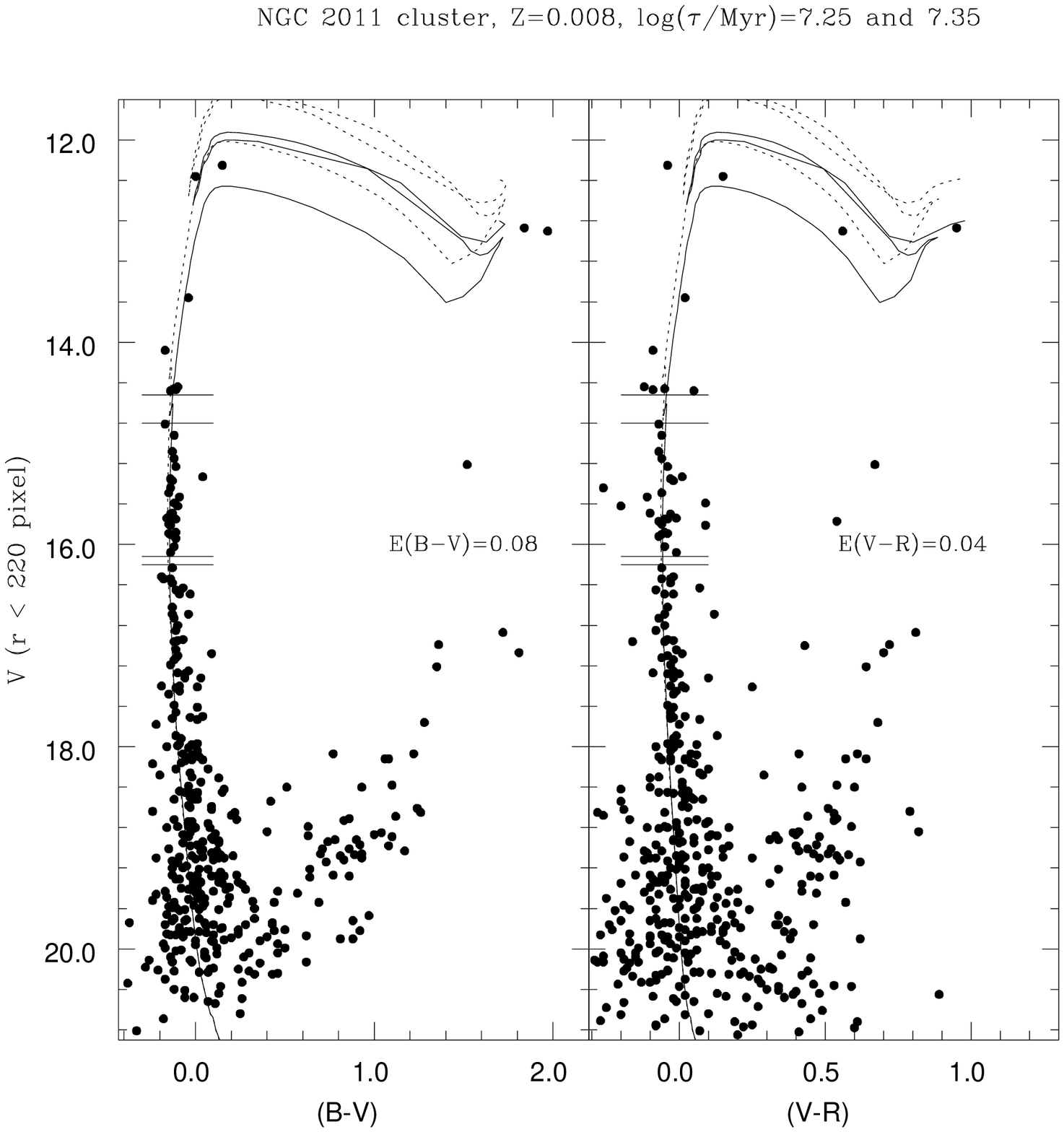}
  \includegraphics[width=7cm]{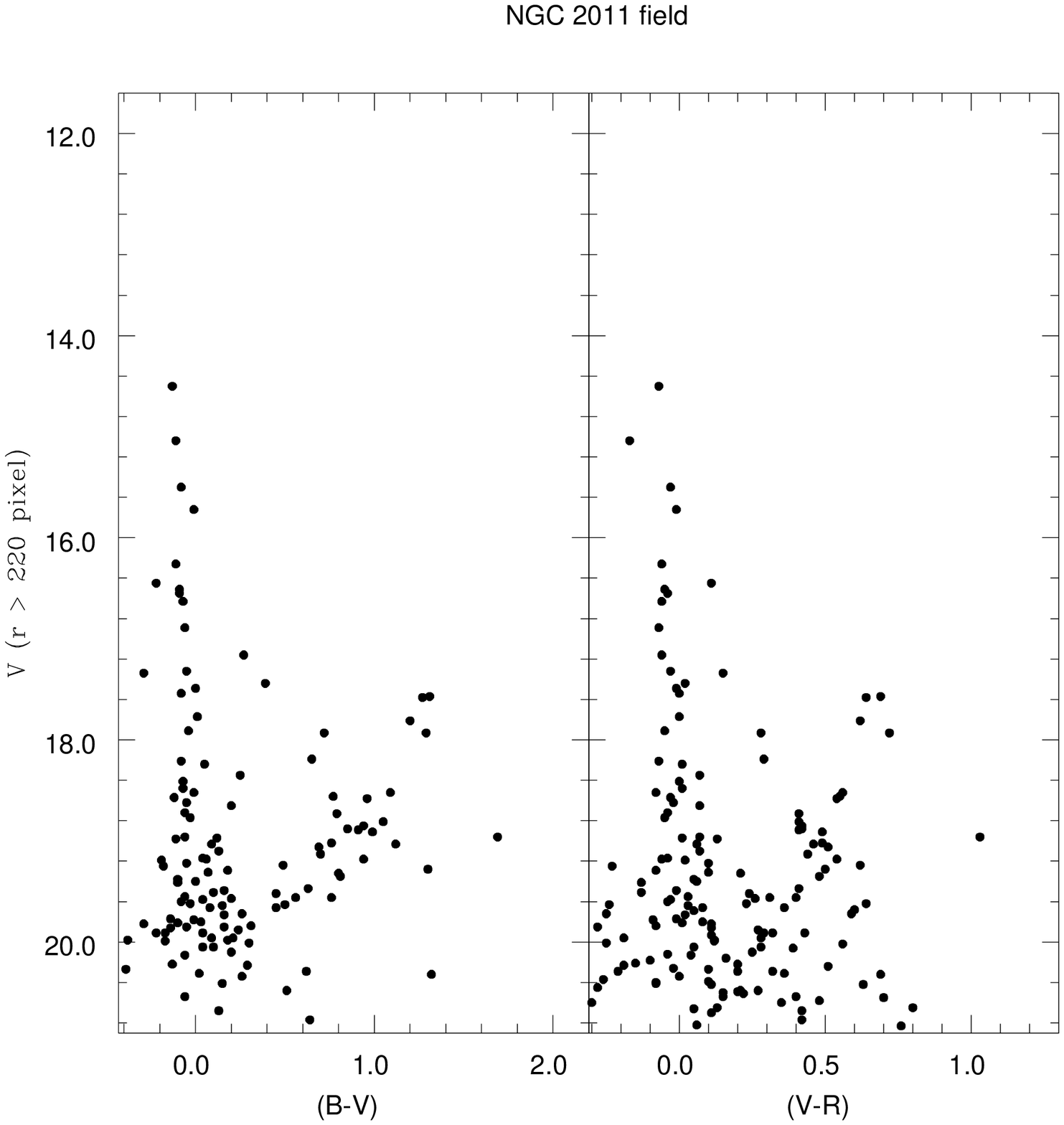}
  \caption{NGC 2011. Other descriptions are the same as for NGC 1767 
           in Fig. \ref{fig:n1767}.}
  \label{fig:n2011}
  \end{figure*}

  \begin{figure*}
  \centering
  \includegraphics[width=7cm]{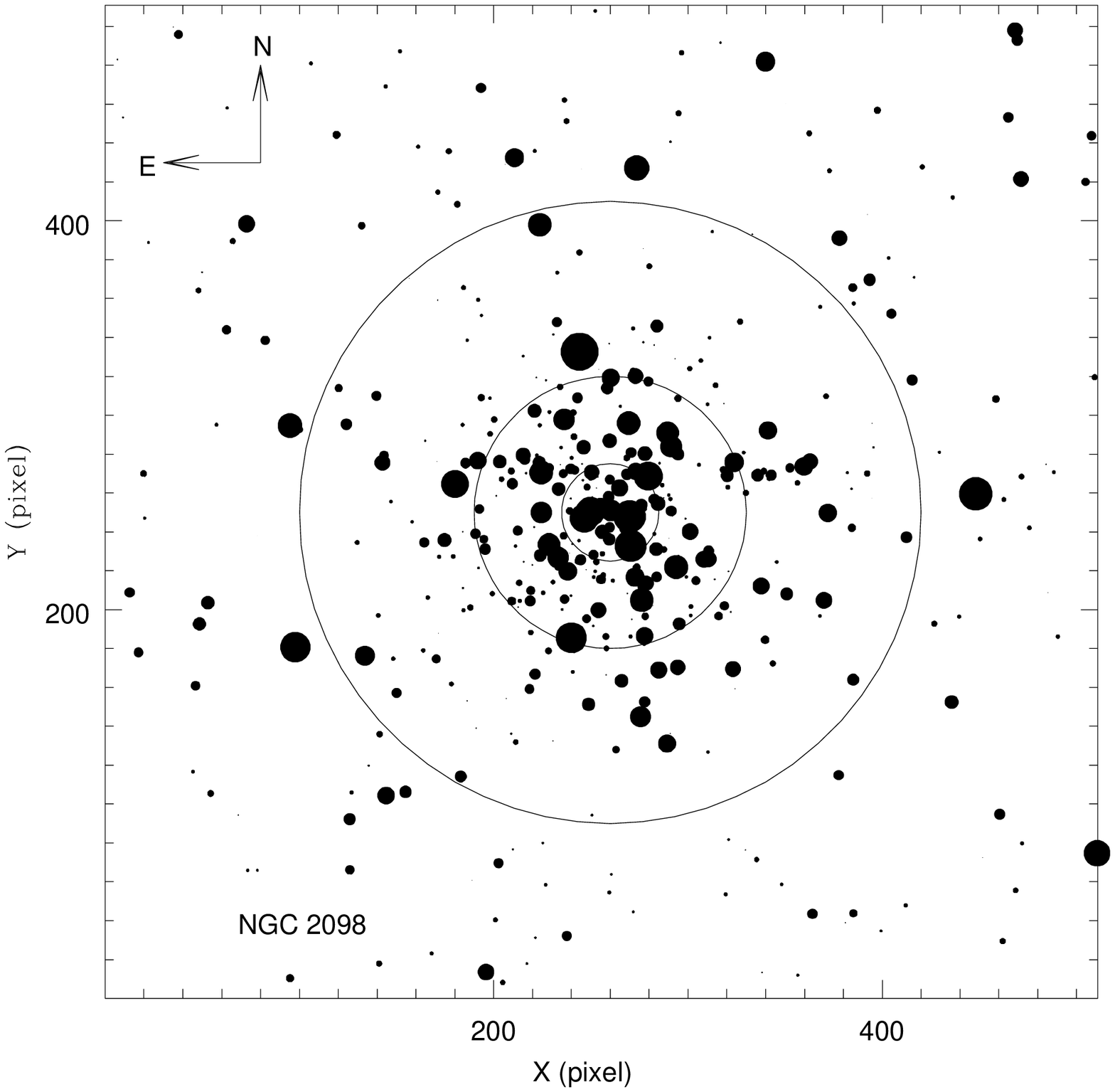}
  \includegraphics[width=7cm]{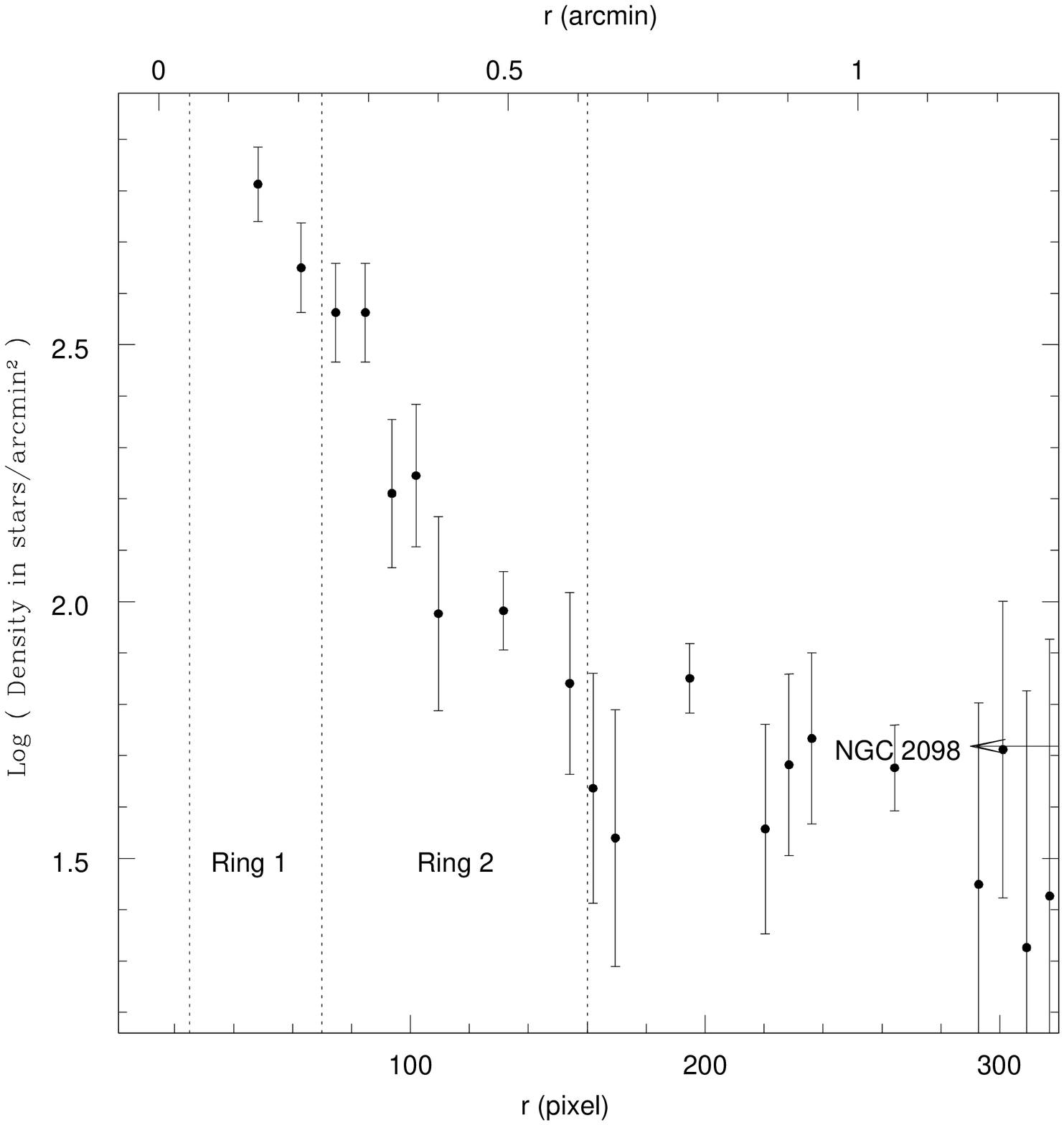}
  \includegraphics[width=7cm]{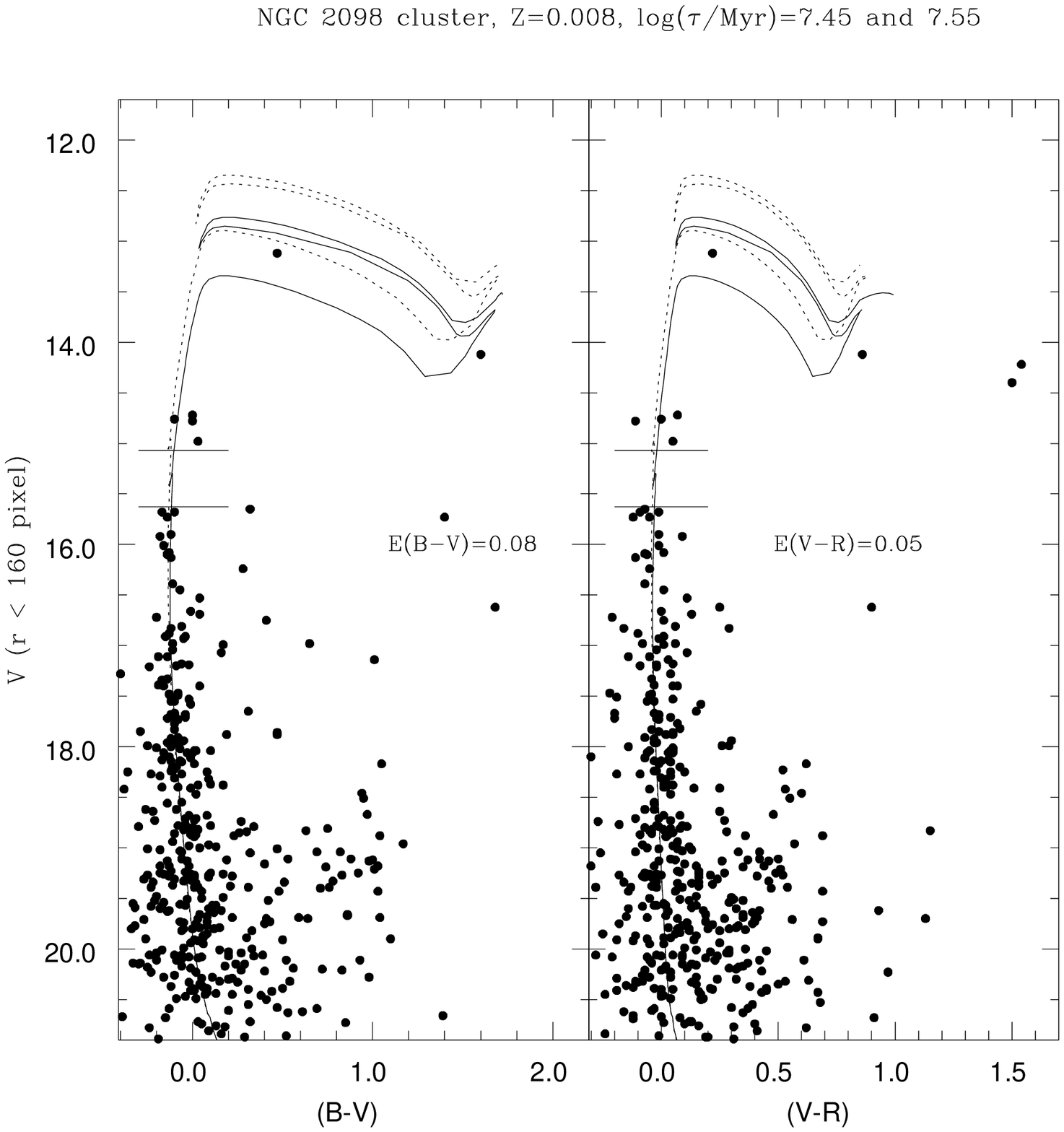}
  \includegraphics[width=7cm]{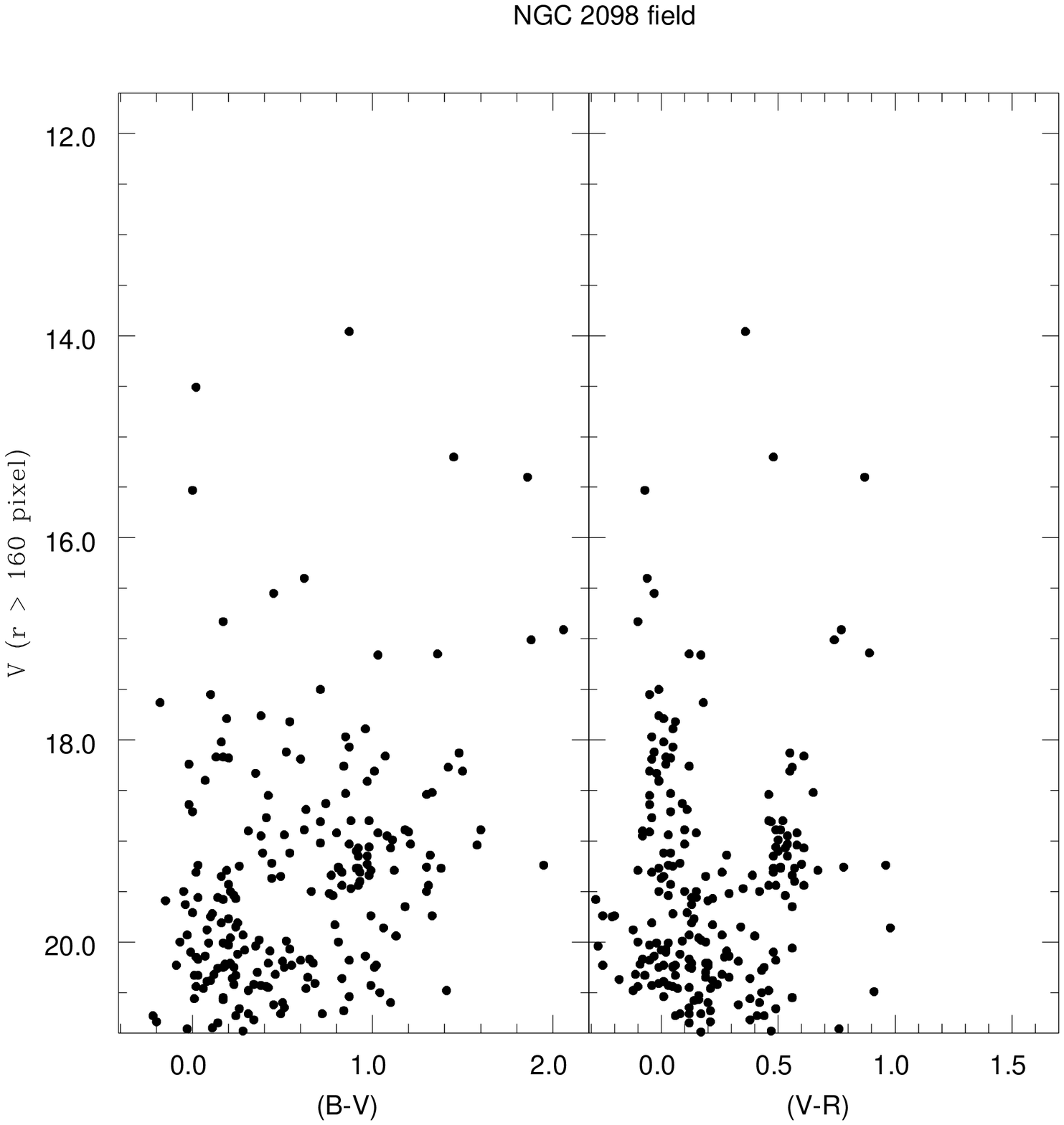}
  \caption{NGC 2098. Other descriptions are the same as for NGC 1767 
           in Fig. \ref{fig:n1767}.}
  \label{fig:n2098}
  \end{figure*}

  \begin{figure*}
  \centering
  \includegraphics[width=7cm]{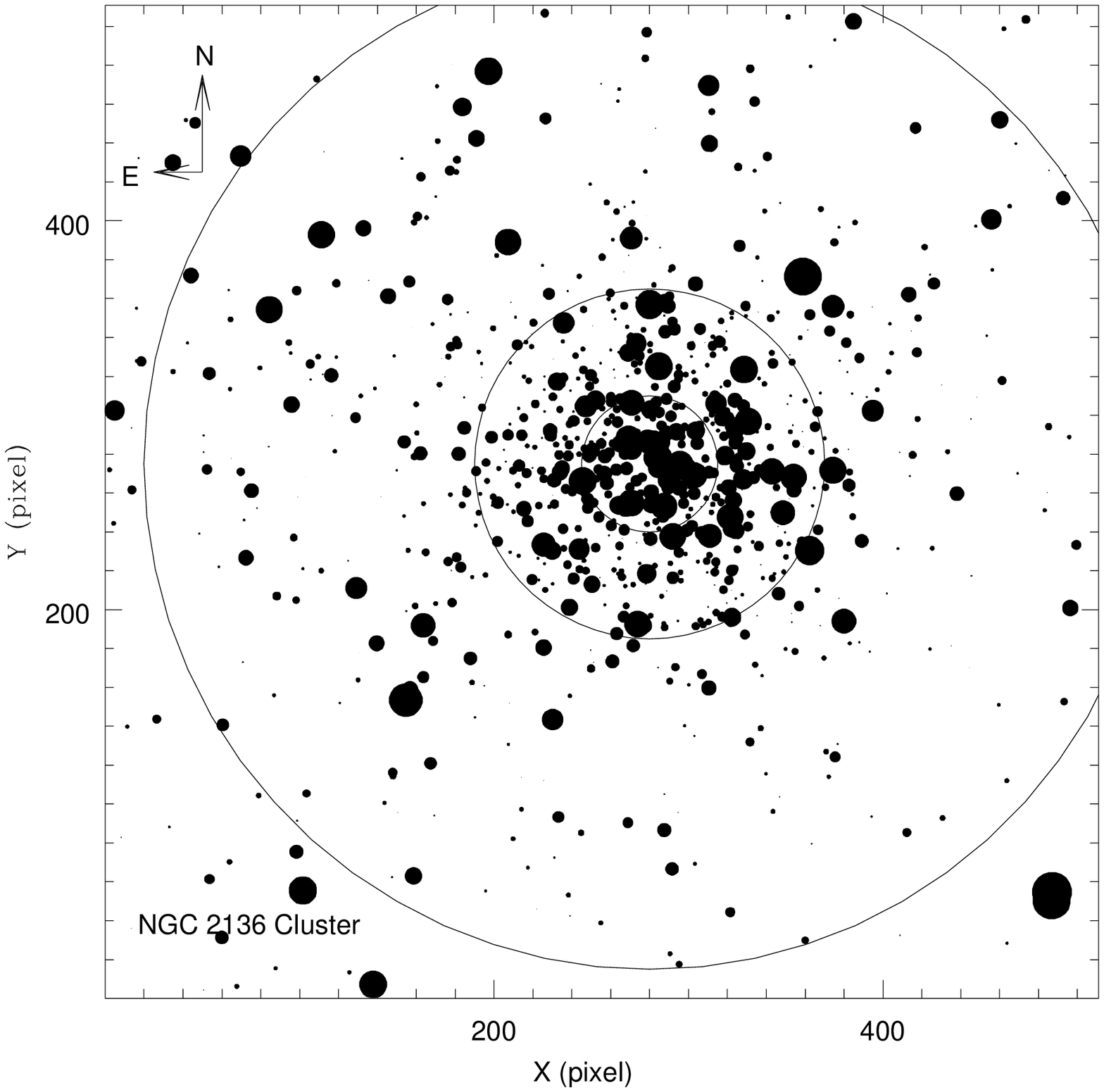}
  \includegraphics[width=7cm]{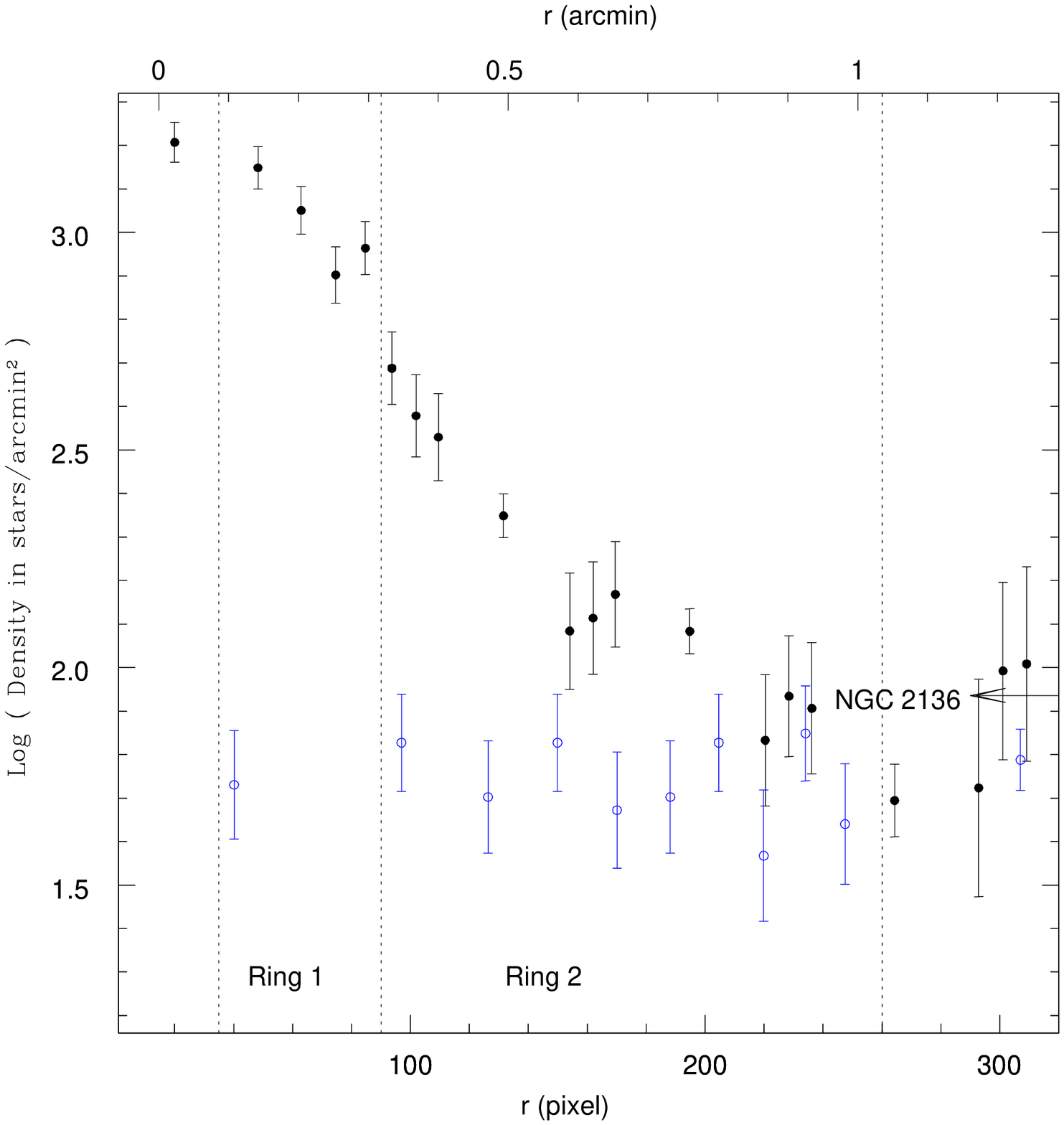}
  \includegraphics[width=7cm]{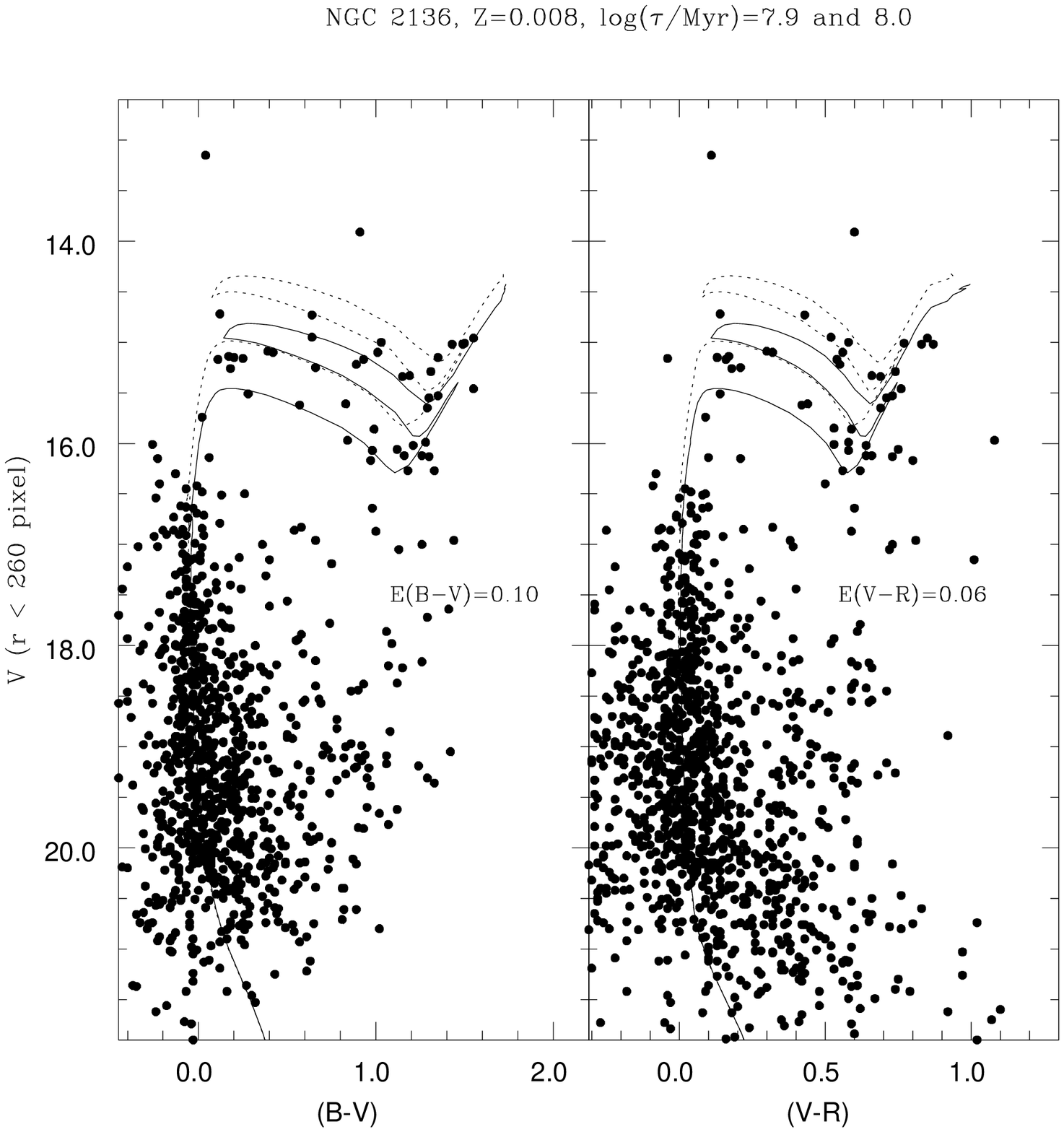}
  \includegraphics[width=7cm]{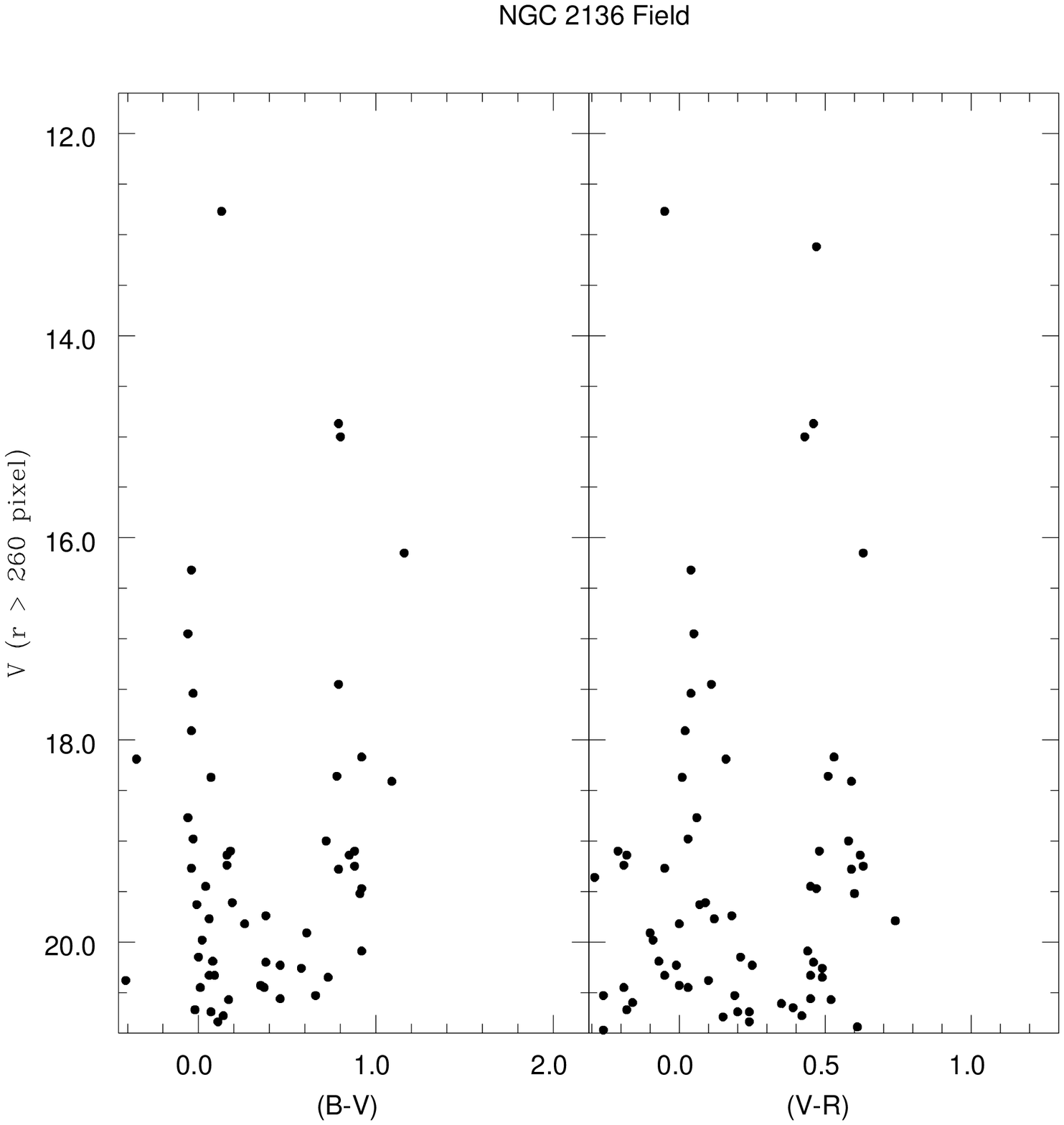}
  \caption{NGC 2136. Other descriptions are the same as for NGC 1767 
           in Fig. \ref{fig:n1767}. The top right figure also shows the
       radial star density (open circles) of the field region imaged
       about 3\arcmin away from the cluster center and shown in 
       Fig. \ref{fig:f2136}, the center is selected arbitrarily at X=256 and
       Y=256.} 
  \label{fig:n2136}
  \end{figure*}

  \begin{figure*}
  \centering
  \includegraphics[width=7cm]{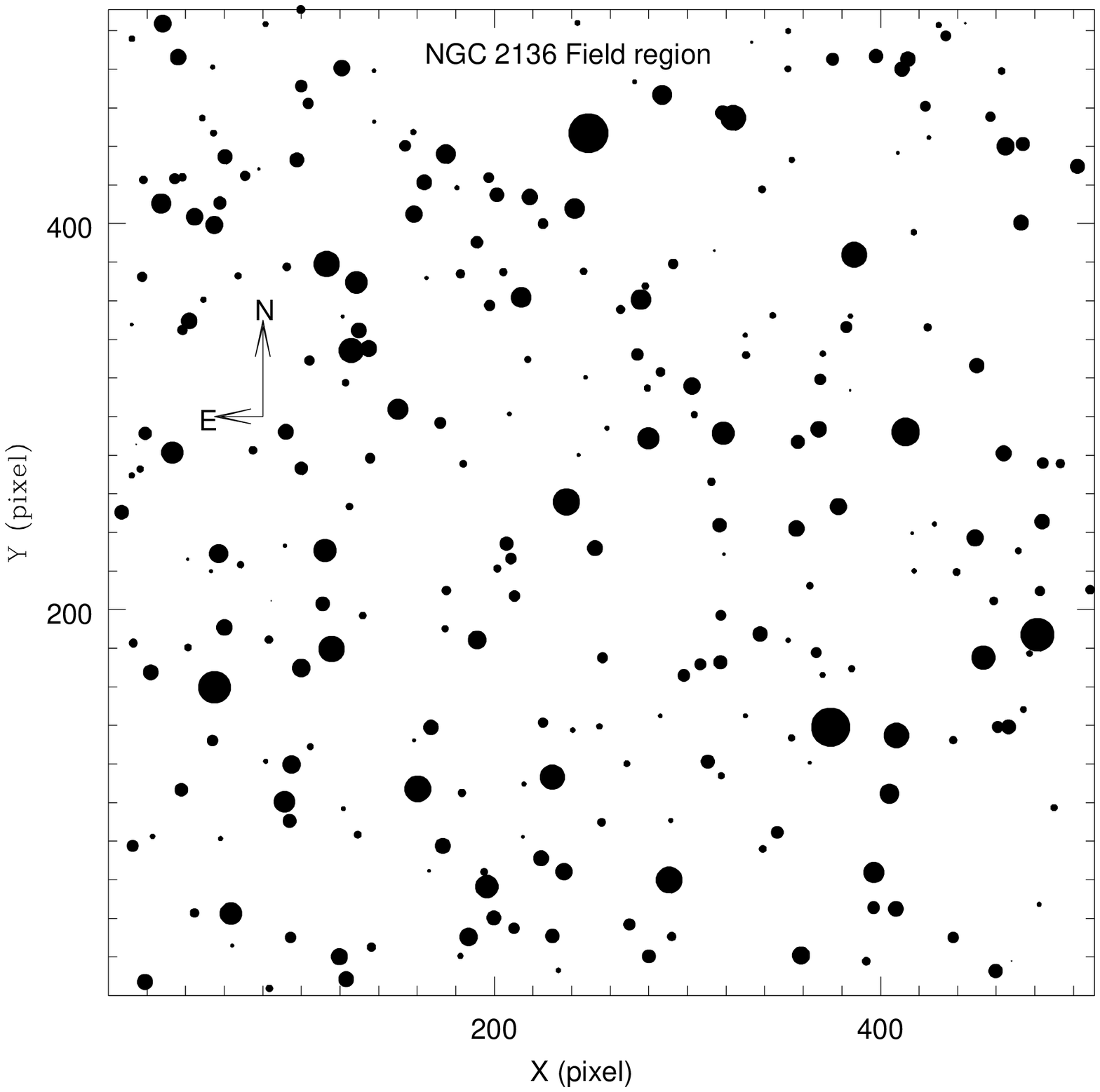}
  \includegraphics[width=7cm]{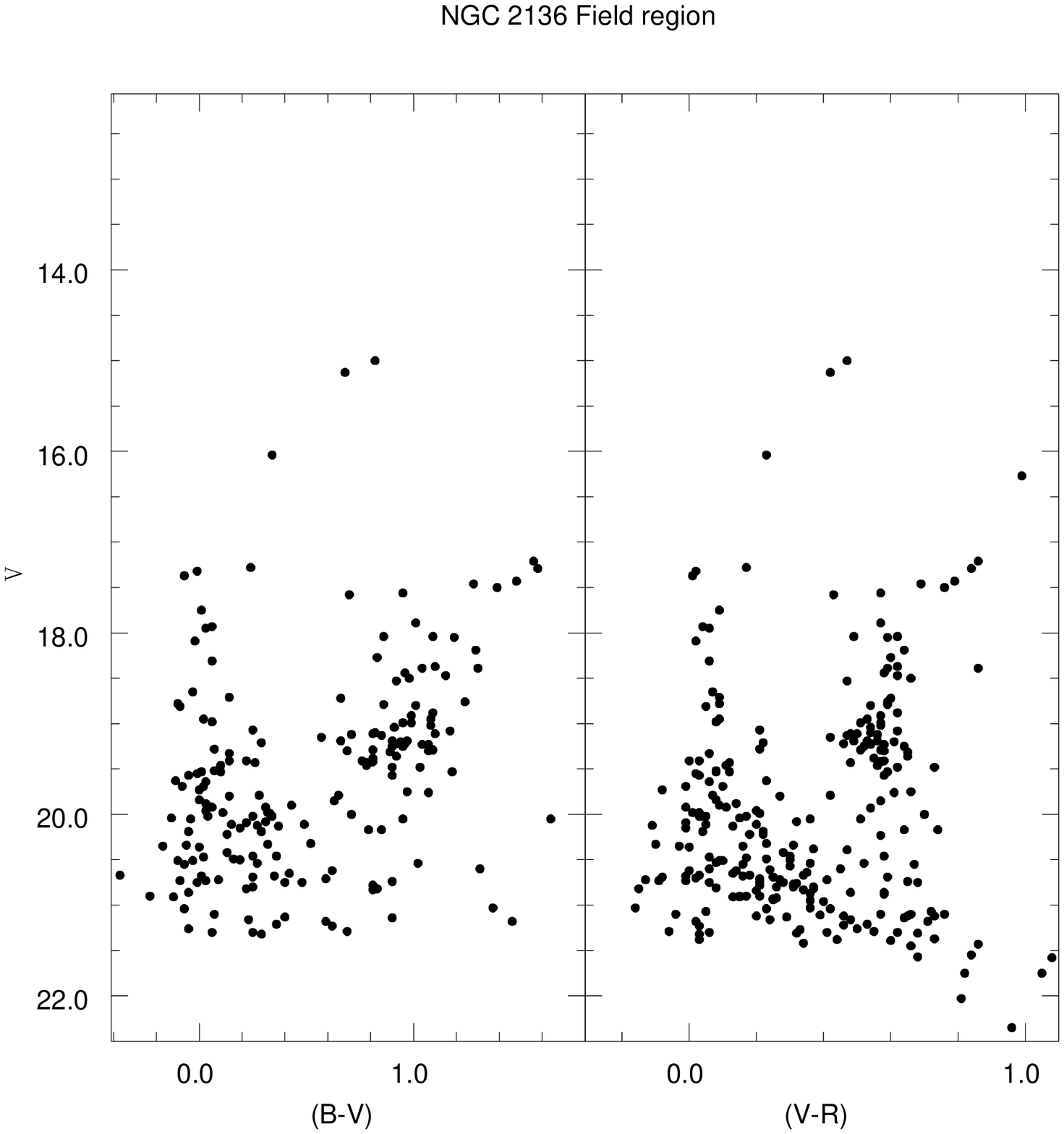}
  \caption{Identification chart and CMDs for NGC 2136 field region.}
  \label{fig:f2136}
  \end{figure*}

  The level of field star density derived from the outer region are also
  shown in these figures. The field star densities up to $V = 20$ mag are 
  in the range of 69 to 102 star arcmin$^{-2}$ with an average value of about 
  80 stars arcmin$^{-2}$. For NGC 2002 and NGC 2136, the field stellar 
  density is also estimated from a $2\arcmin \times 2\arcmin$ region 
  lying $\sim 3\arcmin$ away from cluster center. The derived mean densities 
  are 78 and 56 stars arcmin$^{-2}$ respectively. The corresponding values 
  derived from the outermost region of the clusters ($ r > r_{\rm c}$) are 
  79 and 80 stars/arcmin$^{2}$. Within our statistical uncertainty, they 
  are similar. It can therefore be concluded that the extent of field star 
  contamination is similar in all the clusters under discussion. We also 
  derive the spatial variation of the field star density from the imaged 
  field regions taking $X = 256$ pix and $Y = 256$ pix as its center and 
  the same are shown in Fig. \ref{fig:n2002} and Fig. \ref{fig:n2136}. It 
  is seen that the stellar density of the field region follows the background 
  densities derived from the outermost regions of these two clusters. This 
  indicates that stars with $r > r_{\rm c}$ can be used to estimate field 
  star contamination in the cluster. From these numbers as well as the 
  cluster sequences present in the CM diagrams discussed below, one may 
  say that the field star contamination in the stars brighter 
  than $V \sim 20$ mag is not strong enough to smear the cluster sequences 
  and hence affect the results derived below. 

  As our data are substantially incomplete in the crowded central region 
  of the clusters, we are unable to evaluate the value of stellar density 
  at the cluster center. Consequently, the radius at which central stellar 
  density becomes half can not be determined. However, the radial density 
  profiles of all the clusters clearly indicate the innermost cluster 
  region ($\le$ 30 pixel) where stellar crowding is so large that it cannot 
  be used to determine the cluster MF accurately. Similarly, the outermost 
  regions of the clusters where the stellar density becomes flat is clearly 
  defined and we have considered this as the cluster radius. In Figs 4 to 6 
  and 8 to 12, one can see that for all the clusters beyond a radius 
  $R \sim 0\farcm9$, the number of stars up to 20 mag per unit area drops 
  to a uniform level which might be considered as a good approximation of 
  the background density. The cluster radius for the target clusters range 
  from 120 pixels ($\sim 0\farcm5$) for NGC 2006 \& SL 538 to 260 pixels
  ($\sim 1\farcm0$) for NGC 2136. Its value is about $0\farcm75$ for NGC 1767 
  and NGC 2098 while the remaining clusters have a radius of about $0\farcm92$. 
  In order to see  radial variations of MF, the entire cluster region
  (excluding core) has been divided into two annulus region i.e. Ring 1 and
  Ring 2. The pixel values of the cluster radius, core region, Ring 1 and 
  Ring 2 are listed in Table \ref{tab:radden}.

  \subsection {Color magnitude diagrams} \label{sec:cmd}

  In order to properly analyse the CMDs of our LMC clusters, it is necessary to 
  delineate the cluster sequences from the unavoidable field star contamination.
  We therefore constructed CMDs of stars located at different radial distances
  from the cluster center. This helps us to clearly distinguish cluster 
  features from those characterising the surrounding LMC fields. Consequently,
  two CMDs for each cluster, one representing the features of the clusters
  while other characterising the surrounding field region are constructed.
  Figures \ref{fig:n1767} to \ref{fig:n2002} and \ref{fig:n2003} to 
  \ref{fig:n2136} show the CMDs of the cluster ($r \le r_{\rm c}$) as well 
  as of 
  the field ($r > r_{\rm c}$) regions. For NGC 2002 and NGC 2136, we also show 
  the CMDs of a nearby field ($2\arcmin \times 2$\arcmin) regions in Figs 7 and
  13 respectively. A characteristic MS from $V \sim 14$ mag to 20 mag is seen 
  in all the clusters, except NGC 2136 where it begins at around $V \sim $ 16 
  mag, indicating their youthful (age $\la 25$ Myr) nature. In addition, the 
  brighter end ($V \sim 13$ mag) is also populated by a few blue and red 
  supergiants. This is in contrast to the field regions which is only sparsely 
  populated by stars towards lower MS ($V \ga 16$ mag). Red clump of stars 
  near $V \sim 19.5$ mag, $(B-V) \sim 0.9$ mag and $(V-R) \sim 0.5$ mag in the 
  CMDs, are populated by evolved stars arising from the old age ($\ga$ 1 Gr) 
  stellar populations of the LMC. Such a feature has also been observed in 
  other CCD photometric studies of LMC star 
  clusters (see \citealt{sagar91b} and references therein). 
  These are intermediate age core helium burning stars of the LMC 
  field forming a clump in the CMDs. 
  
  For constructing mass functions of the clusters under study, we need 
  determination of distance, age and reddening for each object and the same are 
  being described below. 
    
  A value of $18.5\pm0.1$ mag for the true distance modulus of the LMC is 
  now well 
  constrained since it is derived using more than two dozen independent 
  measurements (see \citealt{alves04, schaefer08} and references therein). 
  The individual determinations, however, vary from 18.1 to 
  18.8 mag primarily due to different standard candles being 
  used e.g. the Red clump, tip of red giant branch, Cepheids, RR Lyrae 
  stars, Mira variables, SN1987A, eclipsing binaries etc. We adopt
  $18.5 \pm 0.1$ mag as distance modulus for LMC in the present study. 
  Being everything similar, smaller distances result in lower ages 
  and affect the derived mass ranges for MFs. A closer analysis indicates that
  adopting distance modulus 0.4 mag different changes the derived mass 
  functions significantly, although its effect on MF slopes are observed 
  to be negligible \citep{sagar91a}.  

  We use the stellar evolutionary models by \citet{girardi02} to estimate 
   clusters ages and adopt a constant value of metallicity $Z=0.008$ 
  (Fe/H $\sim$ 0.3 dex). Recent estimates on the present day chemical 
  abundance for LMC stellar population converge to a sub-solar metallicity, 
  for example, \citet{rolleston02}
  derive a metallicity index of $-0.31\pm0.4$ for OB-type main sequence stars.
  For many young ($\tau < 100$ Myr) LMC star clusters, the metallicity seems to
  have a plateau around Fe/H $\sim$ 0.4. \citep{mackey03, kerber07}. 
  Effects of metallicity variation on the derived MF slope indicate that it 
  becomes flatter with decreasing value of Z. A change in Z from 0.02 to 0.004 
  has a negligible effect on the MF slope, see Fig. 6 of \citet{sagar91a}. 

  Reddening towards surrounding LMC region is observed to be around 
  $E(B-V) = 0.075$ mag as estimated from all sky maps
  at 100 $\mu m$ \citep{schlegel98}. Based on HI emission 
  map \citep{burstein82}, the predicted $E(B-V)$ towards cluster lie 
  between 0.05 to 0.1 mag. However, the intra-galactic reddening across 
  LMC is observed to vary and it may be as 
  high as 0.3 mag in some regions \citep{bessell91}. We therefore adjusted
  the value of reddening to best fit the isochrones to the MS. Our age 
  estimates are greatly facilitated by the presence of a few blue and red 
  supergiants. The best estimate for age lies between two isochrones identified 
  for each cluster and are shown in Figs. \ref{fig:n1767} to \ref{fig:n2002}
  and \ref{fig:n2003} to \ref{fig:n2136} and accordingly, we adopt mean age 
  ($\tau$/Myr) and uncertainty. The reddening and the adopted ages derived
  in this way are listed in Table \ref{tab:age}. The values
  of $E(B-V)$ are $\le 0.1$ mag for all the clusters except for NGC 1767 and 
  NGC 2002 where its value is 0.18 and 0.2 mag respectively. Thus the present 
  low reddening values are in agreement with those based on HI and dust 
  emission map of sky.

  The present age estimates for NGC 2006 ($25\pm3$ Myr) and SL 538 ($20\pm2$ 
  Myr) are consistent with the corresponding estimates of $22.5\pm2.5$ Myr and 
  $18\pm2$ Myr by \citet{dieball98}. Our age estimate ($32\pm4$ Myr) for NGC 
  2098 is significantly younger than the estimate of 63\,-\,79 Myr 
  by \citet{kontizas98}. 
  For all other clusters this is the first reliable estimate of age using 
  main-sequence turn-off point in the CMDs. However, studies using integrated 
  spectra of star clusters and single population stellar library derive about 
  10 Myr systematically younger ages \citep{wolf07}. Barring NGC 2098 and
  NGC 2136, their age estimates for the remaining 7 clusters of our sample 
  lie between 6 to 8 Myr, while, our estimates range from 15 to 20 Myr. The 
  age estimates derived from integrated spectra using theoretical models seem 
  to be biased towards blue MS stars.
 
  A gap in the MS is defined as a band, not necessarily perpendicular to the MS,
  with no or very few stars. \citet{bohm74} first located a gap in MS around 
  $(B-V)_{0} = 0.27$ mag which arise due to onset of convection in the stellar
  envelope. Gaps seem to appear as statistically distinct features in MS 
  of star clusters \citep{sagar78, kjeldsen91, subramaniam99}. MS Gaps in the 
  present sample were identified visually and the gap parameters are listed in 
  Table \ref{tab:age}. Except NGC 2136, all the target clusters have about 0.3 
  to 0.5 mag wide gap between $M_{V} \sim$ $-3.0$ to $-5.0$ mag. This seems to
  be characteristic feature of stellar evolution as the brighter gap location
  corresponds to the younger clusters. Stellar evolution models do predict 
  paucity of stars around and beyond the MS turn off. However, 
  some clusters seem to have clumpy MS with more than one gaps, for example, 
  NGC 2003 and NGC 2011 show gaps of smaller amplitude ($\Delta V < 0.3$ mag) 
  at fainter magnitudes. Both these clusters have elongated spatial structures 
  and may have star forming history different from the other clusters. 
  It is therefore noted that the gaps may also arise due to stochastic 
  effects of star formation and sampling along the main sequence and it may not 
  represent any genuine astrophysical effects.

  \subsection {Cluster luminosity and mass functions}
  
  The Luminosity functions (LFs) of LMC star clusters and their 
  corresponding field region is derived from star counts in bin width
  of 0.5 mag in $V$ from the $V$, $(V-R)$ diagrams. It has been preferred over
  other CMDs due to fainter limiting magnitude and better data completeness.
  The main factors which limit the precise determinations of cluster MF 
  from present observations are data incompleteness and field star contamination 
  as the central region of the clusters are more likely to suffer from data 
  incompleteness while the outer region is more affected by field star 
  contamination. Moreover, the present photometry is generally not able to 
  resolve the central 30 pixel diameter region of each cluster. We therefore, 
  estimate the LFs for the inner (Ring 1), outer (Ring 2), and entire regions 
  of each cluster excluding the core. These regions
  are marked in the radial density profile of the respective clusters 
  (see Sect \ref{sec:radden}) and also listed in Table \ref{tab:radden}.  

  For data completeness factor (CF), we follow the usual DAOPHOT procedure of 
  adding and recovering the artificially selected stars with known 
  magnitudes and positions in the original $V$ and $R$ frame and the
  effective CF is taken to be smaller value of them.
  We estimate CF separately for inner, outer, and entire regions
  as well as the corresponding field regions
  which is usually defined as $r \ge r_{\rm c}$ (see Table \ref{tab:radden}).  
  In the case of NGC 2002 and NGC 2136, the field region refers to full 
  CCD frames of a nearby field. For each region, the number of stars (NS) 
  lying on the main sequence in the $V, (V-R)$ diagram are counted in a bin 
  width of 0.5 mag. In order to avoid field star
  contamination from intermediate age stellar populations of the LMC, which 
  normally appear as the characteristic red clump near $V \sim 19.5$ mag 
  and $(V-R) \sim 0.5$ mag, the stars are counted in a 0.5 mag strip 
  around the best fitting isochrones. The LF for each bin is calculated as :

   ${\rm LF} = \left( \frac{\rm NS}{\rm CF} \right)_{\rm cluster} - 
    \left( \frac{\rm NS}{\rm CF} \right)_{\rm field} \times {\rm Area Factor}$ 

  \noindent 
  We present the derived LFs for all the clusters in Table \ref{tab:lumf}. 
  Column 1 provides the magnitude bin and the columns 3 to 10 provides 
  CF and NS values for the inner, outer, entire and field regions 
  respectively. The LFs corrected for the data incompleteness and the 
  field star contamination (corrected for area difference between cluster and
  field region) are given in the last three columns while the masses 
  corresponding to the center of mag bin derived 
  from the best fitting isochrones (see Sec. \ref{sec:cmd}) are 
  given in the 2nd column of Table \ref{tab:lumf}. For the outer region of
  NGC 2098, the MF could not be derived due to poor statistics.

  \begin{table*}
  \centering
  \caption{Derived reddening, $E(B-V)$ and age parameters from the CMDs
    are listed in 2nd and 3rd column
    for the sample clusters. MS gap parameters \eg location magnitude 
    ($V_{0}$), gap width ($\Delta V_{0}$) and the color ($(B-V)_{0}$) are
    given in columns 4-6. The MF slopes for the 'Ring 1', 'Ring 2' and 'Total'
    cluster region are listed in remaining columns. MF slopes have been 
    derived using the $V, (V-R)$ diagrams.} 
   \label{tab:age}

  \begin{tabular}{lllllcccc} \hline 
     Cluster &E(B-V)&   Age&  \multicolumn{3}{c}{MS gap parameters}& \multicolumn{3}{c}{MF slopes}\\ \cline{4-6} \cline{7-9} 
             & (mag)& (Myr)& $V_{0}$& $\Delta V_{0}$&  $B-V_{0}$&  Ring 1&    Ring 2&    Total\\ \hline
     NGC 1767& 0.18&$23\pm 3$&$-4.14$& $0.45$&$-0.20$& $-1.46\pm0.36$& $-1.39\pm0.54$ & $-1.23\pm0.27$ \\
     NGC 1994& 0.10&$16\pm 2$&$-5.08$& $0.50$&$-0.18$& $-2.56\pm0.36$& $-2.00\pm0.52$ & $-2.15\pm0.31$ \\
     NGC 2002& 0.20&$18\pm 2$&$-5.07$& $0.50$&$-0.20$& $-1.73\pm0.43$& $-2.50\pm0.20$ & $-2.28\pm0.21$ \\
     NGC 2003& 0.10&$25\pm 3$&$-3.61$& $0.20$&$-0.22$& $-2.37\pm0.24$& $-2.28\pm0.21$ & $-2.22\pm0.18$ \\
             &     &         &$-2.76$& $0.20$&$-0.23$&               &                &                \\
             &     &         &$-1.69$& $0.15$&$-0.21$&               &                &                \\
     NGC 2006& 0.06&$25\pm 3$&$-3.07$& $0.35$&$-0.23$& $-1.25\pm0.16$& $-2.62\pm0.53$ & $-1.90\pm0.16$ \\
     SL 538  & 0.06&$20\pm 2$&$-2.74$& $0.30$&$-0.23$& $-2.33\pm0.32$& $-1.82\pm0.68$ & $-2.03\pm0.41$ \\
     NGC 2011& 0.08&$20\pm 2$&$-4.08$& $0.28$&$-0.21$& $-1.72\pm0.54$& $-2.41\pm0.43$ & $-2.01\pm0.50$ \\
             &     &         &$-2.59$& $0.08$&$-0.22$&               &                &                \\
     NGC 2098& 0.08&$32\pm 4$&$-3.39$& $0.56$&$-0.19$& $-1.73\pm0.13$&       --       & $-2.19\pm0.19$ \\
     NGC 2136& 0.10&$90\pm10$&-&- &-                 & $-2.13\pm0.39$& $-3.00\pm0.49$ & $-2.22\pm0.20$ \\
     \hline
  \end{tabular}

  \noindent
  \end{table*}

  \begin{figure}
  \centering
  \includegraphics[width=8cm]{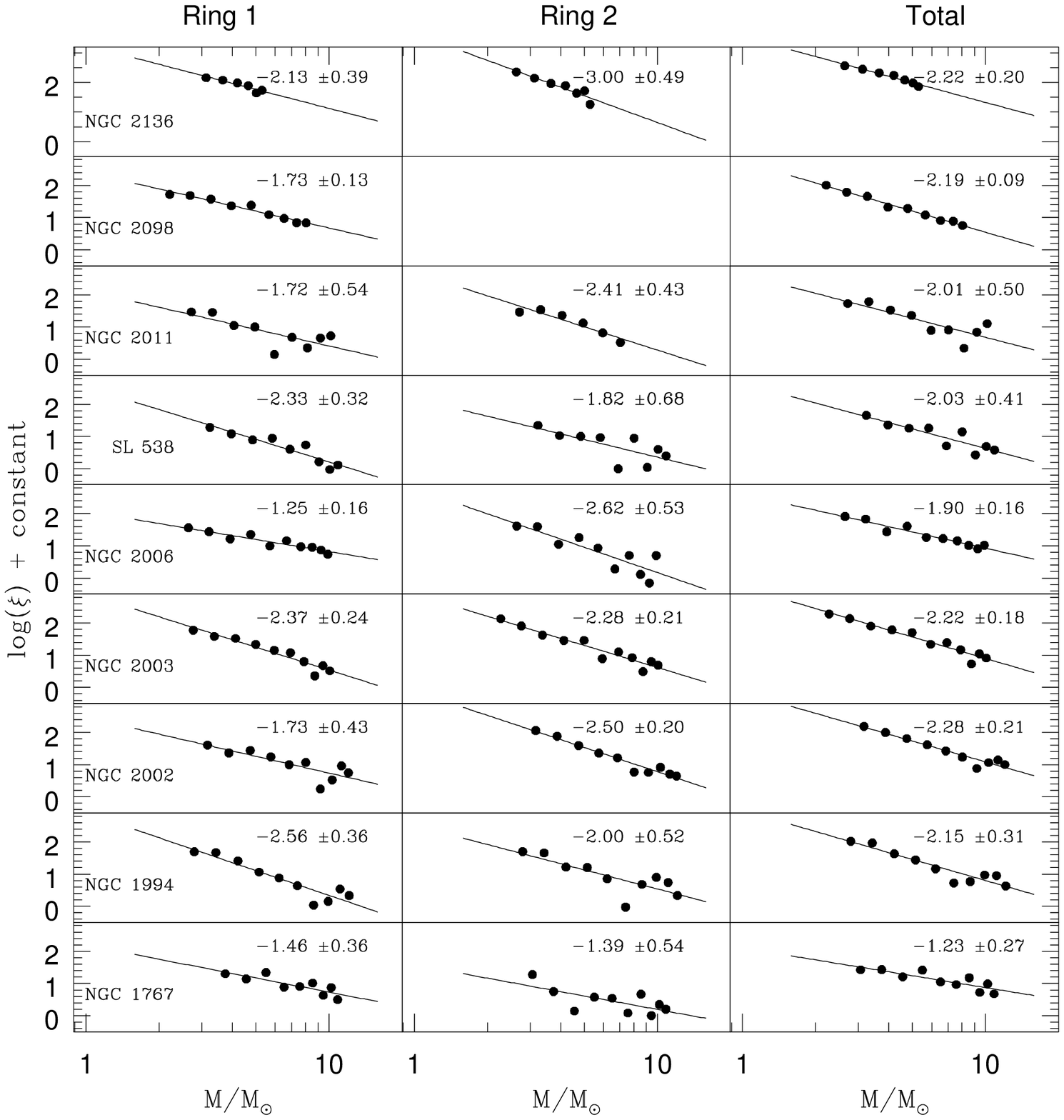}
  \caption{Mass function derived using \citet{girardi02} isochrones
     for 'Ring 1', 'Ring 2' and 'Total' cluster region are shown with open 
     circles. Solid lines denote best fit straight line along with
     the value of slopes displayed in respective panels. Due to low
     statistics, MF could not be derived for the outer region of
     NGC 2098.}
  \label{fig:massf}
  \end{figure}


\section {Results and discussions} \label{sec:results}

  To convert the LFs into mass functions (MFs), we divide the number given
  in Table \ref{tab:lumf} by the mass interval, $\Delta M$, of the magnitude
  bin under consideration. The value of $\Delta M$ is obtained from the 
  mass luminosity relation derived from the appropriate isochrones. The
  resulting cluster MFs are plotted in Fig. \ref{fig:massf} and the slopes
  are given in Table \ref{tab:age}. The quoted uncertainties result from the 
  linear regression solution. The slope is derived from the 
  mass distribution function $\xi(M)$ which is assumed to be a 
  power law with index $\gamma$. If dN denotes the number of stars in a 
  bin with central mass $M$, then the value of $\gamma$ is determined 
  from the linear relation :
  
  $log(dN) = \gamma \times log(M) + constant$

  \noindent
  $\gamma$ is also denoted as $-(1+x)$ in the literature with 
  $\gamma = -2.35$, or $x=1.35$ being \citet{salpeter55} value.

  For NGC 2002, NGC 2006, NGC 2011 and NGC 2136, the MF slopes for inner 
  (Ring 1) and outer (Ring 2) cluster regions differ by about one dex and 
  is shallower for the inner cluster region. 
  In case of NGC 2098 too, the MF slope for the entire region is steeper than 
  the inner region. However, due to poor statistics the MF slope for the outer 
  region could not be derived. \citet{kontizas98} found similar trend for SL 
  566 and NGC 2098, i.e. shallower LF slopes in the inner regions of the 
  clusters. This could be interpreted as mass segregation (high concentration 
  of heavier stars in the central region) and it may arise due to star 
  formation or dynamical evolution processes. As the ages of the clusters under 
  discussion are less than the dynamical relaxation time, the observed 
  variation may be an imprint of star formation. However, we note that the
  combined effect of scatter in MF slope may be as large as one dex and
  hence, we suggest HST observations 
  to resolve the stars of the cluster core region and to confirm the radial 
  variation of MF slope. For further discussion, we consider only the MF slope 
  derived for the total cluster region. 

  The mass range for the sample clusters are similar and vary from $\sim$
  $2$ to $12$ $M_{\odot}$ except for NGC 2136 where it is only $2$ to $6$ 
  $M_{\odot}$. As the ages of all the clusters are less than the dynamical 
  evolution times ($\sim 100$ Myrs), the slope of the present day MFs can be 
  considered as slope of IMFs. Furthermore, we also assume that all the stars
  in the cluster are formed in a single star-forming bursts and hence barring
  most evolved stars, the derived MFs could be least affected by the star 
  formation history of the clusters. Excluding NGC 1767, we get a mean MF slope
  of $-2.13\pm0.14$ for 8 target clusters. This value is not too different 
  from the Salpeter value derived for solar neighbourhood stars and for other 
  young galactic and M33 star clusters in the intermediate mass range (cf. 
  \citealt{sagar00, sagar02, chen07}). The mass function slope for NGC 1767
  was found to be significantly flatter ($\gamma \sim -1.23$) than the 
  Salpeter value.

\begin{figure}
  \centering
  \includegraphics[width=8cm]{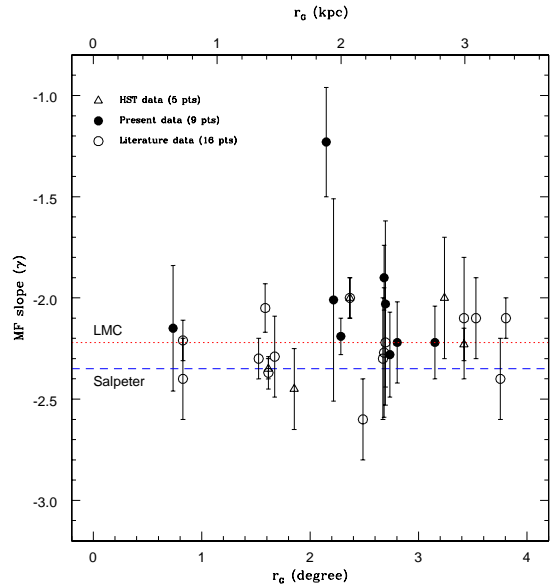}
  \caption{Plot of MF slopes against the 
     galactocentric distance ($r_{\rm G}$) in LMC.  Including estimates
     from the literature (see text), the MF slopes for 26 young ($< 100$ Myr) 
     star clusters and associations are shown. The horizontal dashed line 
     represent the \citet{salpeter55} value of IMF slope for the field stars 
     in solar neighbourhood while dotted line indicate the mean slope for a 
     sample of 25 (excluding one outlier) LMC star clusters and associations.}
  \label{fig:mfslopes}
  \end{figure}

  Our MF slopes $-1.90\pm0.16$ for NGC 2006 and $-2.03\pm0.41$ for SL 538 are 
  consistent with the corresponding values of $2.27\pm0.32$ and $-2.22\pm0.31$ 
  derived by \citet{dieball98}. Fig. \ref{fig:mfslopes}, shows the variation of 
  MF slopes with galactocentric distance for 26 young ($< 100$ Myr) star 
  clusters and associations in the LMC. This includes 9 clusters from the 
  present work, while the data for other objects are taken from the 
  Table 1 of \citet{sagar00}. For four clusters, we have two 
  estimates for the MF slope and these points are also shown in 
  Fig.\ref{fig:mfslopes}. 
  Regarding NGC 1767 as an outlier, remaining sample of 25 has a mean MF 
  slope $\gamma = -2.22\pm0.16$ indicating that the MF slope in LMC clusters 
  are not significantly different from the \citet{salpeter55} value. 
  The scatter seen in Fig. \ref{fig:mfslopes} in MF slope is introduced by many 
  factors, for example, data incompleteness and field star contamination, 
  dynamical and stellar evolutionary state of star clusters, limited range in 
  mass and assumed model to derive mass-luminosity relation and Poisson noise 
  etc.  \citep{kroupa01, sagar91a, sagar02}. The detailed analysis indicate that
  the cumulative effect of the various uncertainties could be as large as 0.4 
  dex for young rich LMC star clusters \citep{sagar02}. We therefore conclude 
  that the scatter seen in Fig. \ref{fig:mfslopes} is real and it does indicate 
  the limitations of MF slope determinations from the ground based observations.
  Despite being situated in different locations of LMC the studied sample of 
  young clusters and associations supports the idea of some universal IMF as a 
  consequence of star formation processes in star clusters and associations.        


\section {Summary} \label{sec:summary}
 
  We present \ubvri CCD data obtained from 3.5-meter ESO NTT/EFOSC2 
  observations for 9 young Large Magellanic Cloud star clusters 
  namely \CLN\ and their nearby field regions reaching down to $V \sim 20$ mag 
  for $\sim$ 6400 stars altogether. They are the first accurate broad band 
  CCD photometric data for all the clusters except for the binary cluster 
  NGC 2006 and SL 538. The observations are made in a region 
  of $\sim 2\arcmin \times 2\arcmin$ around the cluster center. The data 
  were collected during Jan 10 to Jan 13, 1990 in good seeing conditions 
  ranging from 0\farcs7 to 1\farcs0 and reduced using DAOPHOT and MIDAS 
  softwares. Photometric calibrations are done using Landolt (1992) stars 
  and the zero point accuracy is better than $0.02$ mag. Photometric errors 
  become large (\ga 0.1 mag) for stars fainter than $V = 20$ mag. 

  We examine radial density profiles, general features of the main 
  sequence and estimate age and reddening for individual clusters 
  using Padova isochrones. The various CMDs of the clusters under study were 
  used to estimate their MF, age and reddening. In order to study radial 
  variation in MF, the LFs are derived for inner, outer, and entire cluster 
  regions. Due to compactness of the clusters, such study could not be 
  carried out for the core regions of the clusters. The LFs are corrected 
  for both data incompleteness and field contamination. The main conclusions 
  of the present study are as follows.   

  \begin{enumerate}

  \item
   Seven of the nine clusters have ages $\le 25$ Myr, while the 
   remaining two clusters have ages of $32\pm4$ Myr (NGC 2098) 
   and $90\pm10$ Myr (NGC 2136). Our age estimates for NGC 2006 and SL 538 
   were found to be consistent with the previous $BVR$ photometric estimate 
   by \citet{dieball98}. For NGC 2098, our estimates are lower by about 30 Myr 
   than \citet{kontizas98}. Thus, the ages of all the clusters in our sample 
   are significantly lower than their typical dynamical ages of a few 100 Myrs.
   
   \item 
   For younger ($\le$ 25 Myr) clusters, the age estimates based on 
   a recent population synthesis models by \citet{wolf07} and integrated 
   spectra are systematically lower by about 10 Myr than the present age 
   estimates based on CMDs. 
   
   \item Assuming an LMC distance modulus of 18.5 mag, the derived 
   reddening for the  clusters in our sample was found to be consistent 
   with that derived from HI emission and 100 $\mu m$ all sky dust maps.

   \item 
   In the mass range of $2 - 12\ M_{\odot}$, the MF slopes for 8 out of 9 
   sample clusters were found to be similar with values of $\gamma$ ranging 
   from  $-1.90\pm0.16$ to  $-2.28\pm0.21$. For NGC 1767 the slope was 
   found to be significantly shallower with  $\gamma = -1.23\pm0.25$. The 
   present MF values are consistent with those derived by \citet{dieball98} 
   for NGC 2006 and SL 538. \citet{selman05} studied the star formation 
   history and IMF of the field population of 30 Doradus super association 
   and found that it has a Salpeter slope in the mass range of 7 
   to 40 $M_{\odot}$.

   \item 
   Mass function slopes of the inner and outer cluster regions indicates the
   presence of mass segregations in NGC 2002, NGC 2006, NGC 2136 and NGC 2098. 
   For NGC 2098, \citet{kontizas98} derive the dynamical relaxation time, 
   $T_{\rm e}$ 
   between 640 to 1050 Myr. This may indicate that the value of $T_{\rm e}$ for 
   LMC star clusters could be few hundreds of Myr. The ages of LMC star 
   clusters under study are therefore significantly smaller than their 
   dynamical relaxation time. Consequently, observed mass segregation 
   in these clusters is probably  primordial in nature. A compilation of both 
   ground and space based observations of extremely young galactic and MC star 
   clusters (cf.  \citealt{hunter95}; \citealt{sagar88}; 
   \citealt{hillenbrand98}; \citealt{chen07} and references therein) 
   indicates presence of mass 
   segregation in most of them, although to varying degrees. All these indicate
   that in most of the young star clusters located in different galaxies, mass
   segregation effects are observed and most likely they are imprint of star
   formation processes.

   \item 
   A mean MF slope of $\gamma = -2.22\pm0.16$ derived for a sample of 25 
   young ($< 100$ Myr) stellar systems in LMC provide support for the 
   universality of IMF in the intermediate mass range $\sim 2-10\ M_{\odot}$.
   An IMF study of the 30 Doradus star forming region of LMC by 
   \citet{selman05} also support this conclusion.

   \end{enumerate}


\section*{acknowledgments}
     Authors are thankful to the anonymous referee for constructive comments. 
     Useful discussions with Drs. K.S. de Boer, P. Kroupa and T. Richtler 
     are gratefully acknowledged. We thank Dr. Vijay Mohan for help in data 
     reduction. One of us (RS) would like to thank the Alexander von 
     Humboldt Foundation, Bonn for providing financial support to work at 
     the Sterwarte/Argelander Institute of Astronomy in Bonn. BK 
     acknowledge support from the Chilean center for Astrophysics
     FONDAP No. 15010003.   


\clearpage

\onecolumn
  \begin{longtable}{l r r c c c c c c c r r r }

  \caption[width=\textwidth]{\label{tab:lumf}  
  Luminosity functions for the sample LMC star clusters. The columns denote : col.(1) V mag of the bin center - 0.5 mag is
  taken as bin width; col.(2) mass of the bin center
  as read from best fit isochrones in $M_{\odot}$; col.(3) - (10) completeness factor (CF) and number of stars (NS) in the bin 
  for the 'Ring 1', 'Ring 2', 'Total' cluster region as well as the corresponding field region; col. (11) - (13) the 
  derived cluster 
  luminosity function for the above three regions of the cluster after correcting for data incompleteness and field star 
  contamination. For NGC 2098, the LF could not be derived for the outer region due to poor statistics.} \\

  \hline\hline
  \noalign{\smallskip}
     Bin  &     Bin   &  \multicolumn{2}{l}{Ring 1}& \multicolumn{2}{l}{Ring 2}&  \multicolumn{2}{l}{Total}& 
                    \multicolumn{2}{l}{Field Region}& Ring 1& Ring 2&  Total\\ 
        \cline{3-4}
        \cline{5-6}
        \cline{7-8}
        \cline{9-10}
  (V mag)&  ($M_{\odot}$)&      CF&    NS&    CF&  NS& CF& NS& CF&  NS&  LF& LF& LF\\ 
    (1)&       (2)&      (3)&   (4)&    (5)& (6)& (7)& (8)& (9)& (10)& (11)& (12)& (13)\\
  \noalign{\smallskip}
  \hline

  \endfirsthead

  \caption{continued.}\\

  \hline\hline
  \noalign{\smallskip}
        &          &  \multicolumn{2}{l}{Ring 1}& \multicolumn{2}{l}{Ring 2}&  \multicolumn{2}{l}{Total}& 
                    \multicolumn{2}{l}{Field Region}& Ring 1& Ring 2&  Total\\ 
        \cline{3-4}
        \cline{5-6}
        \cline{7-8}
        \cline{9-10}
    Mag.&      Mass&      CF&    NS&    CF&  NS& CF& NS& CF&  NS&  LF& LF& LF\\ 
    (1)&       (2)&      (3)&   (4)&    (5)& (6)& (7)& (8)& (9)& (10)& (11)& (12)& (13)\\
  \noalign{\smallskip}
  \hline

  \endhead
  \hline
  \endfoot

  \noalign{\smallskip}

 \multicolumn{13}{l}{NGC 1767}\\
 15.25& 10.83&    1.00&   2&  1.00&   1&  1.00&   3&  1.00&   0&   2.00&   1.00&   3.00\\
 15.75& 10.19&    1.00&   5&  1.00&   2&  1.00&   7&  1.00&   1&   4.80&   1.42&   6.22\\
 16.25&  9.45&    1.00&   4&  1.00&   2&  1.00&   6&  1.00&   2&   3.60&   0.84&   4.44\\
 16.75&  8.56&    1.00&  10&  1.00&   5&  1.00&  15&  1.00&   1&   9.80&   4.42&  14.22\\
 17.25&  7.57&    1.00&   9&  1.00&   3&  1.00&  12&  1.00&   3&   8.40&   1.26&   9.66\\
 17.75&  6.52&    1.00&   9&  1.00&   6&  1.00&  15&  1.00&   4&   8.21&   3.68&  11.88\\
 18.25&  5.50&    0.99&  22&  1.00&   6&  0.98&  28&  1.00&   4&  21.43&   3.68&  25.45\\
 18.75&  4.56&    0.97&  16&  1.00&  13&  0.96&  29&  0.99&  20&  12.48&   1.26&  14.46\\
 19.25&  3.74&    0.94&  19&  0.99&  19&  0.94&  38&  0.97&  25&  15.10&   4.21&  20.33\\
 19.75&  3.06&    0.89&   8&  0.90&  24&  0.88&  32&  0.88&  23&      -&  11.48&  15.99\\
 \multicolumn{13}{l}{NGC 1994}\\
 15.25& 12.09&    1.00&   2&  1.00&   2&  1.00&   4&  1.00&   0&   2.00&   2.00&   4.00\\
 15.75& 11.08&    1.00&   4&  1.00&   8&  1.00&  12&  1.00&   1&   3.79&   6.02&   9.81\\
 16.25&  9.90&    1.00&   2&  1.00&  12&  1.00&  14&  1.00&   1&   1.79&  10.02&  11.81\\
 16.75&  8.64&    1.00&   2&  1.00&  12&  1.00&  14&  1.00&   3&   1.36&   6.06&   7.42\\
 17.25&  7.40&    1.00&   7&  1.00&  17&  1.00&  24&  1.00&   8&   5.30&   1.15&   6.45\\
 17.75&  6.22&    1.00&   9&  1.00&  12&  1.00&  21&  1.00&   2&   8.57&   8.04&  16.61\\
 18.25&  5.15&    1.00&  14&  1.00&  38&  1.00&  52&  1.00&  11&  11.66&  16.21&  27.87\\
 18.75&  4.21&    0.99&  25&  0.99&  42&  0.98&  67&  0.99&  14&  22.24&  14.41&  37.34\\
 19.25&  3.42&    0.96&  35&  0.97&  67&  0.96& 102&  0.96&  18&  32.47&  31.92&  65.11\\
 19.75&  2.79&    0.82&  29&  0.90&  91&  0.86& 120&  0.92&  34&  27.50&  27.89&  58.45\\
 \multicolumn{13}{l}{NGC 2002}\\
 15.25& 11.99&    1.00&   4&  1.00&   6&  1.00&  10&  1.00&   5&   3.67&   2.88&   6.55\\
 15.75& 11.24&    1.00&   8&  0.96&   6&  1.00&  14&  1.00&   3&   7.80&   4.38&  11.93\\
 16.25& 10.29&    1.00&   4&  1.00&  13&  1.00&  17&  1.00&   7&   3.53&   8.64&  12.17\\
 16.75&  9.19&    0.99&   3&  1.00&  16&  0.99&  19&  1.00&  15&   2.03&   6.65&   8.84\\
 17.25&  8.02&    0.96&  14&  0.98&  15&  0.98&  29&  0.96&  13&  13.68&   6.87&  20.24\\
 17.75&  6.85&    0.90&  11&  0.88&  23&  0.88&  34&  0.94&  11&  11.44&  18.84&  30.56\\
 18.25&  5.75&    0.80&  16&  0.82&  31&  0.80&  47&  0.88&  19&  18.55&  24.35&  43.85\\
 18.75&  4.74&    0.58&  17&  0.62&  40&  0.62&  57&  0.82&  36&  26.37&  37.15&  61.63\\
 19.25&  3.87&    0.45&  11&  0.56&  65&  0.54&  76&  0.76&  67&  18.54&  61.12&  79.89\\
 19.75&  3.16&    0.40&  13&  0.46&  63&  0.45&  76&  0.62&  64&  25.59&  72.62&  97.64\\
 \multicolumn{13}{l}{NGC 2003}\\
 15.25& 10.06&    1.00&   2&  1.00&   3&  1.00&   5&  1.00&   0&   2.00&   3.00&   5.00\\
 15.75&  9.43&    1.00&   3&  1.00&   4&  1.00&   7&  1.00&   0&   3.00&   4.00&   7.00\\
 16.25&  8.73&    1.00&   2&  1.00&   5&  1.00&   7&  1.00&   2&   1.76&   2.38&   4.14\\
 16.75&  7.88&    1.00&   6&  1.00&   9&  1.00&  15&  1.00&   1&   5.88&   7.69&  13.57\\
 17.25&  6.93&    1.00&  12&  1.00&  14&  1.00&  26&  1.00&   1&  11.88&  12.69&  24.57\\
 17.75&  5.94&    1.00&  14&  1.00&   9&  1.00&  23&  1.00&   1&  13.88&   7.69&  21.57\\
 18.25&  4.99&    1.00&  20&  1.00&  28&  1.00&  48&  1.00&   1&  19.88&  26.69&  46.57\\
 18.75&  4.12&    0.99&  28&  1.00&  34&  1.00&  62&  1.00&   8&  27.33&  23.53&  50.58\\
 19.25&  3.37&    0.96&  26&  1.00&  43&  0.99&  69&  0.99&  11&  25.77&  28.45&  53.83\\
 19.75&  2.76&    0.86&  30&  0.98&  68&  0.96&  98&  0.98&  19&  32.59&  44.01&  74.40\\
 20.25&  2.27&    0.78&  20&  0.90&  91&  0.86& 111&  0.96&  30&      -&  60.20&  84.46\\
 \multicolumn{13}{l}{NGC 2006}\\
 15.25&  9.88&    0.99&   3&  0.99&   3&  0.99&   6&  1.00&   2&   2.91&   2.66&   5.57\\
 15.75&  9.27&    0.94&   5&  0.96&   1&  0.95&   6&  1.00&   3&   5.14&   0.48&   5.58\\
 16.25&  8.52&    0.91&   7&  0.92&   2&  0.92&   9&  1.00&   6&   7.33&   1.06&   8.30\\
 16.75&  7.65&    0.88&   8&  0.91&   5&  0.90&  13&  0.99&   4&   8.85&   4.74&  13.45\\
 17.25&  6.68&    0.85&  13&  0.88&   4&  0.84&  17&  0.99&  14&  14.44&   1.91&  16.75\\
 17.75&  5.70&    0.80&   8&  0.83&   8&  0.82&  16&  0.98&   7&   9.57&   8.31&  17.75\\
 18.25&  4.76&    0.73&  16&  0.80&  17&  0.76&  33&  0.96&  26&  20.28&  16.20&  36.74\\
 18.75&  3.92&    0.68&  10&  0.78&  11&  0.73&  21&  0.92&  26&  13.00&   8.84&  21.79\\
 19.25&  3.21&    0.61&  12&  0.70&  23&  0.66&  35&  0.88&  39&  17.00&  24.60&  42.09\\
 19.75&  2.64&    0.45&  10&  0.64&  20&  0.54&  30&  0.82&  44&  18.98&  21.25&  42.31\\
 \multicolumn{13}{l}{SL 538}\\
 15.25& 10.86&    1.00&   1&  1.00&   2&  1.00&   3&  1.00&   2&   0.89&   1.71&   2.61\\
 15.75& 10.06&    0.99&   1&  1.00&   4&  1.00&   5&  1.00&   3&   0.85&   3.57&   4.41\\
 16.25&  9.08&    0.98&   2&  0.99&   2&  1.00&   4&  1.00&   6&   1.72&   1.16&   2.82\\
 16.75&  8.01&    0.97&   6&  0.98&  10&  0.99&  16&  0.99&   4&   5.97&   9.62&  15.36\\
 17.25&  6.90&    0.96&   5&  0.96&   3&  0.95&   8&  0.99&  14&   4.44&   1.10&   5.63\\
 17.75&  5.83&    0.94&   9&  0.94&  10&  0.94&  19&  0.98&   7&   9.19&   9.61&  18.80\\
 18.25&  4.84&    0.91&   8&  0.90&  12&  0.90&  20&  0.96&  26&   7.33&   9.45&  16.87\\
 18.75&  3.96&    0.88&  10&  0.86&  11&  0.87&  21&  0.92&  26&   9.84&   8.74&  18.56\\
 19.25&  3.23&    0.82&  12&  0.78&  16&  0.74&  28&  0.88&  39&  12.24&  14.15&  29.09\\
 \multicolumn{13}{l}{NGC 2011}\\
 15.75& 10.16&    1.00&   5&  1.00&   9&  1.00&  14&  1.00&   2&   4.72&      -&  11.18\\
 16.25&  9.20&    1.00&   5&  1.00&   5&  1.00&  10&  1.00&   2&   4.72&      -&   7.18\\
 16.75&  8.14&    1.00&   3&  1.00&   5&  1.00&   8&  1.00&   4&   2.44&      -&   2.37\\
 17.25&  7.04&    0.99&   6&  1.00&  10&  1.00&  16&  1.00&   5&   5.36&   3.66&   8.96\\
 17.75&  5.96&    0.98&   2&  1.00&  12&  1.00&  14&  1.00&   4&   1.48&   6.93&   8.37\\
 18.25&  4.95&    0.96&  10&  0.99&  20&  0.99&  30&  1.00&   6&   9.58&  12.60&  21.86\\
 18.75&  4.06&    0.95&  10&  0.98&  30&  0.98&  40&  0.99&   9&   9.25&  19.09&  28.02\\
 19.25&  3.31&    0.90&  19&  0.95&  39&  0.95&  58&  0.98&  14&  19.11&  22.94&  40.94\\
 19.75&  2.71&    0.87&  18&  0.91&  54&  0.93&  72&  0.96&  33&  15.88&  15.77&  29.03\\
 \multicolumn{13}{l}{NGC 2098}\\
 15.75&  8.56&    1.00&   6&  1.00&   1&  1.00&   7&  1.00&   7&   5.29&  -&   3.91\\
 16.25&  8.03&    1.00&   5&  1.00&   2&  1.00&   7&  1.00&   8&   4.18&  -&   3.47\\
 16.75&  7.35&    1.00&   7&  1.00&   6&  1.00&  13&  1.00&  16&   5.37&  -&   5.93\\
 17.25&  6.53&    1.00&  10&  1.00&   5&  1.00&  15&  1.00&  18&   8.16&  -&   7.05\\
 17.75&  5.65&    1.00&  14&  1.00&  10&  1.00&  24&  1.00&  30&  10.94&  -&  10.75\\
 18.25&  4.78&    0.99&  24&  0.99&   8&  0.99&  32&  1.00&  36&  20.57&  -&  16.42\\
 18.75&  3.96&    0.98&  21&  0.98&  10&  0.98&  31&  1.00&  35&  17.85&  -&  16.17\\
 19.25&  3.26&    0.96&  28&  0.97&  23&  0.97&  51&  1.00&  54&  23.65&  -&  28.73\\
 19.75&  2.68&    0.83&  26&  0.94&  28&  0.93&  54&  0.99&  59&  25.24&  -&  31.74\\
 20.25&  2.21&    0.74&  22&  0.88&  43&  0.88&  65&  0.96&  66&  22.71&  -&  43.50\\
 \multicolumn{13}{l}{NGC 2136}\\
 16.75&  5.29&    1.00&  12&  1.00&   4&  1.00&  16&  1.00&   0&  12.00&   4.00&  16.00\\
 17.25&  5.02&    1.00&  14&  1.00&  18&  1.00&  32&  1.00&   3&  13.75&  16.19&  29.94\\
 17.75&  4.65&    1.00&  33&  1.00&  20&  1.00&  53&  1.00&   3&  32.75&  18.19&  50.94\\
 18.25&  4.19&    0.99&  49&  1.00&  41&  1.00&  90&  1.00&   2&  49.33&  39.79&  88.63\\
 18.75&  3.65&    0.96&  65&  0.99&  55&  0.98& 120&  1.00&   6&  67.21&  51.94& 118.33\\
 19.25&  3.12&    0.94&  70&  0.97&  74&  0.97& 144&  0.98&   8&  73.78&  71.37& 142.85\\
 19.75&  2.63&    0.86&  59&  0.96& 111&  0.94& 170&  0.97&  21&  66.79& 102.58& 165.99\\
   
  \noalign{\smallskip}

\end{longtable}

\twocolumn

\bsp

\label{lastpage}

\end{document}